\documentclass[a4paper,11pt]{article}

\usepackage{classeJBArxive}

\usepackage{arydshln}
\usepackage{algorithm}
\usepackage[noend]{algpseudocode}

\usepackage{MnSymbol}
\usepackage{movie15}

\usepackage{setspace}

\newcommand*\unit[1]{\bigl[\, \mathsf{#1} \,\bigr]}

\newcommand{\RK}{\textsc{Runge}--\textsc{Kutta}}

\newcommand{\cpu}{\mathrm{cpu}}
\newcommand{\sky}{\medstar}

\newcommand{\set}{\mathrm{set}}

\newcommand{\Bi}{\mathrm{Bi}}
\newcommand{\Fo}{\mathrm{Fo}}

\newcommand{\tf}{t_{\,\mathrm{f}}}

\newcommand{\degC}{^{\,\circ}C}

\newcommand{\eqsvd}{\mathop{\stackrel{\,\mathrm{svd}}{=}\,}}

\newcommand{\bPsi}{\boldsymbol{\Psi}}

\title{
\vspace{-1.5cm}
Assessing the wall energy efficiency design under climate change using POD reduced order model \\	
\vspace{4pt}
}

\author{Julien Berger \textsuperscript{a}$^{\ast}$, Cyrille Allery\textsuperscript{a}, Anaïs Machard\textsuperscript{a}
\date{\today\vspace{-0.5cm}}}

\begin{document}


\maketitle

\begin{center}
\small
\textsuperscript{a} Laboratoire des Sciences de l’Ingénieur pour l’Environnement (LaSIE), UMR 7356 CNRS, La Rochelle Université, CNRS, 17000, La Rochelle, France\\
\end{center}

\begin{abstract}

Within the environmental context, numerical modeling is a promising approach to assess the energy efficiency of building. Resilient buildings need to be designed, capable of adapting to future extreme heat. Simulations are required assuming a one-dimensional heat transfer problem through walls and a simulation horizon of several years (nearly 30). The computational cost associated \revision{with} such modeling is quite significant and model reduction methods are worth \revision{investigating}. The objective is to propose a reliable \revision{reduced-order} model for such \revision{long-term} simulations. For this, an alternative model reduction approach is investigated, assuming a known Proper Orthogonal Decomposition \revision{reduced basis for} time, and not for space as usually. The model enables \revision{computing parametric solutions} using basis interpolation on the tangent space of the \textsc{Grassmann} manifold. Three study cases are considered to verify the efficiency of the \revision{reduced-order} model. Results highlight that the model has a satisfying accuracy of $10^{\,-3}\,$ compared to reference solutions. The last case study focuses on the wall energy efficiency design under climate change according to a \revision{four-dimensional} parameter space. The latter is composed of the load material emissivity, heat capacity, thermal conductivity and thickness insulation layer. \revision{Simulations} are carried over $30$ years considering climate change. The solution \revision{minimizing} the wall work rate is determined with a computational ratio of $0.1\%$ compared to standard approaches.



\end{abstract}

\section{Introduction}

Within the environmental context, it becomes a major issue to focus on reducing \revision{building} energy consumption \cite{us_energy_information_administration_annual_2021}. The latter is responsible \revision{for one-third} of the \revision{world's} global energy consumption. With climate change, resilient buildings need to be designed, capable of adapting to future extreme heat. Numerical tools such as building simulation programs have been elaborated to predict \revision{energy} efficiency and help \revision{building's} design. An extensive review of such programs is proposed in \cite{mendes_numerical_2019,mirzaei_recent_2015}. 

Those numerical tools \revision{intend} to model several physical phenomena. Among others, the heat transfer process through porous walls is one of the most important since building energy consumption emanates mostly from this component. To assess these phenomena, several numerical methods have been developed and included in the building simulation programs as presented in \cite{Delgado_2010}. The methods are based on classical numerical approaches such as finite-differences or finite-volumes. In terms of time integration, both \textsc{Euler} implicit and explicit ~schemes are employed \cite{tariku_transient_2010,mendes_method_2005,kalagasidis_international_2007,MAIA_2021,Guimaraes_2017}. However, those methods \revision{face} important issues. Particularly within the engineering context of building wall design, one aims at performing several parametric studies. Particularly when considering the climate change context, \revision{simulations} are carried \revision{out} over several years and parametric investigations become computationally costly.

Model reduction methods are worth investigations to cut the computational burden of classical approaches. It represents interesting strategies since they intend to compute solutions with a lower computational cost while preserving the whole complexity of the physical phenomena. Lumped models, such as lumped capacitance ones do have a lower computational cost but the \revision{predictions'} reliability are questionable as remarked in \cite{berger_comparison_2020,berger_evaluation_2019-1}. In \revision{recent years}, several works have been published to propose \revision{reduced-order} models for the computation of diffusion problems in building physics \cite{berger_review_2017}. The efficiency of a \revision{reduced-order} model can be stated according to three criteria: \emph{(i)} the accuracy of the solution, \emph{(iii)} the computational reduction and \emph{(iii)} the ability \revision{to perform} parametric computations. 

In \cite{berger_estimation_2020} the heat conduction transfer is computed according to the diffusivity of the layer through a Modal Identification Method (MIM) \revision{reduced-order} model. Other works with this method \revision{arise} from the primer works of \textsc{Girault}, \textsc{Petit}, and \textsc{Videcoq}  in \cite{Petit_1997,Girault_2003}. Other applications \revision{consist} in identifying boundary fluxes or surface temperature in \cite{Girault_2004a,Girault2005a,Girault2005b,Gerardin_2017}. In \cite{berger_proper_2015,berger_2d_2016}, the Proper Generalized Decomposition (PGD) is used to compute the heat and mass transfer in porous walls for one and two-dimensional problems. A Spectral reduced method is employed to solve similar problems in \cite{gasparin_spectral_2019,berger_comparison_2020}. The well-known Proper Orthogonal Decomposition (POD) is also used in \cite{berger_evaluating_2019,Hou_2020}. In \cite{hou_poddeim_2020} the POD is combined with discrete empirical interpolation to solve diffusive \revision{problems}. Parametric solutions are elaborated with both PGD and POD methods. In \cite{berger_innovative_2017}, the wall energy efficiency \revision{is} computed according to the thickness and thermal diffusivity of the layers. A parametric solution is proposed in \cite{azam_mixed_2018} to compute the heat transfer in soil within the urban environment. In \cite{berger_intelligent_2018}, the two-dimensional heat transfer is solved for different climatic boundary conditions. 
All those methods (except the MIM) are based on a separated representation of the solution. In other words, the solution is assumed by a sum up to $N$ modes of \revision{function} product, where each basis depends on one coordinate of the problem (space, time and \revision{sometimes} parameter). All those \revision{reduced-order} models show very satisfying accuracy and computational reduction compared to standard approaches. 

However, one drawback can be highlighted \revision{in} the degree of freedom of the solution (See Section~\ref{sec:metrics} for a precise definition), which is directly related to the computational cost. The above-mentioned model reduction methods require \revision{determining} the \revision{reduced basis} related to the time coordinate. For instance, the POD or Spectral approaches set the space \revision{reduced basis} using preliminary solution (POD) or orthogonal polynomials (Spectral) and only the temporal coefficients need to be determined. Nevertheless, when performing wall design, the diffusion model is generally assumed as one-dimensional and the simulations are carried out for several years. Thus, the solution degree of freedom scales with $\mathcal{O}(\,10^{\,2}\,)$ for space\footnote{considering a sufficiently refined grid to reach accuracy.} and  with $\mathcal{O}(\,10^{\,4}\,)$ for time\footnote{At least $8860$ for one year since the climatic boundary conditions are given with one hour time step.}. Thus, by setting the \revision{reduced basis for} space and determining the temporal coefficient, the degree of freedom of the model reduction solution is still high, at the order $\mathcal{O}(\,10^{\,4}\,)\,$. 

Regarding the \revision{reduced-model} parametric solutions, the PGD focuses on directly computing the \revision{reduced basis} related to the parameter of interest. In the works related to the POD \cite{berger_intelligent_2018,berger_evaluating_2019,Hou_2020}, the solution is not parametric \emph{per se}. Recent works \cite{azam_mixed_2018,berger_intelligent_2018} investigate the accuracy of the POD basis by using the same \revision{reduced basis for space} for computations with different parameters. However, the accuracy of this approach can be very low in some cases as noted in \cite{berger_evaluating_2019}. To overcome this difficulty, it is possible to interpolate the POD basis function according to the parameter. \revision{Direct} standard \revision{approaches} such as \textsc{Lagrange} interpolation are often ineffective for basis interpolation. That is why, in the context of \revision{reduced-order} models in fluid mechanics, \revision{a strategy} based on interpolation on a tangent space of \textsc{Grassmann} manifold has been recently developed to compute basis functions associated \revision{with} new parameters \cite{mosquera_generalization_2021,mosquera_pod_2019,freno_using_2013,oulghelou_parametric_2021,amsallem_interpolation_2008}.

Thus, the objective of this article is twofold. \revision{First,} it explores an alternative strategy where the time \revision{reduced basis} is set and the spatial coefficients are the unknowns to be computed. The reduced basis is built using a preliminary computed solution \revision{for} the classical POD. Then, a boundary value problem is defined to compute the spatial coefficients of the reduced model. The idea originates from recent works on the method of horizontal lines proposed in \cite{gasparin_adaptive_2018}. Then, a step further is proposed compared to approaches from the literature in building physics. Secondly, the parametric dependence of the problem is done by using interpolation based on a tangent space to a \textsc{Grassmann} manifold. 

The article is structured as follows. \revision{Next,} Section presents the physical model of heat transfer in \revision{multi-layers walls}. Then, Section~\ref{sec:numerical_model} presents the model reduction methods and the basis interpolation approach. Section~\ref{sec:validation_case_study} proposes three case study used to verify the \revision{reduced-order} model efficiency \cite{oreskes_verification_1994}. Last, Section~\ref{sec:real_case_study} uses the reduced order model for a parametric wall energy efficiency design over 10 years considering \revision{the} climatic change.

\section{Description of the Physical model}
\label{sec:mathematical_model}

\revision{First, the model describing the physical phenomenon occurring in the wall is described.}

\subsection{Physical model}

\subsubsection{Space and time domains}

\begin{figure}[h!]
\centering
\includegraphics[width=.9\textwidth]{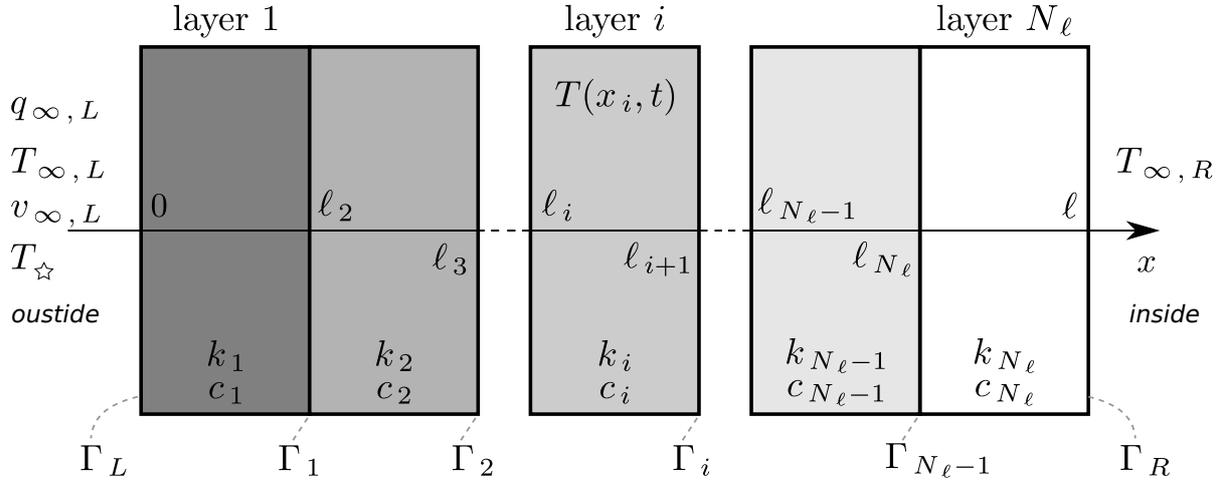}
\caption{Illustration of the multi-layer domain.}
\label{fig:domain}
\end{figure}

The problem involves heat transfer through the facade of a buildings composed of several materials. The domain is defined by $\Omega_{\,x} \egal \bigl[\, 0 \,,\, \ell \,\bigr]\,$, where $\ell \ \unit{m}$ is the length of the wall. The wall is composed of $N_{\,\ell}$ layers as illustrated in Figure~\ref{fig:domain}. Thus, the domain $\Omega_{\,x}$ is the union of several sub-domains:
\begin{align*}
\Omega_{\,x} \egal \bigcup_{i\egal 1}^{\,N_{\,\ell}} \Omega_{\,x}^{\,i} \,,
\end{align*}
where $\Omega_{\,x}^{\,i}$ corresponds to the sub-domains of the layer $i\,$:
\begin{align*}
\Omega_{\,x}^{\,i} \egal \bigl[\, \ell_{\,i} \,,\, \ell_{\,i+1} \,\bigr] \,,
\end{align*}
where by convention $\ell_{\,1} \egal 0$ and $\ell_{\,N_{\,\ell}\plus 1} \egal \ell\,$. Two types of domain boundaries are defined. The first ones corresponds to the interface between the wall and the ambient air: 
\begin{align*}
\Gamma_{\,L} \egal \bigl\{\,x \egal 0 \,\bigr\} \,, \qquad
\Gamma_{\,R} \egal \bigl\{\, x \egal \ell \,\bigr\} \,.
\end{align*}
The second boundary is related to the interface between two layers:
\begin{align*}
\Gamma_{\,i} \egal \bigl\{\, x \egal \ell_{\,i+1} \,\bigr\} \,,  \qquad \forall i \, \in \, \bigl\{\, 1 \,,\, \ldots \,,\, N_{\,\ell}-1 \,\bigr\} \,.
\end{align*}
Regarding, the time domain, the phenomena occur over the interval $\Omega_{\,t} \egal \bigl[\, 0 \,,\, \tf \,\bigr]\,$.

\subsubsection{Heat transfer}

For each layer $i\,$, the one-dimensional heat transfer governing equation is: 
\begin{align*}
c_{\,i} \, \pd{T}{t} \egal \pd{}{x} \biggl(\, k_{\,i} \, \pd{T}{x} \,\biggr) \,,
\quad \forall x \, \in \, \Omega_{\,x}^{\,i} \,, 
\quad t \, \in \, \Omega_{\,t} \,, 
\quad \forall i \, \in \, \bigl\{\, 1 \,,\, \ldots \,,\, N_{\,\ell} \,\bigr\} \,,
\end{align*}
where the material properties of each layer are the thermal conductivity $k_{\,i} \ \unit{W\,.\,m^{\,-1}\,.\,K^{\,-1}}$ and the volumetric heat capacity $c_{\,i} \ \unit{J\,.\,m^{\,-3}\,.\,K^{\,-1}}\,$. At the interface between the wall and the outside air, convective, short and long wave radiation heat transfer occur \cite{sreedhar_development_2016}. Thus, the boundary conditions is formulated as:
\begin{align}
\label{eq:bc_left_NL}
-k_{\,1} \, \pd{T}{x} \plus h_{\,L}(\,t\,) \, T \plus \epsilon \, \sigma \, T^{\,4}
 & \egal h_{\,L}(\,t\,) \, T_{\,\infty\,,\,L}(\,t\,) \plus \epsilon \, \sigma \, T_{\,\sky}^{\,4}(\,t\,)  \plus q_{\,\infty\,,\,L}(\,t\,) \,, \qquad x \, \in \, \Gamma_{\,L} \,,
\end{align}
where $h_{\,L} \ \unit{W\,.\,m^{\,-2}\,.\,K^{\,-1}}$ is the surface heat transfer coefficient at the interface between the wall and the outside air. It depends on the outside air velocity $v_{\,\infty\,,\,L}$ according to \cite{mcadams_heat_1942}:
\begin{align*}
h_{\,L}(\,t\,) \egal 
h_{\,L\,,\,0} \plus h_{\,L\,,\,1} \cdot \frac{v_{\,\infty\,,\,L}(\,t\,)}{v_{\,0}} \,,
\end{align*}
The outside air temperature $T_{\,\infty\,,\,L}$ also varies according to the climatic data.  The long wave radiation are considered in the modeling with $T_{\,\sky}$ being the sky temperature, $\sigma \egal 5.67 \e{-8} \ \mathsf{W\,.\,m^{\,-2}\,.\,K^{\,-4}}$ being the \textsc{Stefan}--\textsc{Boltzmann} constant and $\epsilon \ \unit{-}$ being the emissivity of the surface. In addition, $q_{\,\infty\,,\,L}$ is the incident short wave radiation flux. Since the difference between the surface temperature and sky temperature is expected to be lower than $100\ \mathsf{\degC}\,$, the nonlinear long wave radiation term are linearized according to \cite{sacadura_transferts_1980} and the boundary condition~\eqref{eq:bc_left_NL} becomes:
\begin{align*}
-k_{\,1} \, \pd{T}{x} \plus \bigl(\, h_{\,L}(\,t\,) \plus h_{\,\sky}(\,t\,) \, \bigr) \, T 
 & \egal h_{\,L}(\,t\,) \, T_{\,\infty\,,\,L}(\,t\,) \plus h_{\,\sky}(\,t\,) \, T_{\,\sky}(\,t\,)  \plus q_{\,\infty\,,\,L}(\,t\,) \,, \qquad x \, \in \, \Gamma_{\,L} \,,
\end{align*}
where $h_{\,\sky}$ is the radiative equivalent surface transfer coefficient given by:
\begin{align*}
h_{\,\sky}(\,t\,) \egal 4 \, \epsilon \, \sigma \, T_{\,\sky}^{\,3}(\,t\,) \,.
\end{align*}

On the right boundary condition, at the interface between the wall and the inside ambient air, only convective heat transfer occurs. It is assumed that the radiation between inside surfaces is negligible. Thus, the boundary condition is formulated as \textsc{Robin} type: 
\begin{align*}
k_{\,N_{\,\ell}} \, \pd{T}{x} \plus h_{\,R} \, T & \egal h_{\,R} \, T_{\,\infty\,,\,R}(\,t\,) \,, \qquad x \, \in \, \Gamma_{\,R} \,,
\end{align*}
The inside ambient temperature $T_{\,\infty\,,\,R}$ is varying according to set-up condition of the heating and cooling systems. The inside surface transfer coefficient is assumed as constant. 

At the interface between two layers, the continuity of the heat flux and of the fields is assumed. Thus, those two conditions are formulated as:
\begin{align*}
T(\,x \moins \varepsilon \,,\,t\,) & \egal T(\,x \plus \varepsilon \,,\,t\,) \,, \\[4pt]
k_{\,i} \, \pd{T}{x}\,\biggr|_{\,x \moins \varepsilon} & \egal k_{\,i+1} \, \pd{T}{x}\,\biggr|_{\,x \plus \varepsilon} \,,
\qquad \forall \, \varepsilon \, \rightarrow \, 0 \,,  
\qquad x \, \in \, \Gamma_{\,i}
\qquad \forall \, i \, \in \, \bigl\{\, 1 \,,\, \ldots \,,\, N_{\,\ell}\moins 1 \,\bigr\} \,.
\end{align*}
Last, the initial condition is given by:
\begin{align*}
T\bigl(\,x\,,\,t \egal 0 \,\bigr) \egal T_{\,0}(\,x\,) \,, 
\qquad \forall \, x \, \in \, \Omega_{\,x}^{\,i} \,,
\qquad \forall \, i \, \in \, \bigl\{\, 1 \,,\, \ldots \,,\, N_{\,\ell} \,\bigr\} \,,
\end{align*}
where $T_{\,0}$ is a given function of space. 

\subsubsection{Calculated outputs}

\begin{figure}[h!]
\centering
\includegraphics[width=.9\textwidth]{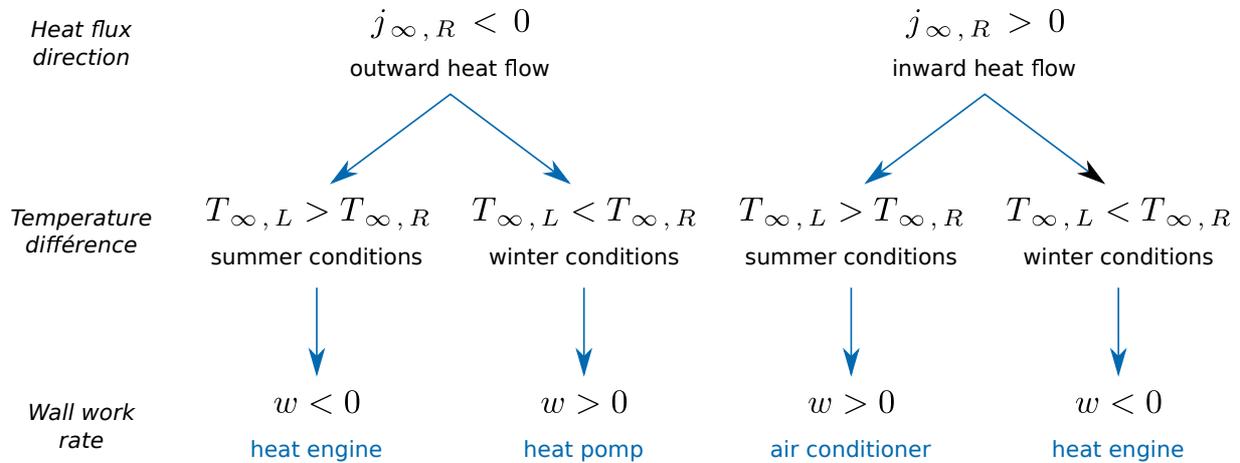}
\caption{Wall operating according to the direction of the flux and the temperature difference, based on \cite[Table ~1]{strub_second_2005}.}
\label{fig:cons_work_sign}
\end{figure}

The calculated outputs, \emph{i.e.} the important quantities computed with the model and used for investigating the physical phenomena are introduced. The first \revision{output} of interest is the temperature $T$ computed in layer $i$ according to time and space. In addition, the heat diffusion flux $j_{\,i} \ \unit{W\,.\,m^{\,-2}}$ is defined as: 
\begin{align*}
j_{\,i} \egal - k_{\,i} \ \pd{T}{x} \,, \qquad \forall \, i \, \in \, \bigl\{\, 1 \,,\, \ldots \,,\, N_{\,\ell}\,\bigr\} \,.
\end{align*}
Based on the thermodynamic analysis of \cite{strub_second_2005,bejan_advanced_2016}, the wall is equivalent to a reversible engine operating between the outside and inside sources. The engine wall dissipates its entire work output into a heat engine that can releases heat to the inside or outside sources. Within this approach, the work rate $w \ \unit{W\,.\,m^{\,-2}}$ is computed according to:
\begin{align*}
w (\,t\,) \egal j_{\,\infty\,,\,R} \ \biggl(\, \frac{T_{\,\infty\,,\,L}}{T_{\,\infty\,,\,R}} \moins 1 \,\biggr) \,,
\end{align*}
where $j_{\,\infty\,,\,R}$ is the flux received by the ambient zone:
\begin{align*}
j_{\,\infty\,,\,R} \egal j\,\bigl(\, x\egal \ell \,,\, t\,\bigr) \,, 
\end{align*}
The so-called net consumed work $\mathcal{W} \ \unit{J\,.\,m^{\,-2}}$ is:
\begin{align}
\label{eq:consumed_work}
\mathcal{W} \egal \int_{\,\Omega_{\,t}} \ w(\,t\,) \, \mathrm{d}t \,.
\end{align}
This approach can be used for thermal insulation design problems, as for building walls. for such problems, the objective is to minimize the heat leaks towards outside during winter conditions and inside during summer ones. This feature are enhanced in Figure~\ref{fig:cons_work_sign} according to the sign of the consumed work. For the first case, the heat flux is directed toward the outside ambient air, \emph{i.e.} the wall is cooling the inside zone. It is combined with summer conditions where the outside temperature is higher than the inside one. Thus, the wall do not require an additional air conditioner system to cool the inside ambient air. The consumed work is defined as negative and the wall works as an heat engine. On the contrary, when the heat flux is still negative and winter conditions are reached. The wall is cooling the inside zone so the inside ambient zone requires a heat pump to maintain occupants' comfort. Interested readers are invited to consult \cite{strub_second_2005} for further details and description of this approach.


\subsection{Dimensionless model}

Before developing any numerical algorithm to compute the governing equations, a dimensionless model is formulated  \cite{berger_dimensionless_2021,berger_evaluation_2019-1,berger_comparison_2020}. For each layer $i$, the dimensionless temperature is defined: 
\begin{align*}
u_{\,i} \egal 
\begin{cases} 
\ \frac{T \moins T^{\,\min}}{T^{\,\max} \moins T^{\,\min}} \,, & \quad x \, \in \, \Omega_{\,x}^{\,i} \,, \\
\ 0  \,, & \quad x \, \notin \, \Omega_{\,x}^{\,i} \,,
\end{cases} 
\qquad \forall i \, \in \, \bigl\{\, 1 \,,\, \ldots \,,\, N_{\,\ell} \,\bigr\} \,,
\end{align*}
where $T^{\,\max}$ and $T^{\,\min}$ are chosen reference temperature values. Unit-less space and time coordinates are set: 
\begin{align*}
\biggl(\,\chi_{\,i} \egal \frac{x \moins \ell_{\,i}}{\ell_{\,i+1} \moins \ell_{\,i}}\,,\,\tau \, \egal \frac{t}{t_{\,0}}\,\biggr) \, \in \, \bigl[\,0 \,,\, 1 \,\bigr] \, \times \, \bigl[\,0 \,,\, \tau_{\,\mathrm{f}} \,\bigr]\,, \qquad \forall i \, \in \, \bigl\{\, 1 \,,\, \ldots \,,\, N_{\,\ell} \,\bigr\} \,.
\end{align*}
The dimensionless temperature for each layer $i$ is solution of the following governing equation:
\begin{align}
\label{eq:heat_1D_dimensionless}
\pd{u_{\,i}}{\tau} \moins 
\Fo_{\,i} 
\, \pd{^{\,2} u_{\,i}}{\chi_{\,i}^{\,2}} \egal 0 \,,
\qquad \forall i \, \in \, \bigl\{\, 1 \,,\, \ldots \,,\, N_{\,\ell} \,\bigr\} \,.
\end{align}
By convention, for the layer $i\,$, the left side boundary condition corresponds to the  \textsc{Robin} boundary condition of $\Gamma_{\,L}$ or to the continuity of the field: 
\begin{subequations}
\label{eq:BC_left_dimensionless}
\begin{align}
-\,\pd{u_{\,i}}{\chi_{\,i}}  \plus \Bigl(\, \Bi_{\,L} \plus \Bi_{\,\sky} \,\bigr) \, u_{\,i} & \egal \Bi_{\,L} \, u_{\,\infty\,,\,L}  \plus \Bi_{\,\sky} \, u_{\,\sky} \plus \varrho_{\,\infty\,,\,L} \,, \qquad \chi_{\,i} \egal 0 \,, \qquad i \egal 1 \,, 
\label{eq:BC_left_dimensionless_a} \\[4pt]
u_{\,i}(\,\chi_{\,i} \egal 0 \,,\,\tau\,) & \egal u_{\,i-1}(\,\chi_{\,i-1} \egal 1  \,,\,\tau\,) \,, 
\qquad \forall \, i \, \in \, \bigl\{\, 2 \,,\, \ldots \,,\, N_{\,\ell}\moins 1 \,\bigr\} \,.
\label{eq:BC_left_dimensionless_b}
\end{align}
\end{subequations}
Respectively, the right side boundary condition of the layer $i$ can be the \textsc{Robin} boundary condition of $\Gamma_{\,R}$ or the continuity of the flux:
\begin{subequations}
\label{eq:BC_right_dimensionless}
\begin{align}
\pd{u_{\,i}}{\chi_{\,i}}\,\biggr|_{\,\chi_{\,i} \egal 1} & \egal \kappa_{\,i+1} \, \pd{u_{\,i+1}}{\chi_{\,i+1}}\,\biggr|_{\,\chi_{\,i+1} \egal 0} \,,
\qquad \forall \, i \, \in \, \bigl\{\, 1 \,,\, \ldots \,,\, N_{\,\ell}\moins 1 \,\bigr\} \,, 
\label{eq:BC_right_dimensionless_a}\\[4pt]
\pd{u_{\,i}}{\chi_{\,i}}  \plus \Bi_{\,R}\, u_{\,i} & \egal \Bi_{\,R}\, u_{\,\infty\,,\,R} \plus \varrho_{\,\infty\,,\,R} \,, \qquad  \chi_{\,i} \egal 1 \,, \qquad i \egal N_{\,\ell} \,. \label{eq:BC_right_dimensionless_b}
\end{align}
\end{subequations}
The boundary fields are transformed according to:
\begin{align*}
& \qquad \qquad \qquad \qquad \qquad u_{\,\infty}  \egal \frac{T_{\,\infty} \moins T^{\,\min}}{T^{\,\max} \moins T^{\,\min}} \,, \\[4pt]
\varrho_{\,\infty\,,\,L} & \egal \frac{\ell_{\,2}}{k_{\,1} \, \bigl(\, T^{\,\max} \moins T^{\,\min} \,\bigr) } \, q_{\,\infty\,,\,L} \,,
\qquad \varrho_{\,\infty\,,\,R} \egal \frac{\ell \moins \ell_{\,N_{\,\ell}}}{k_{\,N_{\,\ell}} \, \bigl(\, T^{\,\max} \moins T^{\,\min} \,\bigr) } \, q_{\,\infty\,,\,R} \,.
\end{align*}
Last the initial condition is:
\begin{align*}
u_{\,i}\bigl(\,\chi_{\,i}\,,\,t \egal 0 \,\bigr) \egal u_{\,0}(\,\chi_{\,i}\,) \,, 
\qquad \forall \, \chi_{\,i} \, \in \, \Omega_{\,x}^{\,i} \,,
\qquad \forall \, i \, \in \, \bigl\{\, 1 \,,\, \ldots \,,\, N_{\,\ell} \,\bigr\} \,.
\end{align*}
As remarked in Eqs~\eqref{eq:heat_1D_dimensionless}, \eqref{eq:BC_left_dimensionless} and \eqref{eq:BC_right_dimensionless}, the dimensionless formulation enables to highlight important scaling parameters as the \textsc{Fourier} $\Fo$ and \textsc{Biot} $\Bi$ numbers:
\begin{align*}
\Fo_{\,i}  \egal & \frac{k_{\,i} \ t_{\,0}}{c_{\,i} \,  \Bigl(\, \ell_{\,i+1} \moins \ell_{\,i} \,\Bigr)^{\,2}}  \,, 
\quad \forall \, i \, \in \, \bigl\{\, 1 \,,\, \ldots \,,\, N_{\,\ell} \,\bigr\} \,, \\[4pt]
\kappa_{\,i+1} \egal & \frac{k_{\,i+1}}{k_{\,i}} \, \biggl(\, \frac{\ell_{\,i} \moins \ell_{\,i-1}}{\ell_{\,i+1} \moins \ell_{\,i}} \,\biggr) \,, 
\quad \forall \, i \, \in \, \bigl\{\, 1 \,,\, \ldots \,,\, N_{\,\ell}-1 \,\bigr\} \,, \\[4pt]
\Bi_{\,L}  \egal \frac{h_{\,L} \ \ell_{\,2}}{k_{\,1} } &  \,, \qquad
\Bi_{\,\sky}  \egal \frac{h_{\,\sky} \ \ell_{\,2}}{k_{\,1} } \,, \qquad
\Bi_{\,R} \egal \frac{h_{\,R} \ \Bigl(\, \ell \moins \ell_{\,N_{\,\ell}} \,\Bigr)}{k_{\,N_{\,\ell}} } \,.
\end{align*}
where $t_{\,0}$ is a chosen reference time scale. 

\section{Numerical models}
\label{sec:numerical_model}

The numerical models are used to solve the governing equation and assess the prediction of the physical phenomena are now described. In the literature, Eq.~\eqref{eq:heat_1D_dimensionless} can be solved using approaches such as finite-difference method \cite{gasparin_innovative_2019}, finite-volume method \cite{mendes_method_2005} or finite-element method \cite{thomas_fully_1980}. For the sake of compactness, the finite-difference method is briefly presented as background of traditional approaches. It is referenced as the Complete Original Model (COM). Then, the model reduction methods are presented. For the sake of clarity, the methods are described for one layer only $\Omega_{\,x}^{\,i}\,, \forall \, i \, \in \, \bigl\{\, 1 \,,\, \ldots \,,\, N_{\,\ell} \,\bigr\} \,$.

\begin{figure}[h!]
\begin{center}
\includegraphics[width=.9\textwidth]{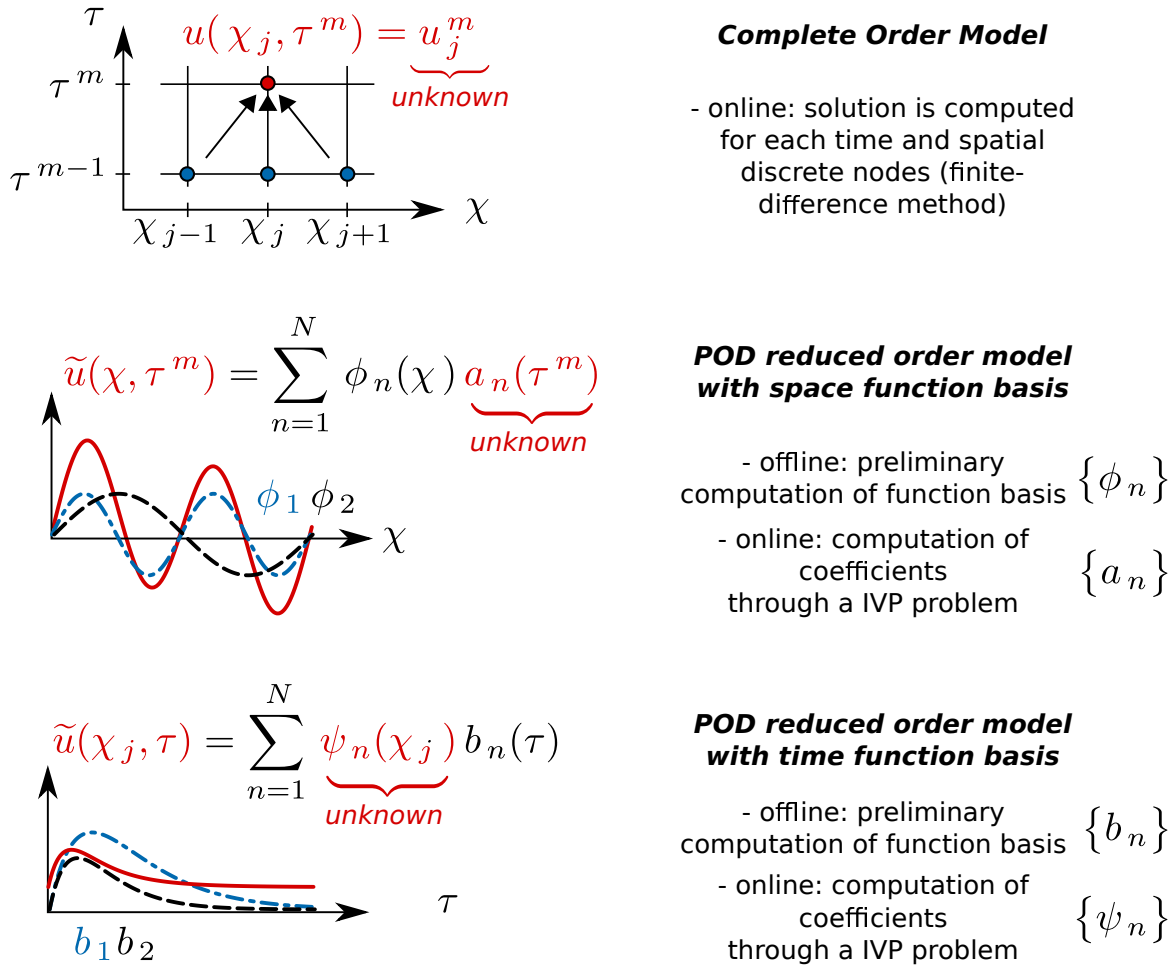}
\caption{Illustration of the construction of the \revision{reduced-order} models based on POD approach compared to the complete original model obtained by finite-difference methods and \textsc{Euler} time scheme.}
\label{fig:Rom_secret_sauce}
\end{center}
\end{figure}

Figure~\ref{fig:Rom_secret_sauce} illustrates briefly the methodology. Traditional methods as finite-differences seek to compute the value of the solution at each time and space grid using approximation of the operators. One understands that the number of unknowns  (or degree of freedom as defined in Section~\ref{sec:metrics}) to compute is $N_{\,\chi} \,\times \, N_{\,\tau}\,$. For the  \revision{reduced-order} model (ROM), the solution is assumed as decomposed on space and time functions. Usually, the POD is built with space \revision{reduced basis}. So the \revision{reduced basis} $\bigr\{\,\phi_{\,n}\,\bigr\}$ is preliminary computed during the offline phase using a solution of the problem. Then, during the online phase, knowing the function $\phi_{\,n}(\,\chi\,)$ only the temporal coefficients  $\bigr\{\,a_{\,n}\,\bigr\}$ need to be computed. The degree of freedom is $N \, \times \, N_{\,\tau}$ (at each time step $N$ coefficients are determined). The order of the ROM $N$ is expected to be low (a few tens) so that the degree of freedom is lower than complete order model. Thus, faster computations are hoped. Here, an alternative approach is investigated. Instead of determining the space functions, a time \revision{reduced basis} $\bigr\{\,b_{\,n}\,\bigr\}$ is preliminary built. This alternative is justified by the fact that the problem (simulation over $30$ years) has higher time scale than spatial one. Knowing the time \revision{reduced basis}, only the coefficients $\psi_{\,n}$ require to be computed. The degree of freedom is $N \, \times \, N_{\,\chi}\,$.

\newpage

\subsection{Classical approach: Large original model}
\label{ssec:LOM}

A uniform discretisation in space is adopted, where $\Delta \chi$ is the mesh parameter with $\chi_{\,i\,,\,j} \egal j \cdot \Delta \chi $ with $j \ \in \ \bigl\{\, 1 \,, \ldots \,, N_{\,\chi} \,\bigr\}$ and $N_{\,\chi}$ being the total number of space points. Following the methods of horizontal line proposed by \cite{berezin_computing_1965}, the semi-discrete values of function $u_{\,i}$ are denoted by $u_{\,i\,,\,j} \egal u_{\,i}(\,\chi_{\,j}\,,\,\tau\,)\,$. Recalling that index $i$ stands for the layer where equation are computed. The semi-discretisation of Eq.~\eqref{eq:heat_1D_dimensionless} yields to:
\begin{align}
\label{eq:ivp}
\od{u_{\,i\,,\,j}}{\tau} \egal \frac{\Fo}{\Delta \chi^{\,2}}\,\biggl(\, u_{\,i\,,\,j+1} \moins 2 \, u_{\,i\,,\,j} \plus u_{\,i\,,\,j-1} \,\biggr)\,.
\end{align}
Similarly, a second-order semi-discretisation is performed for the boundary conditions, Eqs. ~\eqref{eq:BC_left_dimensionless_a} and \eqref{eq:BC_right_dimensionless_b}, and for the interface condition~\eqref{eq:BC_right_dimensionless_a}. The obtained initial value problem is solved using an explicit \RK of order $4$ solver with a set tolerance \cite{shampine_matlab_1997}. This model is denoted by the Large (or complete) Original Model (LOM). The degree of freedom of the solution (See definition in Section~\ref{sec:metrics}) scales with:
\begin{align*}
\mathfrak{D}(\,\mathrm{LOM}\,) \egal N_{\,\chi}\, \times N_{\,\tau}\,,
\end{align*}
where $N_{\,\tau}$ is the number of time steps where the solution $u_{\,i}(\,\chi_{\,i\,,\,j}\,,\,t\,)$ is computed.

\subsection{POD \revision{reduced-order} model with time \revision{reduced basis}}

The issue is to approximate the solution of the governing equation with a lower degree of freedom. It aims in building a \revision{reduced-order} model (ROM). Several approaches are used in the literature \cite{berger_review_2017,gasparin_spectral_2019} and the Proper Orthogonal Decomposition (POD) is one of the most used and widely known. Interested readers are invited to consult \cite{tallet_pod_2015,tallet_fast_2017,sempey_fast_2009,sirovich_turbulence_1987,holmes_turbulence_1996} a primary overview. It assumes that the solution is approximated by:
\begin{align}
\label{eq:PODt_solution}
u_{\,i} \, (\,\chi_{\,i}\,,\,\tau\,) \, \approx \, \widetilde{u}_{\,i} \, (\,\chi_{\,i}\,,\,\tau\,) \egal \sum_{n \egal 1}^{\,N} \, \phi_{\,n}^{\,i} (\,\chi_{\,i}\,) \, a_{\,n}^{\,i}(\,\tau\,) \,,
\end{align}
where $\bigl\{\,\phi_{\,n}^{\,i}\,\bigr\}_{\,n = 1}^{\,N}$ is the POD \revision{reduced basis} that depend on space coordinate $\chi_{\,i} \, \in \, \Omega_{\,x}^{\,i}\,$. Here, a basis is built per layer indicated by the index $i\,$. Using such approximation, the temporal coefficient $\bigl\{\,a_{\,n}^{\,i}\,\bigr\}_{\,n = 1}^{\,N}$ are solution of a set of $N$ ordinary differential equations, obtained by \textsc{Galerkin} projection of the governing equations onto each basis function $\phi_{\,n}^{\,i}\,$. This ROM is then solved using any time-integrator solver such as \RK. Such model is denoted by PODt, as ROM based on POD with temporal coefficients to compute. Given the time horizon of the problem, the degree of freedom scales with:
\begin{align*}
\mathfrak{D}(\,\mathrm{PODt}\,) \egal N\, \times N_{\,\tau}\,,
\end{align*}
where $N_{\,\tau}$ is the number of time steps required by the time solver and $N$ the model order which scales with a few ten $N \egal \mathcal{O}(\,10\,)\,$. Thus, even if $N \, << \, N_{\,\chi}$, $N_{\,\tau}$ is still high for long-term simulation problem. 

Another method is explored following the works on horizontal lines applied to complete models in \cite{meyer_time-discrete_2015,gasparin_adaptive_2018}. The proposed approximation of the solution is the following:
\begin{align}
\label{eq:PODx_solution}
u_{\,i} \, (\,\chi_{\,i}\,,\,\tau\,) \, \approx \, \widetilde{u}_{\,i} \, (\,\chi_{\,i}\,,\,\tau\,) \egal \sum_{n \egal 1}^{\,N} \, \psi_{\,n}^{\,i}(\,\tau\,) \, b_{\,n}^{\,i} (\,\chi_{\,i}\,)\,,
\end{align}
Here the $\bigl\{\,\psi_{\,n}^{\,i}\,\bigr\}_{\,n = 1}^{\,N}$ is the POD \revision{reduced basis} that depends on time coordinate $\tau\,$. Again, one basis is built per layer $i\,$. The model is denoted by PODx, as POD ROM with spatial coefficients to compute. It is built using an \emph{a posteriori} approach  assuming $N_{\,s}$ snapshots of the solution of the governing equation~\eqref{eq:heat_1D_dimensionless} previously computed \cite{sirovich_turbulence_1987}. With those snapshots, the correlation matrix is evaluated by:
\begin{align*}
C & \egal \bigl[\, C_{\,m\,p} \,\bigr] \,, \qquad \forall \, (\,m\,,\,p\,) \, \in \, \bigl\{\, 1 \,,\, \ldots \,,\, N_{\,s}\,\bigr\}^{\,2} \,, \\[4pt]
C_{\,m\,p} & \egal \int_{\,0}^{\,1} u_{\,i} (\,\chi_{\,i}\,,\,\tau_{\,m}\,) \, u_{\,i} (\,\chi_{\,i}\,,\,\tau_{\,p}\,) \, \mathrm{d}\chi \,.
\end{align*}
The vector basis $\bigl\{\,\psi_{\,n}^{\,i}\,\bigr\}_{\,n = 1}^{\,N}$ equals to the eigenvectors of matrix $C$ corresponding to the $N$ highest eigenvalues. Note that the vectors are orthogonal and that the solution can be approximated on this basis of small size $N$ by $\displaystyle u_{\,i} \egal \sum_{n \egal 1}^{\,N} \, \psi_{\,n}^{\,i} \, b_{\,n}^{\,i} \,$. Then, the spatial coefficient coefficients $\bigl\{\,b_{\,n}^{\,i}\,\bigr\}_{\,n = 1}^{\,N}$ are computed by inserting the approximation~\eqref{eq:PODx_solution} into Eq.~\eqref{eq:heat_1D_dimensionless}: 
\begin{align}
\label{eq:POD_heat_equation}
\sum_{n \egal 1}^{\,N} \od{\psi_{\,n}^{\,i}}{\tau} \, b_{\,n}^{\,i} \moins \Fo \, \sum_{n \egal 1}^{\,N} \psi_{\,n}^{\,i}  \, \od{^{\,2}b_{\,n}^{\,i}}{\chi_{\,i}^{\,2}}  \egal 0\,,
\end{align}
By projecting Eq.~\eqref{eq:POD_heat_equation} on each vector $\psi_{\,m}^{\,i}$ and using the orthogonality property of the vector basis, a system of $N$ boundary value problems is obtained for each layer $i\,$:
\begin{align}
\label{eq:POD_equation}
\sum_{m \egal 1}^{\,N} \alpha_{\,n\,m} \, b_{\,m}^{\,i} \moins
\Fo_{\,i} \, \od{^{\,2}b_{\,n}^{\,i}}{\chi_{\,i}^{\,2}}   \egal 0 
 \,, \qquad \forall n \, \in \, \bigl\{\,1 \,,\, \ldots \,,\, N \,\bigr\} \,,
 \,, \qquad \forall i \, \in \, \bigl\{\,1 \,,\, \ldots \,,\, N_{\,\ell} \,\bigr\} 
\end{align}
with
\begin{align*}
\alpha_{\,n\,m}  \egal  \int_{\,0}^{\,\tau_{\,\mathrm{f}}} \ \psi_{\,n}^{\,i}  \, \od{\psi_{\,m}^{\,i} }{\tau} \   \, \mathrm{d}\tau \,.
\end{align*}
Similarly, the approximation~\eqref{eq:PODx_solution} is introduced into the boundary conditions equations~\eqref{eq:BC_left_dimensionless} and \eqref{eq:BC_right_dimensionless}. After projection, one obtains the left boundary conditions for the first layer ($i \egal 1$) and the layers $2$ to $N_{\,\ell}$:
\begin{align}
\label{eq:POD_left_BC}
\begin{cases}
\ \displaystyle - \, \od{b_{\,n}^{\,i} }{\chi_{\,i}} \plus \sum_{m \egal 1}^{\,N} \delta_{\,n\,m}^{\,L} \, b_{\,m}^{\,i}  \egal \beta_{\,n}^{\,L} \,, \quad \chi_{\,i} \egal 0 \,, \qquad & i \egal 1 \,, \\[4pt]
\ b_{\,n}^{\,i} (\,\chi_{\,i} \egal 0\,) 
\egal \sum_{m \egal 1}^{\,N} \gamma_{\,n\,m}^{\,L} \ b_{\,m}^{\,i-1} (\,\chi_{\,i-1}\egal 1\,) \,,
\qquad & \forall \, i \, \in \, \bigl\{\, 2 \,,\, \ldots \,,\, N_{\,\ell}\,\bigr\} \,,
\end{cases}
\end{align}
and the right boundary conditions for layers $1$ to $N_{\,\ell} \moins 1$ and for the last layer ($i \egal N_{\,\ell}$):
\begin{align}
\label{eq:POD_right_BC}
\begin{cases}
\ \displaystyle \od{b_{\,n}^{\,i} }{\chi_{\,i}} \,\biggr|_{\,\chi_{\,i} \egal 1} \egal \kappa_{\,i+1} \, \sum_{m \egal 1}^{\,N} \gamma_{\,n\,m}^{\,R} \
\od{b_{\,m}^{\,i+1} }{\chi_{\,i+1}} \,\biggr|_{\,\chi_{\,i+1} \egal 0} \,,
\qquad & \forall \, i \, \in \, \bigl\{\, 1 \,,\, \ldots \,,\, N_{\,\ell}\moins 1 \,\bigr\} \,, \\[6pt]
\ \displaystyle \od{b_{\,n}^{\,i} }{\chi_{\,i}} \plus \sum_{n \egal 1}^{\,N} \delta_{\,n\,m}^{\,R} \, b_{\,m}^{\,i}  \egal \beta_{\,n}^{\,R} \,, \quad \chi_{\,i} \egal 1 \,, \qquad & i \egal N_{\,\ell} \,.
\end{cases}
\end{align}
The coefficients $\beta_{\,n}^{\,L}\,$, $\beta_{\,n}^{\,R}\,$, $\delta_{\,n\,m}^{\,L}\,$, $\delta_{\,n\,m}^{\,R}\,$, $\gamma_{\,n\,m}^{\,L}$ and $\gamma_{\,n\,m}^{\,R}$ are defined according to:
\begin{subequations}
\label{eq:coeff_beta_delta_gamma}
\begin{align}
\beta_{\,n}^{\,L} & \egal  \int_{\,0}^{\,\tau_{\,\mathrm{f}}} \psi_{\,n}^{\,i}  \, 
\bigl(\, \Bi_{\,L} \ u_{\,\infty\,,\,L} \plus \Bi_{\,\sky} \ u_{\,\sky} \plus \varrho_{\,\infty \,,\,L} \,\bigr) \, \mathrm{d}\tau \,, &&
\beta_{\,n}^{\,R} \egal \int_{\,0}^{\,\tau_{\,\mathrm{f}}} \psi_{\,n}^{\,i}  \, 
\bigl(\, \Bi_{\,R} \ u_{\,\infty\,,\,R}  \plus \varrho_{\,\infty \,,\,R} \,\bigr)  \, \mathrm{d}\tau \,, \\[4pt]
\delta_{\,n\,m}^{\,L} & \egal  \int_{\,0}^{\,\tau_{\,\mathrm{f}}} \bigl(\, \Bi_{\,L} \plus \Bi_{\,\sky} \,\bigr) \ \psi_{\,n}^{\,i}  \, \psi_{\,m}^{\,i} \, \mathrm{d}\tau \,, &&
\delta_{\,n\,m}^{\,R} \egal \int_{\,0}^{\,\tau_{\,\mathrm{f}}} \Bi_{\,R} \, \psi_{\,n}^{\,i}  \,\psi_{\,m}^{\,i}  \, \mathrm{d}\tau \,, \\[4pt]
\gamma_{\,n\,m}^{\,L} & \egal \int_{\,0}^{\,\tau_{\,\mathrm{f}}} \psi_{\,n}^{\,i} \ \psi_{\,m}^{\,i-1}  \,   \, \mathrm{d}\tau \,, &&
\gamma_{\,n\,m}^{\,R} \egal \int_{\,0}^{\,\tau_{\,\mathrm{f}}} \psi_{\,n}^{\,i} \ \psi_{\,m}^{\,i+1}  \,   \mathrm{d}\tau \,.
\end{align}
\end{subequations}
Note that if \textsc{Biot} coefficients $\Bi_{\,L}\,$, $\Bi_{\,\sky}$ and $\Bi_{\,R}$ do not depend on time, then we have:
\begin{align*}
\delta_{\,n\,m}^{\,L} \egal \delta_{\,n\,m}^{\,R}
\egal 
\begin{cases} 
\ 1 \,, & \quad n \egal m \,, \\
\ 0  \,, & \quad n \, \neq \, m \,,
\end{cases} 
\end{align*}
due to the orthogonality of the basis vectors. Equations~\eqref{eq:POD_equation}, \eqref{eq:POD_left_BC} and~\eqref{eq:POD_right_BC} enable to compute the spatial coefficients $\bigl\{\,b_{\,n}^{\,i}\,\bigr\}_{\,n = 1}^{\,N}$ for each layer $i\,$. With this approach, the problem to solve scale with $N_{\,\chi} \, \times \, N\,$. It is computed using a boundary value problem solver based on the three--stage \textsc{Lobatto} formula with a fourth--order accuracy. It is implemented in the \texttt{Matlab} environment with the \texttt{bvp4c} function and a set tolerance \cite{kierzenka_bvp_2001,shampine_matlab_1997}. Two important advantages can be remarked. First, the implementation using \texttt{Matlab} solver enables to obtain straightforwardly the spatial derivative of the coefficients, namely $\displaystyle \pd{b_{\,n}^{\,i}}{\chi_{\,i}}\,,\, \forall n  \, \in \, \bigl\{\,1 \,,\, \ldots \,,\, N \,\bigr\}\,$. Thus, the heat flux can be easily obtained through the space derivative of the solution:
\begin{align*}
\pd{u_{\,i}}{\chi_{\,i}} \egal \sum_{n \egal 1}^{\,N} \, \psi_{\,n}^{\,i}(\,\tau\,) \, \pd{b_{\,n}^{\,i}}{\chi_{\,i}}\,.
\end{align*}
The second advantage is that the numerical method faces no numerical stability regarding the time discretisation. The scheme is unconditionally stable.

\subsection{\revision{Reduced basis} interpolation}

Since the POD is an \emph{a posteriori} model reduction method, a preliminary computation of the field is required to compute the vector basis. As a consequence, the basis $\psi$ from Eq.~\eqref{eq:PODx_solution} is constructed for a given parameters $p$ of the model such as \textsc{Fourier} or \textsc{Biot} coefficients. Thus, to have a parametric solution, a mapping is required between the \revision{reduced basis} obtained for a parameter $p$ and another one $\widetilde{p}\,$. For this, we denote by $\Omega_{\,p}$ the set of parameter:
\begin{align*}
p \egal \bigl(\, p_{\,1} \,,\,\ldots \,,\, p_{\,N_{\,p}} \,\bigr) \, \in \, \Omega_{\,p} \,, \qquad
\Omega_{\,p} \egal \bigcup_{q \egal 1}^{\,N_{\,p}} \Omega_{\,p}^{\,q} \,, \qquad
\Omega_{\,p}^{\,q} \egal \bigl[\, p_{\,q}^{\,\min} \,,\, p_{\,q}^{\,\max} \, \bigr] \,,
\end{align*}
where $N_{\,p}$ is the total number of varying parameters and $p_{\,q}^{\,\min}\,$, $p_{\,q}^{\,\max}$ are the lower and upper bound of the parameter $p_{\,q}\,$. Then, the parametric solution is defined by:
\begin{align*}
u \,: \,\qquad   \Omega_{\,x} \, \times \, \Omega_{\,t} \, \times \, \Omega_{\,p} \,  
& \longrightarrow \,  \mathbb{R}  \\[4pt]
 \bigl(\, \chi \,,\, \tau \,,\, p\,\bigr) 
&\longmapsto \, u \, (\,\chi\,,\,\tau\,,\,p\,) \,.
\end{align*}
Note that the subscript $i$ corresponding to the layer is omitted for the sake of clarity and without loss of generality. To achieve it, we propose to interpolate the preliminary computed POD basis $\Bigl\{\,\psi_{\,n}^{\,b} \,\Bigl\}$ by using the interpolation on a tangent space of \textsc{Grassmann} manifold. This approach is specially designed and adapted for basis interpolation. In the context of ROM, it was introduced by \textsc{Amsallem} and \textsc{Farha} \cite{amsallem_interpolation_2008} to solve parametric aeroelasticity   problem. The approach was also applied successfully to solve compressible flow \cite{freno_using_2013}, \textsc{Navier}--\textsc{Stokes} equations or control problems \cite{oulghelou_non-intrusive_2020,oulghelou_non_2021,oulghelou_fast_2018}. In this paper, we use the methodology proposed by \cite{amsallem_interpolation_2008} so, in the following, we give only the algorithm. More details can be founded in \cite{amsallem_interpolation_2008}. Note that some alternative approaches exist as presented in \cite{oulghelou_parametric_2021,mosquera_generalization_2021,mosquera_pod_2019}. 

Consider a set of $N_{\,b}$ POD basis (previously) computed for parameters in the parameter space $\Omega_{\,p}\,$:
\begin{align*}
\bigl\{\,\bPsi_{\,n}^{\,j}  \,\bigr\}_{\,n = 1\,,\,j=1}^{\,N\,,\, N_{\,b}} \egal 
\bigl\{\,\psi_{\,n}(\,\tau\,,\,p_{\,1}\,)\,,\, \ldots \,,\,\psi_{\,n}(\,\tau\,,\,p_{\,N_{\,b}}\,) \,\bigr\}_{\,n \egal 1}^{\,N} \,,
\end{align*}
and $ \bigl[\,\bPsi^{\,1}  \,\bigr] \,,\, \ldots \,,\, \bigl[\,\bPsi^{\,N_{\,b}}  \,\bigr]$ the associated vector subspace belonging to the \textsc{Grassmann} manifold. The interpolation method is as follows: \\~
\emph{(1)} Choose a reference point $p_{\,0} \, \in \, \bigl\{\,p_{\,1} \,,\, \ldots \,,\, p_{\,N_{\,b}} \,\bigr\}\,$, with its associated vector subspace $ \bigl[\,\bPsi^{\,j_{\,0}}  \,\bigr]\,$, \\~
\emph{(2)} Each remaining $ \bigl[\,\bPsi^{\,j}  \,\bigr]$ is mapped to the tangent space at the reference point by using the logarithmic application:
\begin{align*}
\boldsymbol{\Gamma}_{\,j} \egal \boldsymbol{U}_{\,j} \ \mathrm{arctan} \, \bigl(\, \boldsymbol{\Sigma}_{\,j} \,\bigr) \ \boldsymbol{V}_{\,j}^{\,\mathsf{T}} \,, \quad \forall \, j \, \in \, \bigl\{\,1 \,,\, \ldots \,,\, N_{\,b} \,\bigr\} \,, \quad j \, \neq \, j_{\,0} \,,
\end{align*}
where $\boldsymbol{U}_{\,j} \,$, $\boldsymbol{\Sigma}_{\,j}$ and $\boldsymbol{V}_{\,j}$ are obtained by singular value decomposition (SVD) according to:
\begin{align*}
\Bigl(\, \boldsymbol{I} \moins \bPsi_{\,j_{\,0}} \, \bPsi_{\,j_{\,0}}^{\,\mathsf{T}} \,\Bigr) \, \bPsi_{\,j} \, 
\Bigl(\, \bPsi_{\,j_{\,0}}^{\,\mathsf{T}} \bPsi_{\,j} \,\Bigr)^{\,-1} 
\ \eqsvd \ \boldsymbol{U}_{\,j} \, \boldsymbol{\Sigma}_{\,j} \, \boldsymbol{V}_{\,j}^{\,\mathsf{T}} \,, \qquad
\forall \, j \, \in \, \bigl\{\,1 \,,\, \ldots \,,\, N_{\,b} \,\bigr\} \,, \quad j \, \neq \, j_{\,0} \,,
\end{align*}
noting that super script $^{\mathsf{T}}$ stands for the conjugate transpose operation.\\
\emph{(3)} On the tangent space, interpolate $\boldsymbol{\Gamma}_{\,1}\,,\,\ldots\,,\,\boldsymbol{\Gamma}_{\,N_{\,b}}$ to obtain $\boldsymbol{\widetilde{\Gamma}}$ associated to a new parameter $\tilde{p}\,$. Since this tangent space is a vector space, the standard interpolation methods can be used. In this paper, the Radial Basis Function (RBF) method is employed. Then, 
\begin{align*}
\boldsymbol{\widetilde{\Gamma}} & \egal 
\begin{bmatrix}
\widetilde{\Gamma}_{\,m\,n}
\end{bmatrix} \,, \qquad
\bigl(\,m\,,\,n\,\bigr) \, \in \, \bigl\{\, 1 \,,\, \ldots \,,\, N_{\,\tau} \bigr\} \, \times \, \bigl\{\, 1 \,,\, \ldots \,,\, N \bigr\} \,, \\[4pt]
\widetilde{\Gamma}_{\,m\,n} & \egal  \sum_{b \egal 1}^{\,N_{\,b}} \, \alpha_{\,b} \ \phi_{\,b} \Bigl(\, d(\,p_{\,b} \moins \widetilde{p}\,)\,\Bigr) \,,
\end{align*}
where $\alpha_{\,b}\,\in\, \mathbb{R}$ are the weighting coefficients determined using $\bigl\{\, \Gamma_{\,j\,,\,m\,n} \,\bigr\}_{\,j=1}^{\,N_{\,b}}\,$, $\phi$ is the radial basis function and $d$ stands for the \textsc{Euclidean} distance in the parameter space.\\~
\emph{(4)} Compute the interpolated subspace vector $\bigl[\,\widetilde{\bPsi}\,\bigr]$ by using the exponential mapping  such as:
\begin{align*}
\widetilde{\bPsi} \egal 
\bPsi_{\,j_{\,0}} \, \boldsymbol{\widetilde{V}} \, \cos \bigl(\, \boldsymbol{\widetilde{\Sigma}} \,\bigr)
\plus 
\boldsymbol{\widetilde{U}} \, \sin \bigl(\, \boldsymbol{\widetilde{\Sigma}} \, \bigr) \,,
\end{align*}
where $\boldsymbol{\widetilde{U}} \,$, $\boldsymbol{\widetilde{\Sigma}}$ and $\boldsymbol{\widetilde{V}}$ are obtained by SVD of $\boldsymbol{\widetilde{\Gamma}}\,$:
\begin{align*}
\boldsymbol{\widetilde{\Gamma}} \ \eqsvd \ 
\boldsymbol{\widetilde{U}} \, \boldsymbol{\widetilde{\Sigma}} \, \boldsymbol{\widetilde{V}}^{\,\mathsf{T}} \,.
\end{align*}

\section{Verification of model's efficiency}
\label{sec:validation_case_study}

The POD numerical solution \revision{requires} a verification with a reference one according to ASME standard \cite{asme_vv_asme_2009}. This Section presents the metrics to evaluate the efficiency and three case study considered for verification.

\subsection{Metrics to evaluate the numerical efficiency}
\label{sec:metrics}

Several metrics are defined to judge the efficiency of a numerical model. First, the degree of freedom of the solution $\mathfrak{D}$ is defined as the number of unknowns to be computed to obtain a solution for a certain accuracy. The second criteria is the accuracy of the computed numerical solution $y$ against a reference solution denoted by the superscript $\mathrm{ref}$. Two errors metrics are defined according to:
\begin{align*}
\varepsilon_{\,2} \, \circ \, y \, (\,\tau\,) & \egal
\Biggl(\, \int_{\,\Omega_{\,x}} \, 
\Bigl(\,
y(\,\chi  \,, \tau\,) \moins y^{\mathrm{\, ref}}(\,\chi  \,, \tau \,)
\,\Bigr)^{\,2}
\mathrm{d}\chi \
\,\Biggr)^{\,\half} \,, \\[4pt]
\varepsilon_{\,\infty} \, \circ \, y \,  & \egal \max_{\,\Omega_{\,t}} \, \varepsilon_{\,2}(\,\tau\,) \,.
\end{align*}
When focusing on parametric solution that depends on space $\chi\,$, time $\tau$ and parameter $p$ coordinates, the following error is defined:
\begin{align*}
\varepsilon_{\,\bowtie} \, \circ \, y \egal \max_{\,\Omega_{\,p}} \, \Bigl(\, \varepsilon_{\,\infty}\, \circ \, y \,(\,p\,) \,\Bigr)\,.
\end{align*}
Last, the computational run time $t_{\,\cpu} \ \unit{s}$ is measured using the \texttt{Matlab\texttrademark} environment with a computer equipped with \texttt{Intel} Xeon(R) Gold ($2.2 \ \mathsf{GHz}$ $6^{\,th}$) and $125.6$ GB of RAM. It is evaluated with the mean of $20$ runs of the numerical model.

\subsection{Mono-layer case with no basis interpolation}
\label{ssec:mono_layer}

\begin{figure}[h!]
\begin{center}
\subfigure[]{\includegraphics[width=.45\textwidth]{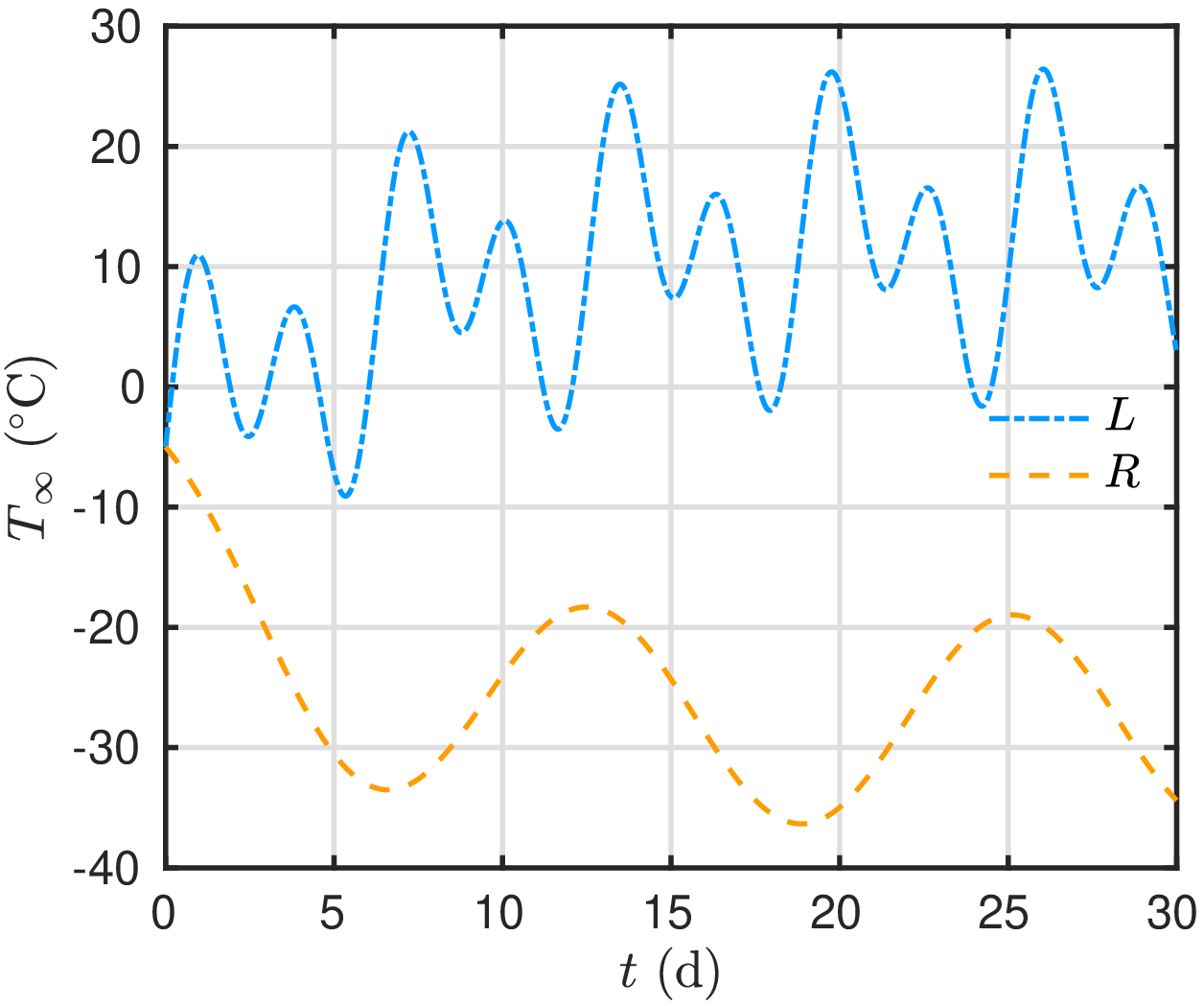}} \hspace{0.2cm}
\subfigure[]{\includegraphics[width=.45\textwidth]{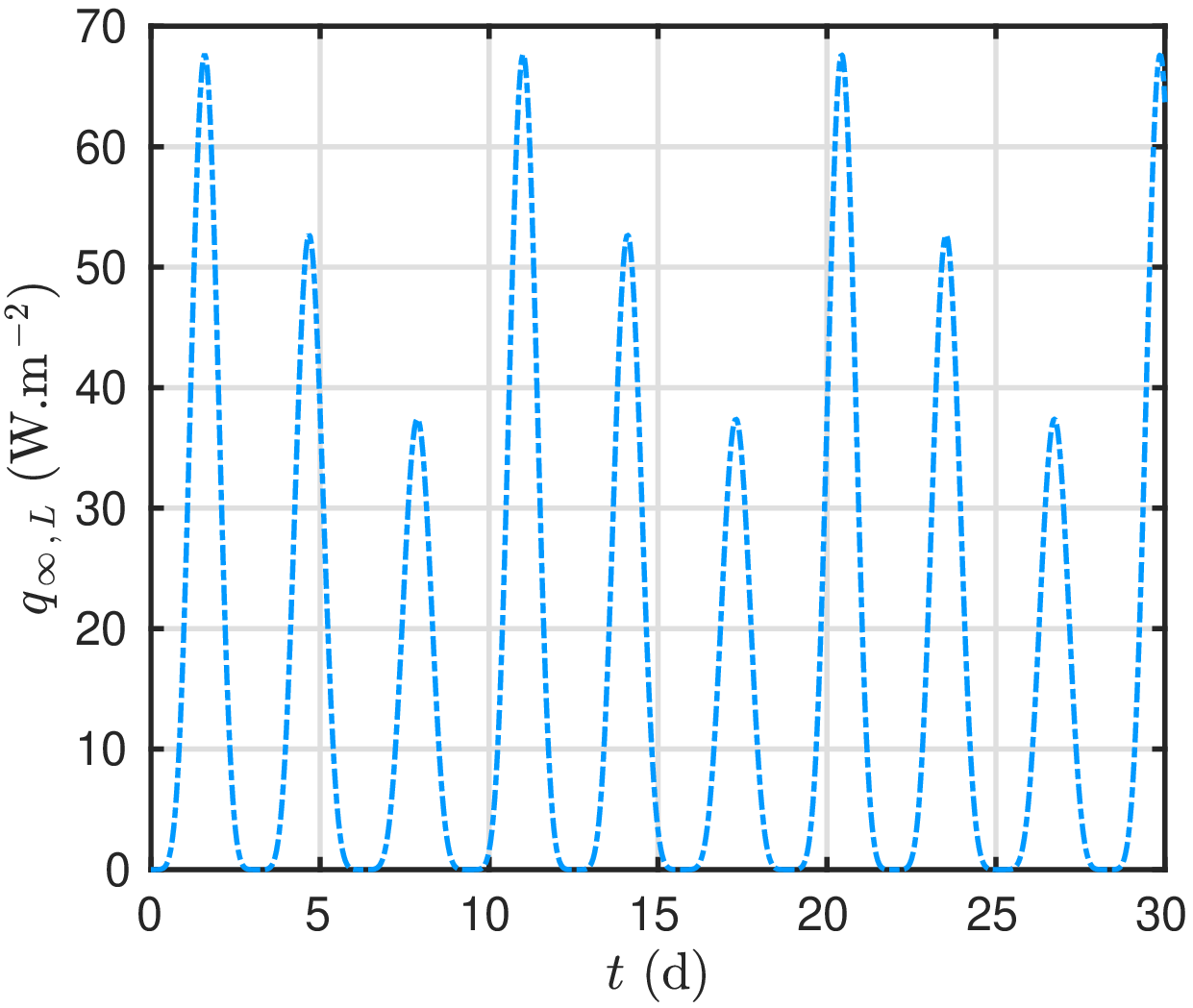}} \\
\caption{Time evolution of the boundary conditions for ambient temperature \emph{(a)} radiation flux \emph{(b)}.}
\label{fig:monolayer_BC}
\end{center}
\end{figure}

The first case study aims at verifying the \revision{reduced-order} model and assessing its efficiency compared to standard approaches. No interpolation of \revision{reduced basis} is considered here. The reduced basis is built using a preliminary solution computed for the same inputs parameters. The case considers a mono-layer wall and $N_{\,\ell} \egal 1\,$. The length of the wall is $\ell \egal 10 \ \mathsf{cm}\,$. The thermal conductivity and heat capacity are $k_{\,1} \egal 0.5 \ \mathsf{W\,.\,m^{\,-1}\,.\,K^{\,-1}}$ and $c_{\,1} \egal 1.5 \e{5} \ \mathsf{J\,.\,m^{\,-3}\,.\,K^{\,-1}}\,$. Note that Table~\ref{tab:mat_pro_verif} synthesizes the material properties for the three verification cases. Initially the wall is at $T_{\,0} \egal -\,5 \ \mathsf{\degC} \,$. It is stressed according to the boundary condition illustrated in Figure~\ref{fig:monolayer_BC}. The radiation flux on the right side boundary is null. The surface transfer coefficient are fixed in time $h_{\,\infty\,,\,L} \egal 6 \ \mathsf{W\,.\,m^{\,-2}\,.\,K^{\,-1}}$ and $k_{\,\infty\,,\,R} \egal 7.5 \ \mathsf{W\,.\,m^{\,-2}\,.\,K^{\,-1}}\,$. No long wave radiation is assumed for this case. The time horizon is $\tf \egal 30 \ \mathsf{d}\,$.

The temperature in the material is computed using three numerical models. First, the classical one (COM) is used with a space discretisation $\Delta \chi \egal 10^{\,-2}\,$. Then, the two \revision{reduced-order} models with space (PODt) and time (PODx) \revision{reduced basis}. For both of them, the preliminary solutions is computed with the COM and $30 \, \times \, 24$ snapshots equally space in time ($\Delta \tau \egal 1$) are considered to build the POD basis. The order of the ROM is $N \egal 6\,$. All solver tolerances are set to $10^{\,-4}\,$. A fourth solution is computed using a numerical pseudo--spectral approach obtained with the \texttt{Matlab\texttrademark} open source toolbox \texttt{Chebfun} \cite{driscoll_chebfun_2014}. It is chosen as reference solution to compute the accuracy of the three first models and abbreviated by CF. 

\begin{figure}[h!]
\begin{center}
\subfigure[
\label{fig:monolayer_T_ft}]{\includegraphics[width=.45\textwidth]{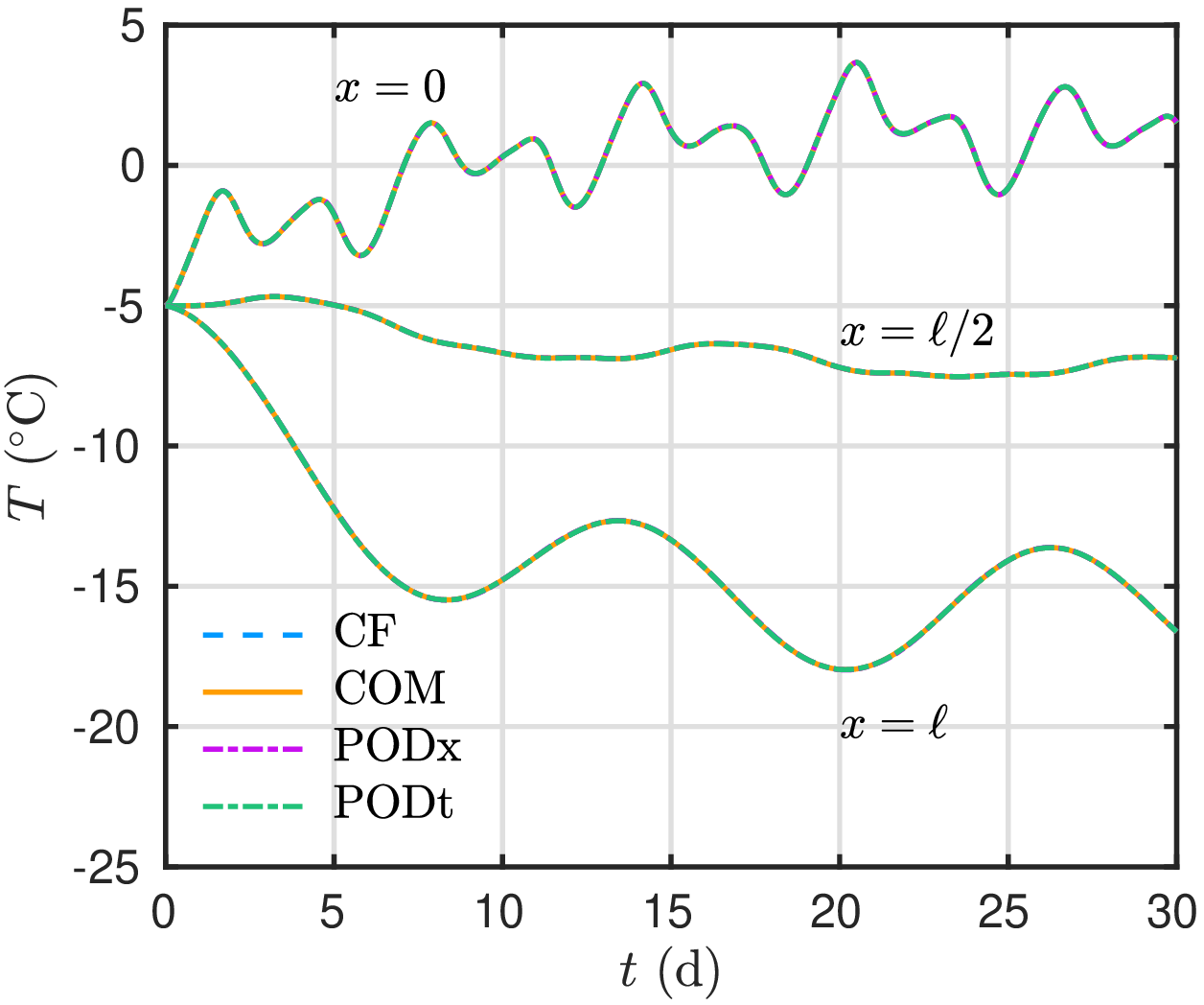}} \hspace{0.2cm}
\subfigure[
\label{fig:monolayer_T_fx}]{\includegraphics[width=.45\textwidth]{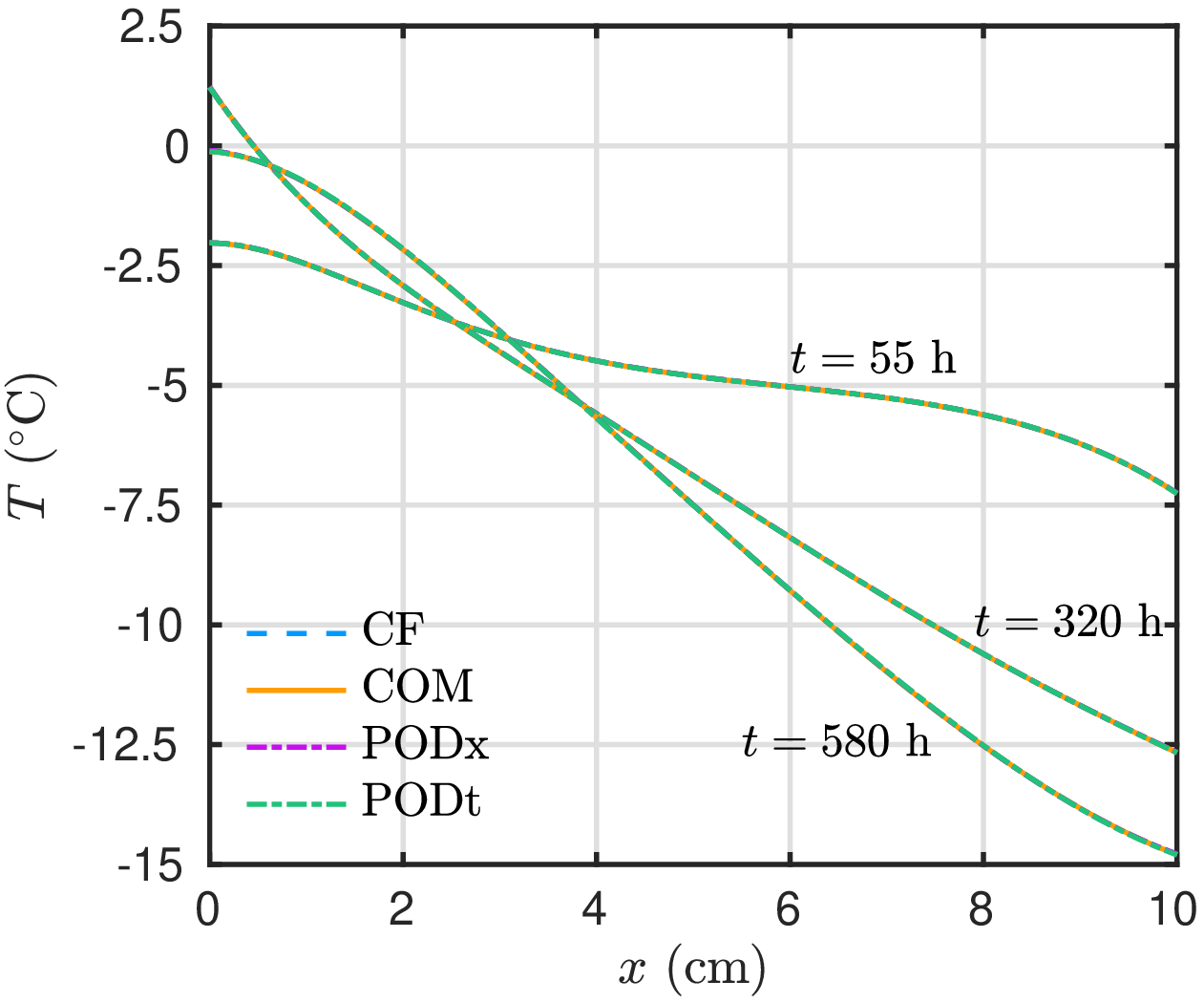}} \\
\caption{Time evolution of the computed temperature inside the material \emph{(a)}. Profiles of the the computed temperature inside the material \emph{(b)}.}
\end{center}
\end{figure}

The time evolution and profiles of the temperature are shown in Figures~\ref{fig:monolayer_T_ft} and \ref{fig:monolayer_T_fx}. It can be remarked that the four solution are overlapped, indicating a good agreement between them. The accuracy of the numerical solutions is evaluated for the solution $u$ and its first space partial derivative $\displaystyle \pd{u}{x}\,$ used for the computation of the density heat flux. All solutions have an error scaling with $\mathcal{O}(\,10^{\,-4}\,)\,$. For the derivative, the error is also very satisfying at the order of $\mathcal{O}(\,10^{\,-3}\,)\,$. An order of accuracy is lost for the spatial flux computation. 

Last, with the chosen parameters, the degree of freedom of the models are:
\begin{align*}
\mathfrak{D}(\,\mathrm{COM}\,) & \egal 101 \, \times \, 721 \egal 72821\,, \\[4pt]
\mathfrak{D}(\,\mathrm{PODt}\,) & \egal 6 \, \times \, 721 \egal 4326 \,, \\[4pt]
\mathfrak{D}(\,\mathrm{PODx}\,) & \egal 101 \, \times \, 6 \egal 106  \,.
\end{align*}
One can observe the reduction of the degree of freedom of the solution with the POD \revision{reduced-order} model using a time basis functions. 

\begin{figure}[h!]
\begin{center}
\subfigure[
\label{fig:monolayer_eps_u_fN}]{\includegraphics[width=.45\textwidth]{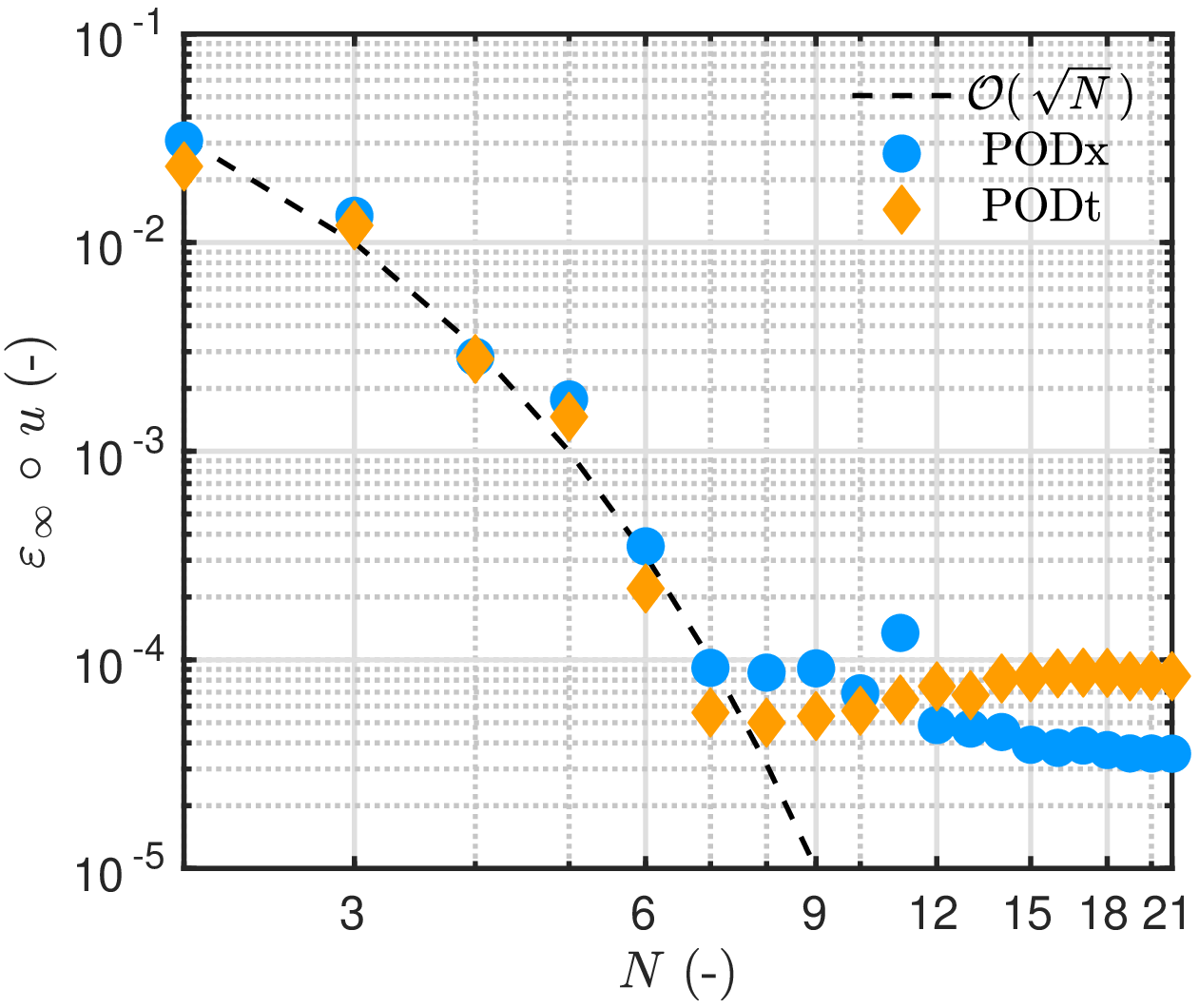}} \hspace{0.2cm}
\subfigure[
\label{fig:monolayer_eps_ux_fN}]{\includegraphics[width=.45\textwidth]{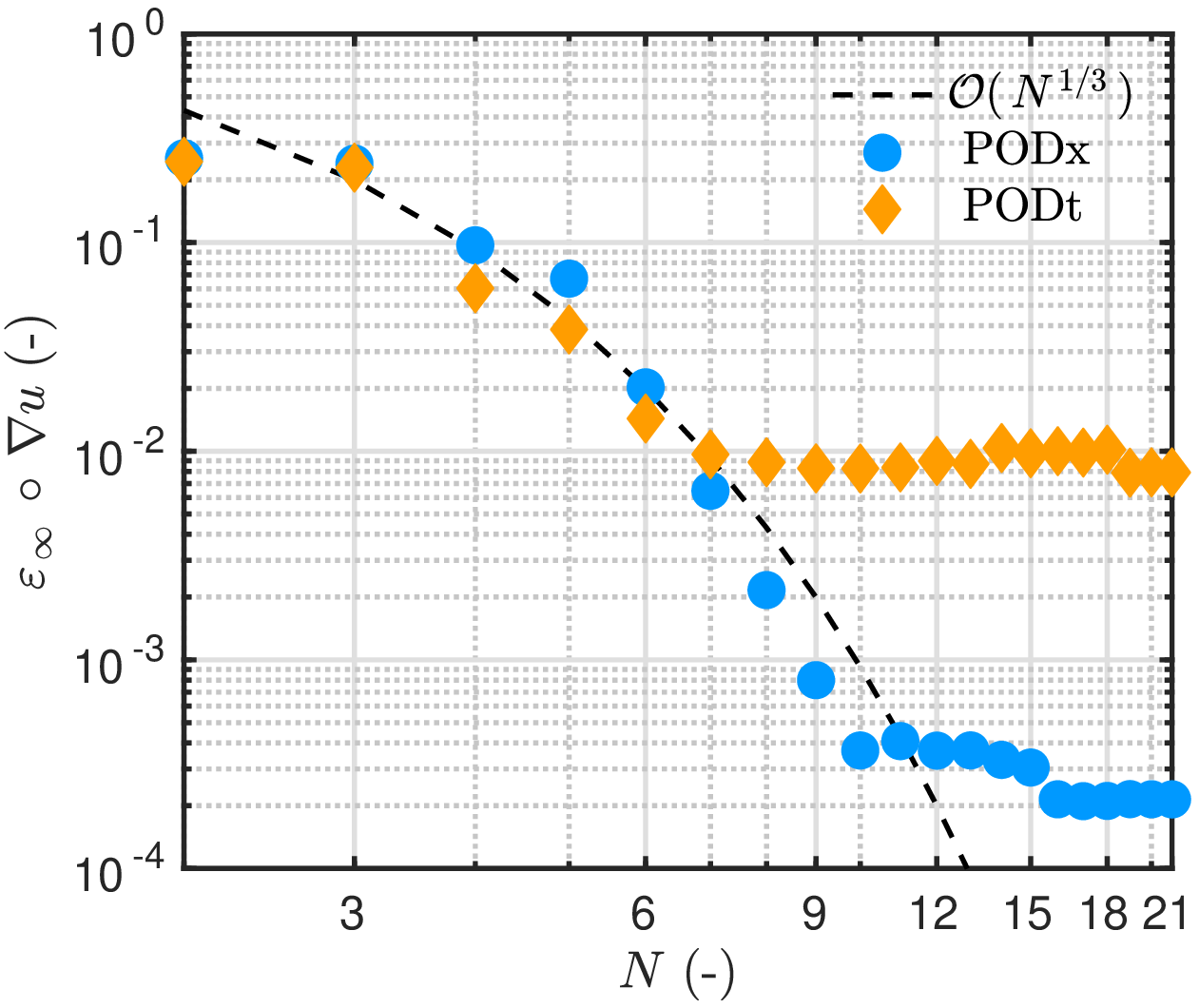}} \\
\caption{Evolution of the error on the computed temperature \emph{(a)} and heat flux \emph{(b)} according to the modes number of the \revision{reduced-order} models.}
\label{fig:monolayer_eps_fN}
\end{center}
\end{figure}

\begin{figure}[h!]
\begin{center}
\includegraphics[width=.95\textwidth]{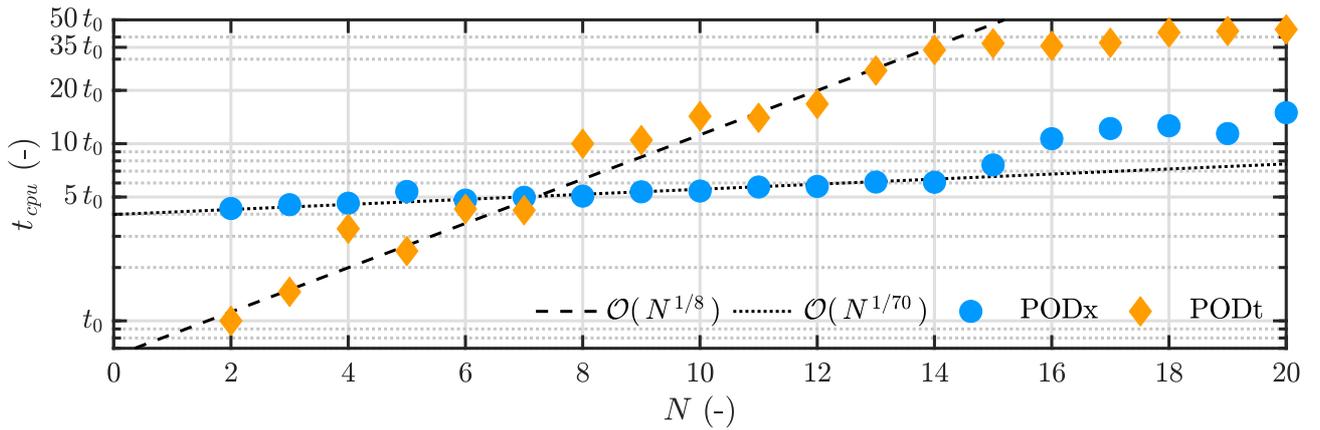}
\caption{Evolution of the computational time required by the \revision{reduced-order} models to compute the solution according to their modes number.}
\label{fig:monolayer_tcpu_fN}
\end{center}
\end{figure}

Complementary computations are realized to carry a parametric study according to the mode number. The accuracy and computational costs are evaluated for PODt and PODx according to $N\,$. Other parameters remains at the same values. Figure~\ref{fig:monolayer_eps_fN} shows the error variation for the solution and its space derivative. The error decrease similarly with the mode number for both ROM. For the solution, the error reaches a threshold at $\mathcal{O}(\,10^{\,-5}\,)\,$. For the space derivative, the error decreases further for the PODx until $\mathcal{O}(\,10^{\,-4}\,)$ while the PODt is constant at $\mathcal{O}(\,10^{\,-2}\,)$. The evolution of the computational cost is presented in Figure~\ref{fig:monolayer_tcpu_fN}, considering a reference time $t_{\,0} \egal 0.2 \ \mathsf{s}\,$. For $N \, < \, 8\,$, the computational cost is five times higher for the PODx. It switches after for $N \, \geqslant \, 8\,$. The increase of the computational time is slower for the PODx model. From this parametric analysis, it can be deduced that, for a mode number around ten, the PODx has a better accuracy for the solution, particularly for the computation of the heat flux. Furthermore, the computational cost compared to the classical PODt approach is two times smaller. 

\begin{figure}[h!]
\begin{center}
\subfigure[
\label{fig:monolayer_e2_fx}]{\includegraphics[width=.45\textwidth]{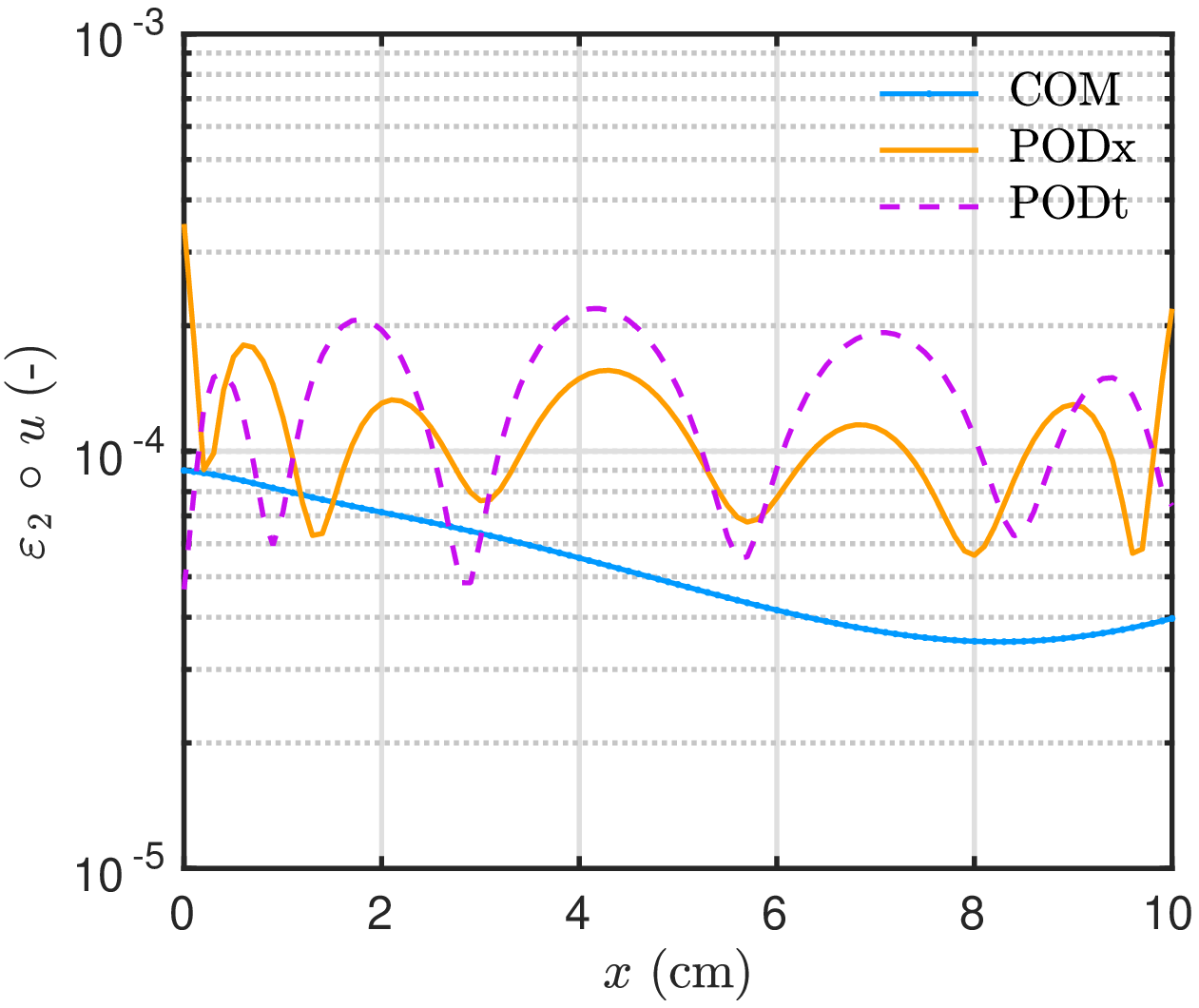}} \hspace{0.2cm}
\subfigure[
\label{fig:monolayer_e2diff_fx}]{\includegraphics[width=.45\textwidth]{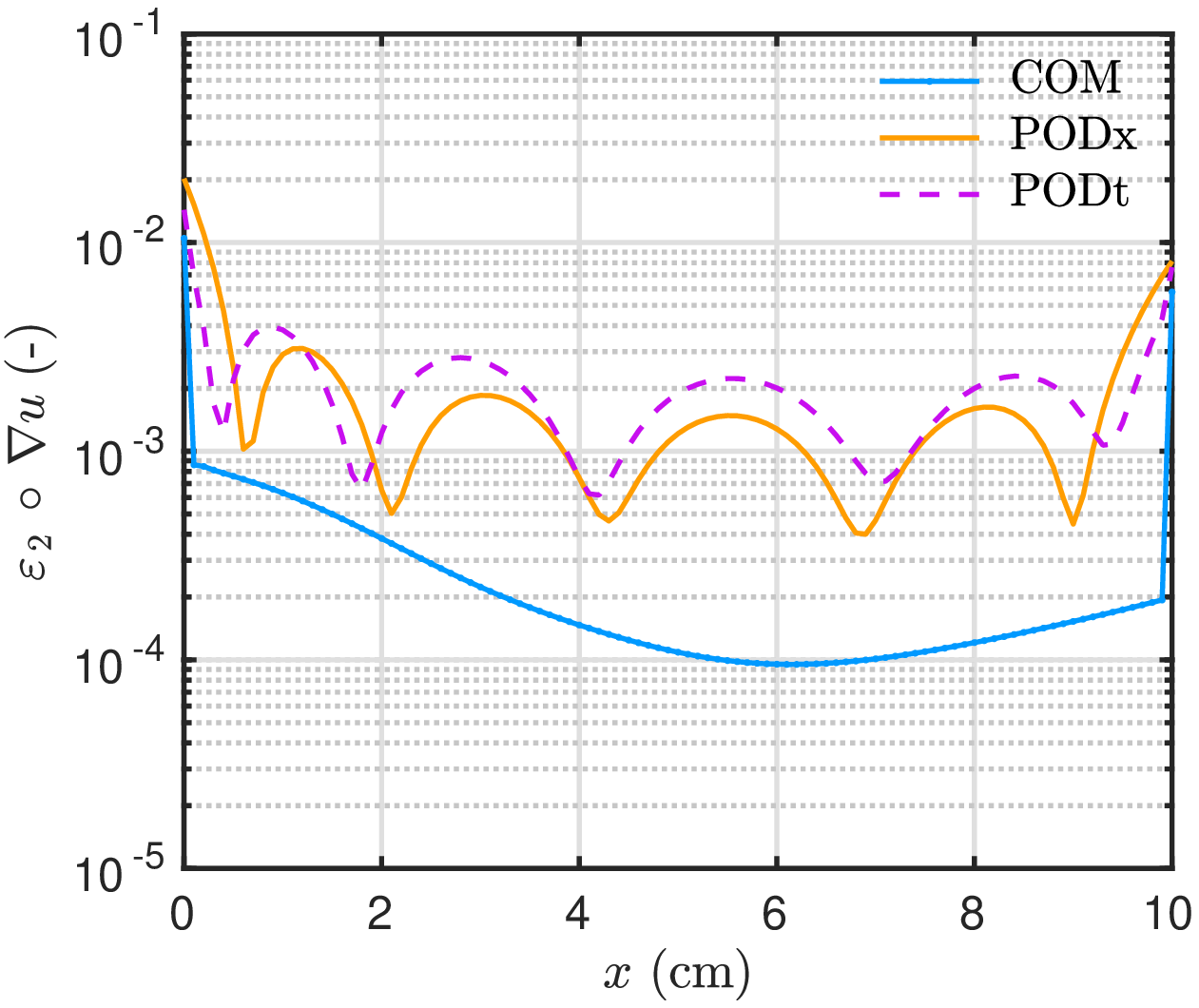}} \\
\caption{Evolution of the error on the computed temperature \emph{(a)} and heat flux \emph{(b)}.}
\end{center}
\end{figure}

\begin{table}
\centering
\caption{Material properties considered for the three verification case.}
\label{tab:mat_pro_verif}
\setlength{\extrarowheight}{.5em}
\begin{tabular}[l]{@{} c cccc}
\hline
\hline
\multirow{2}{*}{\textit{Case study}} & \multirow{2}{*}{\textit{Layers id.}} & \textit{Layers length} & \textit{Thermal conductivity} & \textit{Volumetric heat capacity}  \\
 & 
& $\ell_{\,i} \unit{cm}$ 
& $k_{\,i} \unit{W\,.\,m^{\,-1}\,.\,K^{\,-1}}$
& $c_{\,i} \unit{J\,.\,m^{\,-3}\,.\,K^{\,-1}}$ \\
Section~\ref{ssec:mono_layer}
& $1\,$ & $10$ & $0.5$ & $1.5 \e{5}$ \\ \hline
\multirow{2}{*}{Section~\ref{ssec:multi_layer}}
& $1$ & $18$ & $1.0$ & $1.3 \e{6}$ \\
& $2$ & $60$ & $0.5$ & $2.45 \e{5}$ \\\hline
\multirow{2}{*}{Section~\ref{ssec:multi_layer_param}}
& $1$ & $12$ & $1.0$ & $1.85 \e{6}$ \\
& $2$ & $\bigl[\,10 \,,\, 30\,\bigr]$ & $\bigl[\,0.1 \,,\, 2\,\bigr]$ & $1.0 \e{6}$ \\
\hline 
\hline
\end{tabular}
\end{table}

\subsection{Multi-layer case with no basis interpolation}
\label{ssec:multi_layer}

Transfer in a multi-layer wall is now investigated to verify the reliability of the proposed \revision{reduced-order} model. Such case is challenging in terms of computation at the interface between two layers. For this, an analytical solution is built based on a separated representation of the solution. The details of the solution construction are given in Appendix~\ref{sec:analytical_solution}.  Again, no interpolation of \revision{reduced basis} is considered here. The reduced basis is built using a preliminary solution computed for the same inputs parameters. The problem considers a wall with two layers. The interface position is at $\ell_{\,2} \egal 18 \ \mathsf{cm}$ and the total length of the wall is $\ell_{\,3} \egal 60 \ \mathsf{cm}$. The properties are $k_{\,1} \egal 1 \ \mathsf{W\,.\,m^{\,-1}\,.\,K^{\,-1}}$ and $c_{\,1} \egal 1.3 \e{6} \ \mathsf{J\,.\,m^{\,-3}\,.\,K^{\,-1}}$ for the layer $1$ and $k_{\,2} \egal 0.5 \ \mathsf{W\,.\,m^{\,-1}\,.\,K^{\,-1}}$ and $c_{\,2} \egal 2.45 \e{5} \ \mathsf{J\,.\,m^{\,-3}\,.\,K^{\,-1}}$ for the layer $2\,$. \textsc{Dirichlet} homogeneous condition are assumed at the left and right boundaries. The initial conditions is:
\begin{align*}
T_{\,0}(\,x\,) \egal
\begin{cases} 
\ 35 \ \sin \Bigl(\, \frac{\pi}{2} \, \frac{x}{\ell_{\,2}}\,\Bigr) \moins 5 \,, & \quad \forall \, x \, \in \, \Omega_{\,x}^{\,1} \,, \\
\ 35 \ \sin \Bigl(\, \frac{\pi}{2} \, \frac{\ell_{\,3} \moins x}{\ell_{\,3} \moins \ell_{\,2}}\,\Bigr) \moins 5 \,, & \quad \forall \, x \, \in \, \Omega_{\,x}^{\,2}\,,
\end{cases} 
\qquad \unit{\degC} \,.
\end{align*}
The final simulation time is $\tf \egal 6 \ \mathsf{h}\,$.

\begin{figure}[h!]
\begin{center}
\subfigure[
\label{fig:multilayer_u_ft}]{\includegraphics[width=.45\textwidth]{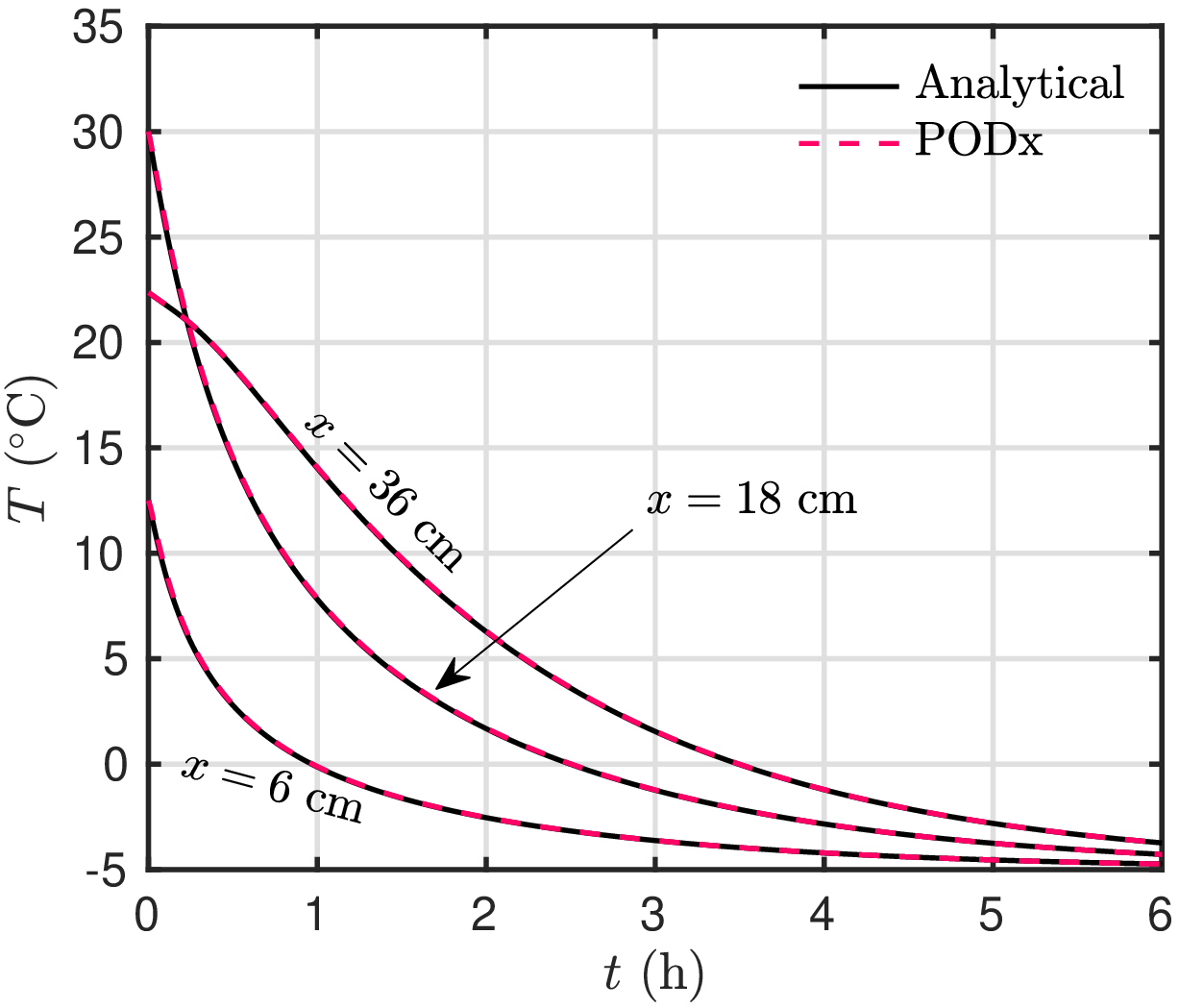}} \hspace{0.2cm}
\subfigure[
\label{fig:multilayer_j_ft}]{\includegraphics[width=.45\textwidth]{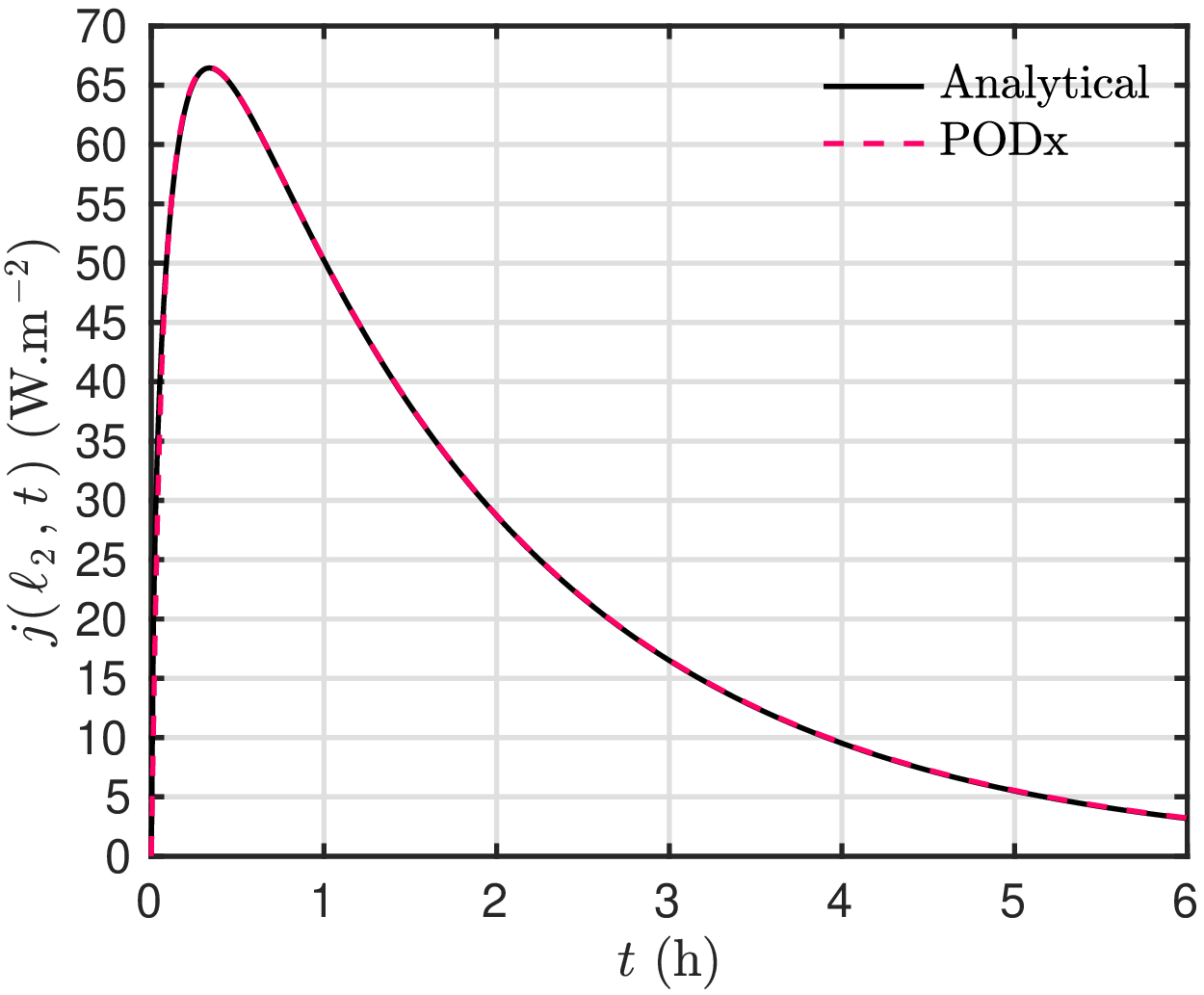}} \\
\subfigure[
\label{fig:multilayer_u_fx}]{\includegraphics[width=.45\textwidth]{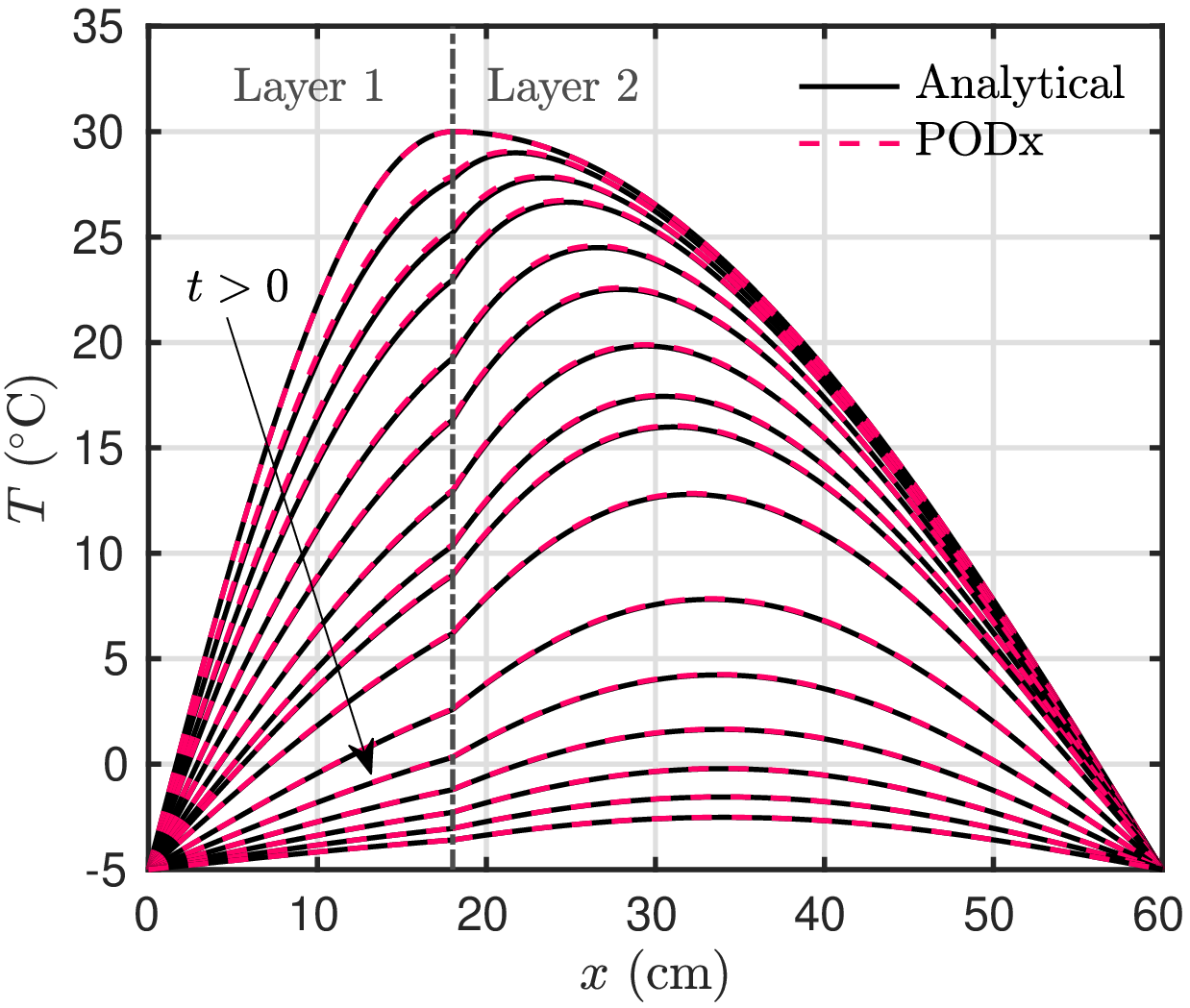}} \hspace{0.2cm}
\subfigure[
\label{fig:multilayer_e2_fx}]{\includegraphics[width=.45\textwidth]{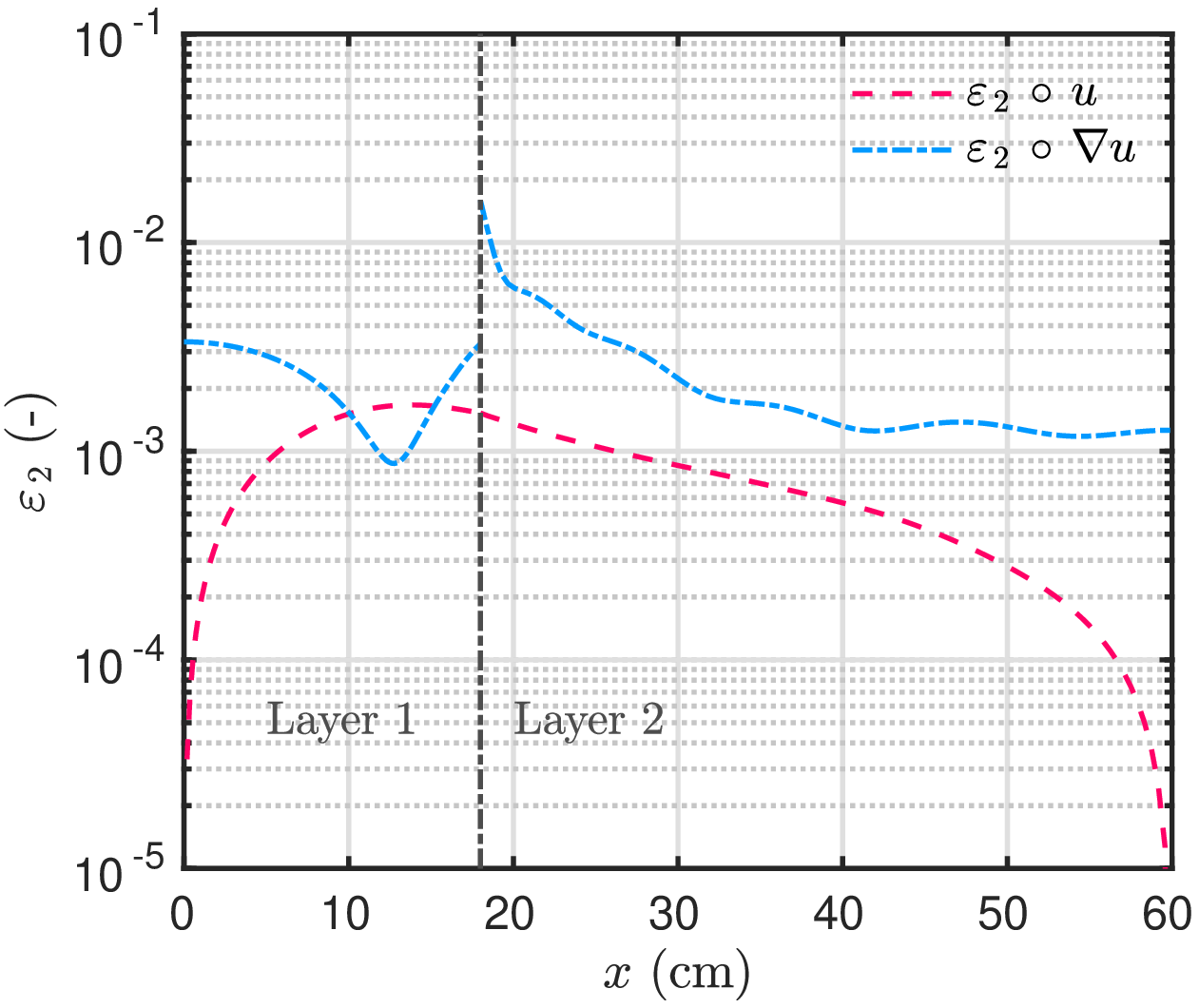}} \\
\caption{Time evolution of the computed temperature \emph{(a)} and heat flux \emph{(b)}. Evolution of the computed temperature \emph{(c)} and of the error \emph{(d)} according to space.}
\end{center}
\end{figure}

The \revision{reduced-order} model PODx using a time \revision{reduced basis} is built using a preliminary solution computed with the COM using $\Delta \chi \egal 10^{\,-2}\,$. Then, $6 \e{3} $ snapshots equally spaced in time ($\Delta \tau \egal 10^{\,-3}$) are used to build the POD basis. The order of the PODx is $N \egal 5$  for each layer. The analytical solution is computed using $30$ modes. Figures~\ref{fig:multilayer_u_ft} and \ref{fig:multilayer_j_ft} show the time evolution of the temperature and heat flux for both models. It highlights a satisfying agreement between both. The profiles of temperature are given in Figure~\ref{fig:multilayer_u_fx}. The temperature and the flux at the interface between both materials $x \egal \ell_{\,2}$ are precisely predicted by the PODx. Last, Figure~\ref{fig:multilayer_e2_fx} presents the error on the both solution ($\varepsilon_{\,2} \,\circ\,u$) and its space derivative ($\varepsilon_{\,2} \,\circ\,\nabla \,u$). The error scales with $\mathcal{O}(\,10^{\,-3}\,)$ for both fields. Those results demonstrates the good reliability of the \revision{reduced-order} model to predict the heat transfer in walls with multiple layers.

\subsection{Multi-layer case: parametric model}
\label{ssec:multi_layer_param}

The first and second case study verified the efficiency and accuracy of the \revision{reduced-order} model for the computation of the temperature in multi-layer walls. The last case study focuses on the reliability of parametric solutions. In other words, the accuracy of the the \revision{reduced-order} model with basis interpolation is now evaluated. 

A wall with $N_{\,\ell} \egal 2$ layers is analyzed. The first layers is composed of bricks with the following properties: $\ell_{\,2} \egal 12 \ \mathsf{cm}\,$, $k_{\,1} \egal 1.0 \ \mathsf{W\,.\,m^{\,-1}\,.\,K^{\,-1}}$ and $c_{\,1} \egal 1.85 \ \mathsf{MJ\,.\,m^{\,-3}\,.\,K^{\,-1}}\,$. The parametric solutions is built according to the thermal conductivity $k_{\,2}$ and length of the second layers $d_{\,2} \egal \ell_{\,3} \moins \ell_{\,2}\,$, as well as the outside (left) surface transfer coefficient $h_{\,L}\,$. The domain of variation are the following: 
\begin{align*}
\Omega_{\,k_{\,2}} \egal \bigl[\,0.1 \,,\, 2\,\bigr]  \ \mathsf{W\,.\,m^{\,-1}\,.\,K^{\,-1}} \,, \qquad
\Omega_{\,d_{\,2}} \egal \bigl[\,0.1 \,,\, 0.3\,\bigr] \ \mathsf{m} \,, \qquad 
\Omega_{\,h_{\,L}} \egal \bigl[\,10 \,,\,30 \,\bigr]  \ \mathsf{W\,.\,m^{\,-1}\,.\,K^{\,-2}} \,, \qquad
\end{align*}
The heat capacity of the second layer is set to $c_{\,2} \egal 1  \ \mathsf{MJ\,.\,m^{\,-3}\,.\,K^{\,-1}}\,$. No long wave radiation is assumed for this case. The right surface transfer coefficient is set to $h_{\,R} \egal 7 \ \mathsf{W\,.\,m^{\,-1}\,.\,K^{\,-2}}\,$. The boundary conditions are set as follows: 
\begin{align*}
T_{\,\infty\,,\,L} & \egal 20 \plus 5 \ \sin \, \biggl(\, \frac{t}{24 \cdot 3600}\,\biggr)
\quad \unit{\degC} \,, \\[4pt]
q_{\,\infty\,,\,L} & \egal 150 \ \Biggl(\, \sin \, \biggl(\, \frac{t}{6 \cdot 3600}  \,\biggr) \,\Biggr)^{\,6} \quad \unit{W\,.\,m^{\,-2}} \,, \\[4pt]
T_{\,\infty\,,\,R} & \egal 20 \plus 2.5 \ \Bigl(\, 1 \moins \cos \, \biggl(\, \frac{t}{24 \cdot 3600}  \,\biggr) \, \Biggr) \quad \unit{\degC} \,.
\end{align*}
The initial condition is $T_{\,0} \egal 20 \ \mathsf{\degC}\,$. The model order is set to $N \egal 10\,$. The horizon of the simulation is $\tf \egal 0.5 \ \mathsf{h}\,$.

\begin{figure}[h!]
\begin{center}
\includegraphics[width=.45\textwidth]{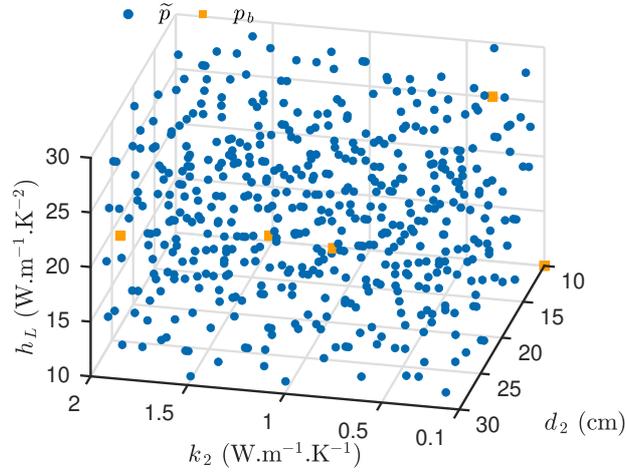}
\caption{Samples in the parameter space $\Omega_{\,p} \egal \Omega_{\,k_{\,2}} \, \bigcup \, \Omega_{\,c_{\,2}} \, \bigcup \, \Omega_{\,h_{\,L}} \,$.}
\label{fig:param_Pspace}
\end{center}
\end{figure}

\begin{figure}[h!]
\begin{center}
\subfigure[
\label{fig:param_epsu_fp}]{\includegraphics[width=.45\textwidth]{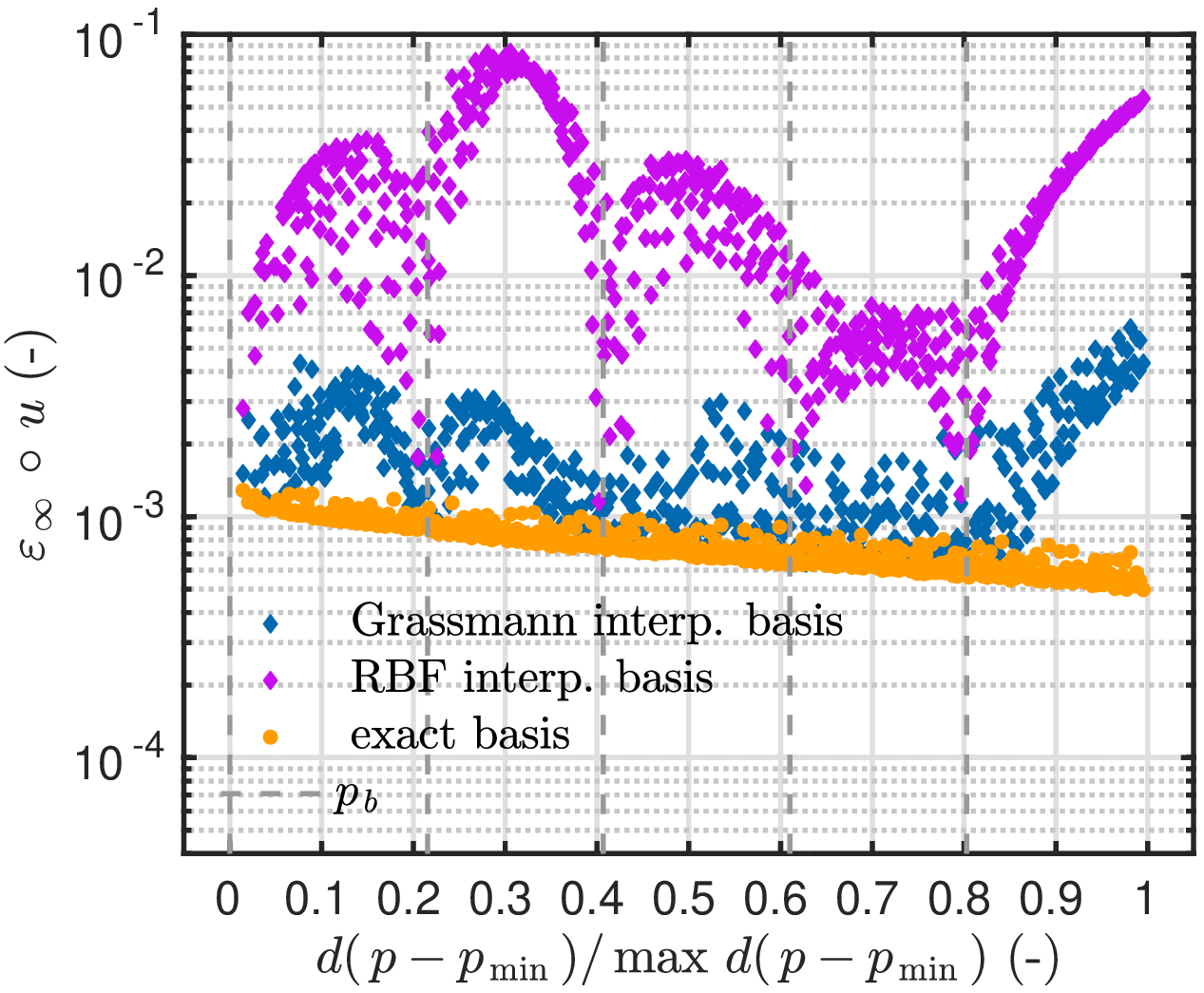}} \hspace{0.2cm}
\subfigure[
\label{fig:param_epsux_fp}]{\includegraphics[width=.45\textwidth]{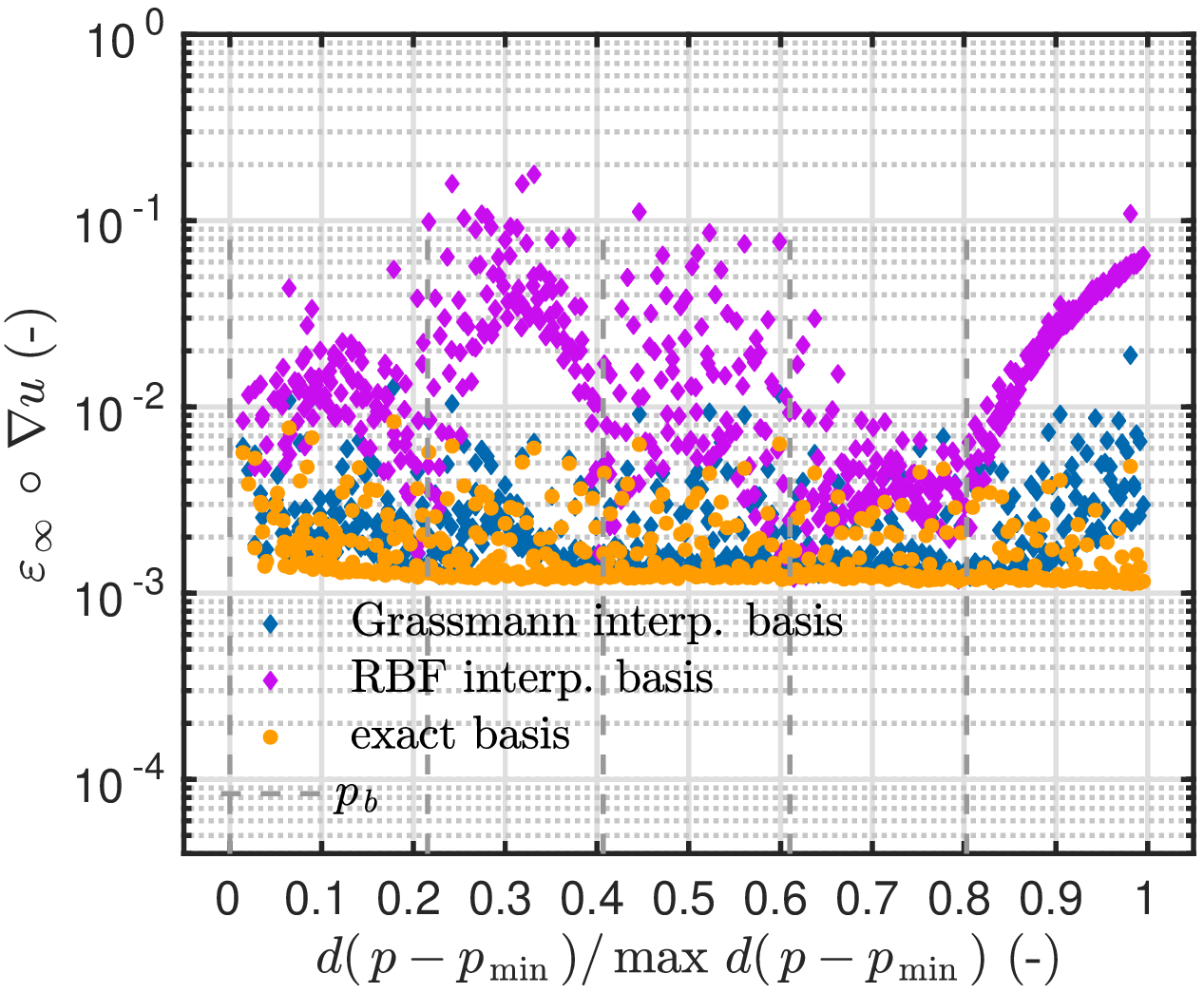}} 
\caption{Variation of the error on the solution \emph{(a)} and its space derivative \emph{(b)} according to the distance with $p_{\,\min}$.}
\end{center}
\end{figure}

Computations are first carried with a $N_{\,b} \egal 5$ basis chosen for reduced basis interpolation in the parameter space. The parameters selected of the related basis are chosen randomly and are illustrated in Figure~\ref{fig:param_Pspace}. The five basis are generated using the COM model with a tolerance set to $\mathrm{Tol} \egal 10^{\,-4}$ using $\Delta \chi \egal 10^{\,-2}\,$. Then, $500$ snapshots equally spaced in time ($\Delta \tau \egal 10^{\,-3}$) are used to build the POD basis. A total of $500$ parameters are randomly chosen in the parameter space as shown in Figure~\ref{fig:param_Pspace}. For each parameter, the solution and its partial derivative are computed for a model order $N \egal 10$ for each layer. Three computations are carried out: one using the reduced basis interpolated on the \textsc{Grassmann} manifold, a second one with the exact basis (generated with the COM for the same parameters) and a last one with a simple RBF interpolation of the reduced basis given the preliminary computed basis. Note that for the first and third cases, the same five basis are used to carry out the interpolations. The errors for the solution and its space derivative are presented in Figures~\ref{fig:param_epsu_fp} and \ref{fig:param_epsux_fp}, respectively. For the \textsc{Euclidean} distance computations, the origin point is $p_{\,\min} \egal \bigl(\,k_{\,2} \egal 0.1\,,\, d_{\,2} \egal 0.1 \,,\, h_{\,L} \egal 10 \,\bigr)\,$. It corresponds to the point with minimum thermal conductivity, length and surface transfer coefficient. The error for the solution $u$ scales with $\mathcal{O}(\,10^{\,-3}\,)$ for the solution with exact basis. For the case with \textsc{Grassmann} interpolated basis, the error is a bit higher but still very acceptable. It oscillates with minimum when the distance corresponds to the one of the parameters used for generating the basis. Note that the error increases when the distance becomes important, near $1\,$. Indeed, no basis have been generated for parameters with $d(\,p \moins p_{\,\min}\,) \, \geqslant \, 1\,$. Regarding the simple RBF interpolation of the reduced basis, the accuracy of the solution is greatly depreciated. The error becomes only acceptable when the new parameter is closed to the one from the preliminary computed basis. This results justify the use of interpolation on a tangent space of \textsc{Grassmann} manifold.
Last, considering the spatial derivative of the solution, the results are similar for cases one and two. The error is very satisfying below $10^{\,-2}\,$. There is no apparent variation according to the distance in the parameter space.

\begin{figure}[h!]
\begin{center}
\subfigure[
\label{fig:param_u_fx_p1}]{\includegraphics[width=.45\textwidth]{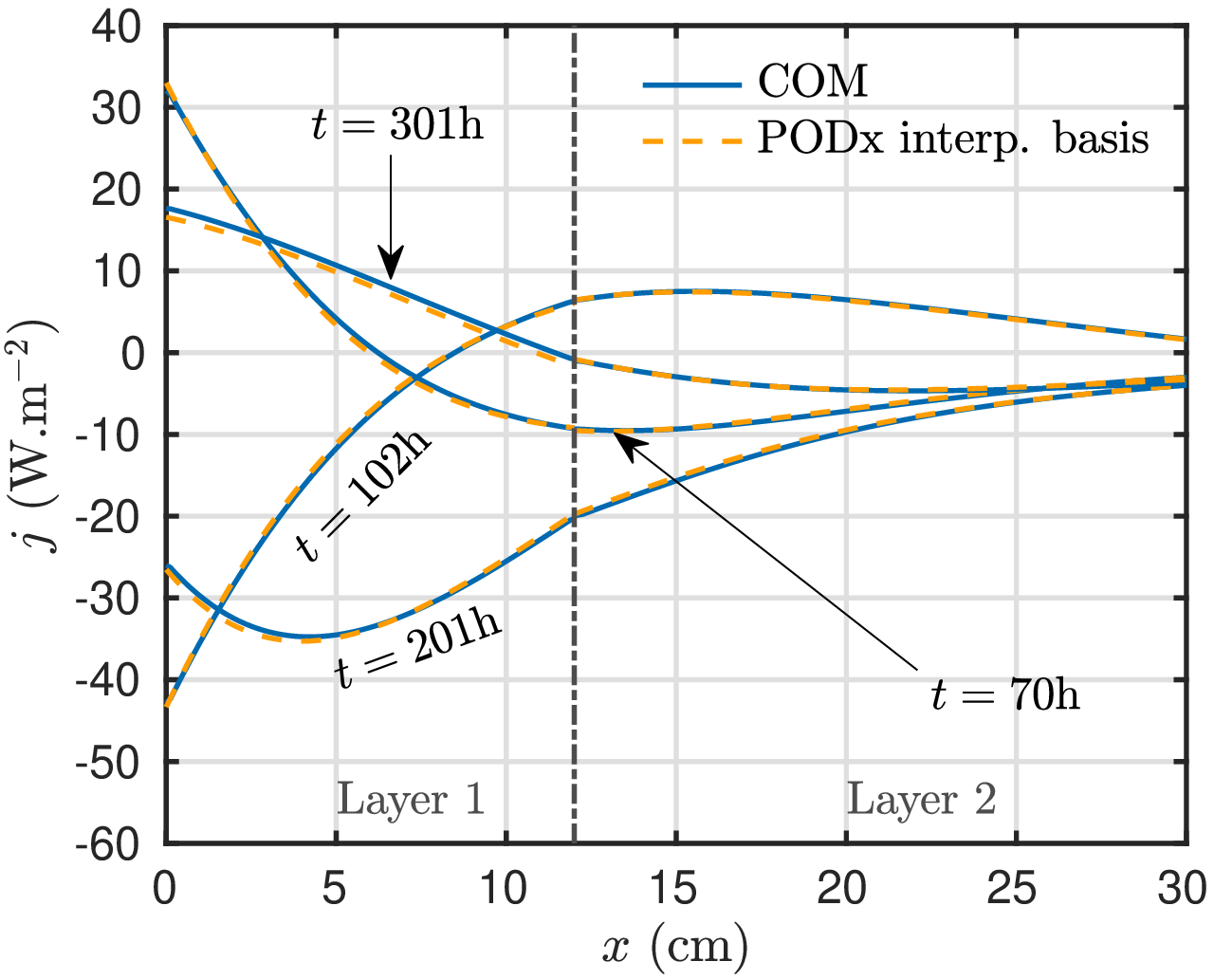}} \hspace{0.2cm}
\subfigure[
\label{fig:param_u_fx_p2}]{\includegraphics[width=.45\textwidth]{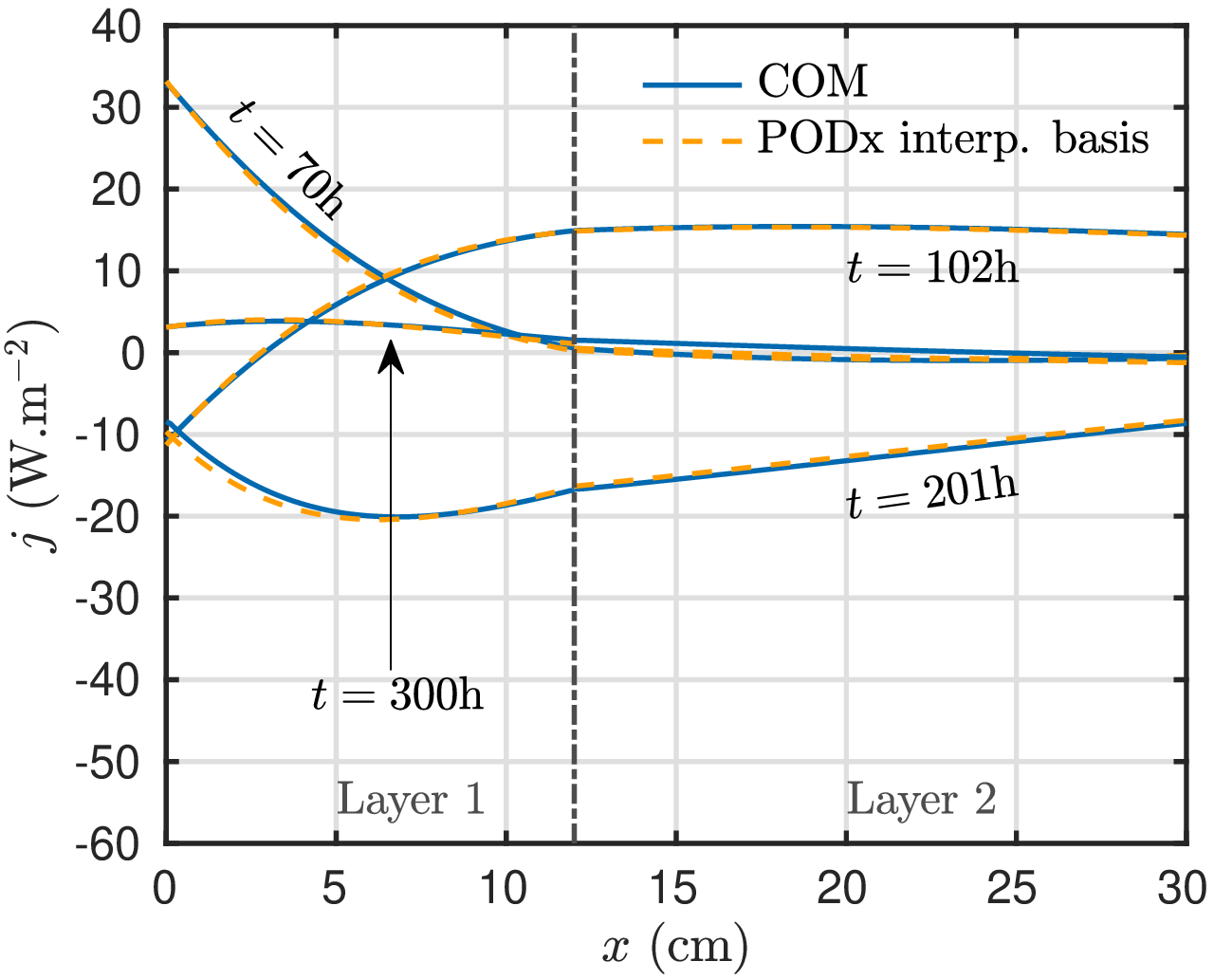}} 
\caption{Time evolution of the computed heat flux for $p_{\,1}$ \emph{(a)} and $p_{\,2}\,$ \emph{(b)}.}
\label{fig:param_u_fx}
\end{center}
\end{figure}

The heat flux is computed with the interpolated basis \revision{reduced-order} model for two parameters
\begin{align*}
p_{\,1} \egal \bigl(\,k_{\,2} \egal 0.45\,,\, d_{\,2} \egal 0.26 \,,\, h_{\,L} \egal 10.2 \,\bigr) \,, \quad d(\,p_{\,1} \moins p_{\,\min}\,) \egal 0.02 \cdot \max d(\,p \moins p_{\,\min}\,) \,,\\[4pt]
p_{\,2} \egal \bigl(\,k_{\,2} \egal 1.58\,,\, d_{\,2} \egal 0.18 \,,\, h_{\,L} \egal 28.4 \,\bigr) \,,\quad d(\,p_{\,2} \moins p_{\,\min}\,) \egal 0.92 \cdot \max d(\,p \moins p_{\,\min}\,) \,.
\end{align*}
The solution is compared to the one obtained with the COM in Figure~\ref{fig:param_u_fx}. A very good agreement is observed between the two solutions. Even for a parameter $p_{\,2}$ with a high distance, the error between both solution is very satisfying. Note that the flux continuity at the interface between two layers is verified. 

\begin{figure}[h!]
\begin{center}
\subfigure[
\label{fig:param_cpu_fp}]{\includegraphics[width=.45\textwidth]{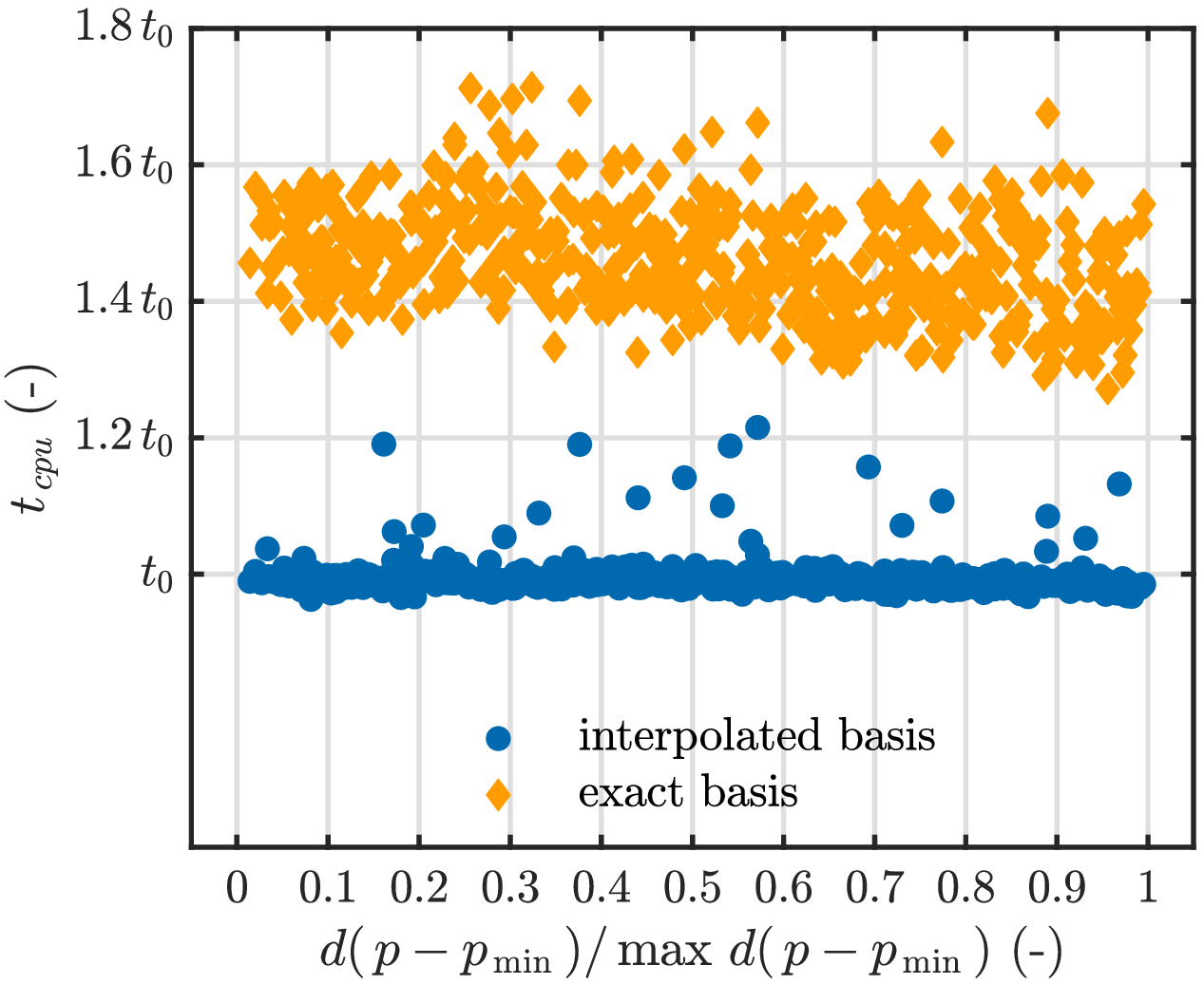}} \hspace{0.2cm}
\subfigure[
\label{fig:param_eps_fNphi}]{\includegraphics[width=.45\textwidth]{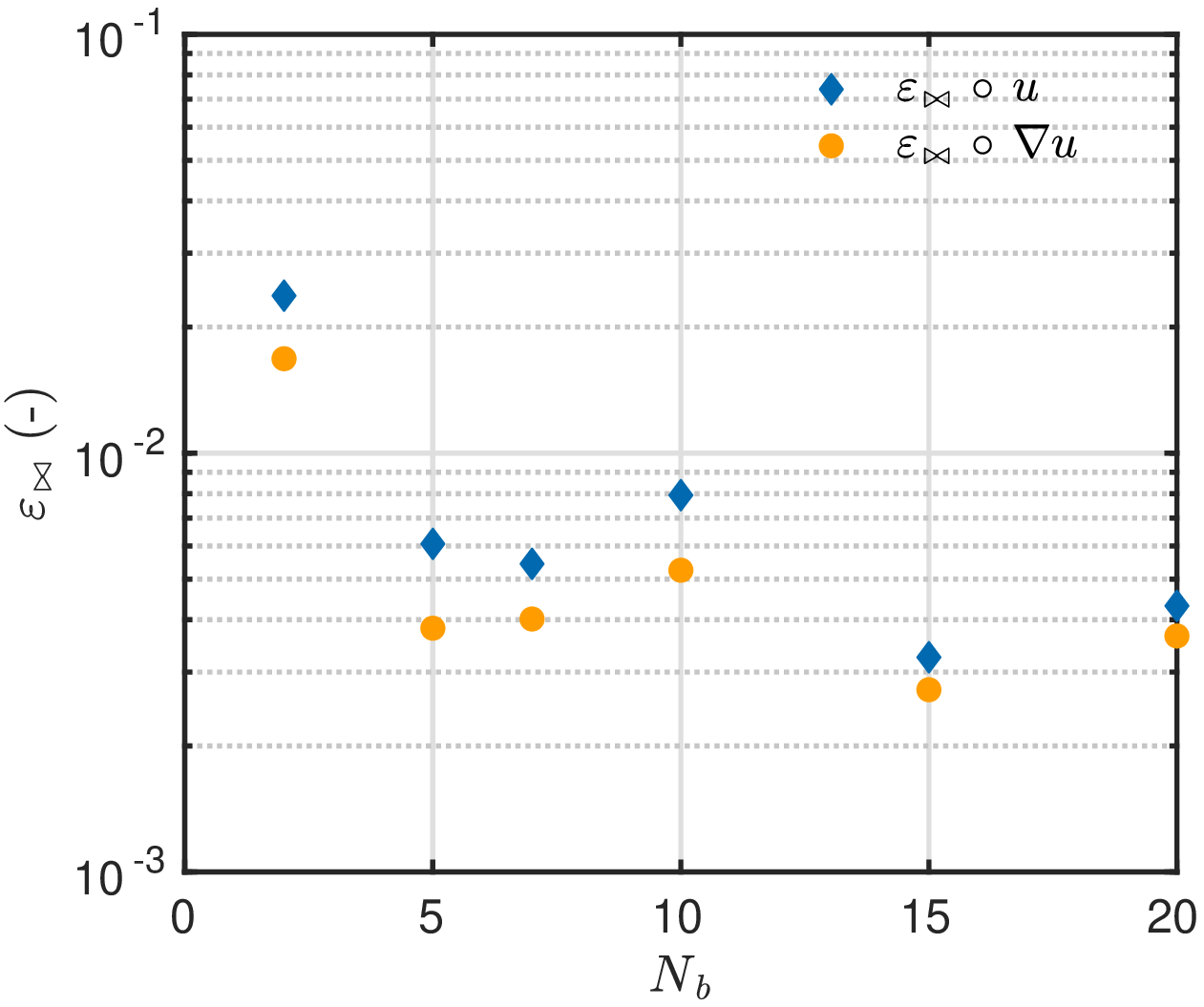}} 
\caption{Variation of the computational time according to the distance with $p_{\,\min}$ \emph{(a)} and of the error according to the number of basis $N_{\,b}$ used for the basis interpolation among the parameter space.}
\end{center}
\end{figure}

Regarding the computational time. A comparison is carried between the PODx with interpolated and exact reduced basis in Figure~\ref{fig:param_cpu_fp}. The model using interpolated basis is faster than the one using exact basis. The latter requires between $40$ and $60\ \%$ more time to generate the basis using the solution of the COM with the exact parameter and then compute the solution with the reduced model. It is also important to remark that the PODx with interpolated basis requires less than $1.5 \ \%$ of the COM. 

Last, additional computations are performed according to the number of basis $N_{\,b}$ used for the interpolation of others. The error is presented in Figure~\ref{fig:param_eps_fNphi}. For a small number of basis, $N_{\,b} \egal 2\,$, the model lack of accuracy. For $N_{\,b} \, \geqslant 5\,$, the accuracy of the PODx is very satisfying around $\mathcal{O}(\,10^{\,-3}\,)\,$. Those results highlight the reliability of the PODx with interpolated basis to compute with high precision the temperature and heat flux in multi-layers wall. 

\section{Case study}
\label{sec:real_case_study}

\subsection{Description}

\begin{figure}[h!]
\centering
\includegraphics[width=.9\textwidth]{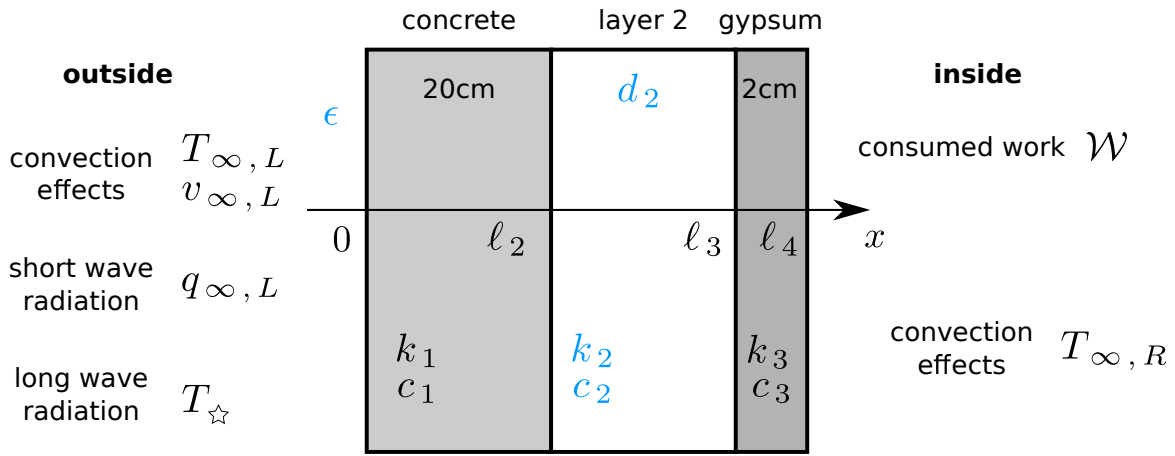}
\caption{Illustration of the realistic case study. The parameters highlighted in blue are considered for the parametric simulations.}
\label{fig:real_case}
\end{figure}

Since the model's efficiency has been demonstrated in previous Section, a case study is now considered to perform numerical predictions using the POD ROM. The objective is to perform parametric simulations to determine which wall configuration is the best in terms of energy consumption. Such configuration minimizes the wall consumed work $\mathcal{W}$ defined in Eq.~\eqref{eq:consumed_work}. As illustrated in Figure~\ref{fig:real_case} the wall is composed of three layers. The outside and inside layers are $20 \ \mathsf{cm}$ concrete and $2 \ \mathsf{cm}$ gypsum, respectively. Their material properties are $k_{\,1} \egal 1.65 \ \mathsf{W\,.\,m^{\,-1}\,.\,K^{\,-1}}\,$, $c_{\,1} \egal 2.2 \ \mathsf{MJ\,.\,m^{\,-3}\,.\,K^{\,-1}}\,$, $k_{\,3} \egal 0.25 \ \mathsf{W\,.\,m^{\,-1}\,.\,K^{\,-1}}$ and $c_{\,3} \egal 0.8 \ \mathsf{MJ\,.\,m^{\,-3}\,.\,K^{\,-1}}\,$. The parametric simulations are computed according to the thermal properties of the second layer, namely $k_{\,2}$ and $c_{\,2}$, its thickness $d_{\,2} \egal L_{\,3} \moins L_{\,2}$ and the emissivity of the outside surface $\epsilon\,$. Thus, the dimension of the parameter space is four. The  lower and upper bounds of each parameter are given in Table~\ref{tab:real_p_space}. The initial condition is set to the steady state regime at $T_{\,0} \egal 19 \ \mathsf{\degC}\,$.

\begin{figure}[h!]
\begin{center}
\subfigure[
\label{fig:real_Tinf_ft}]{\includegraphics[width=.95\textwidth]{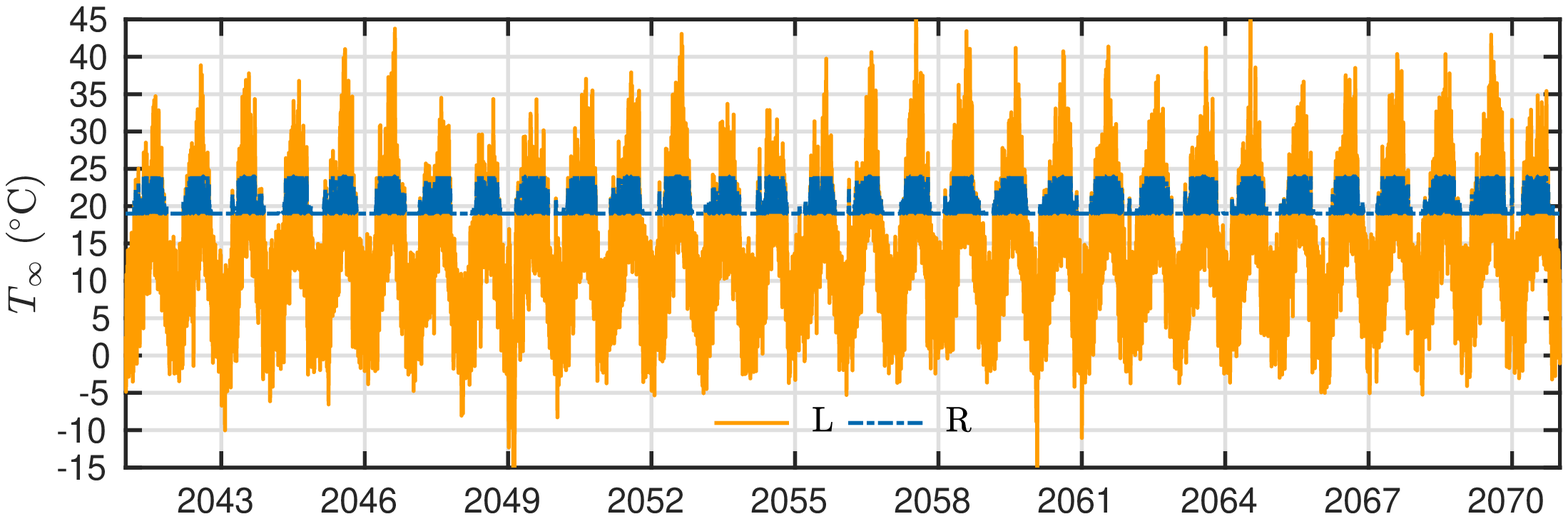}}  \\
\subfigure[
\label{fig:real_qinf_ft}]{\includegraphics[width=.95\textwidth]{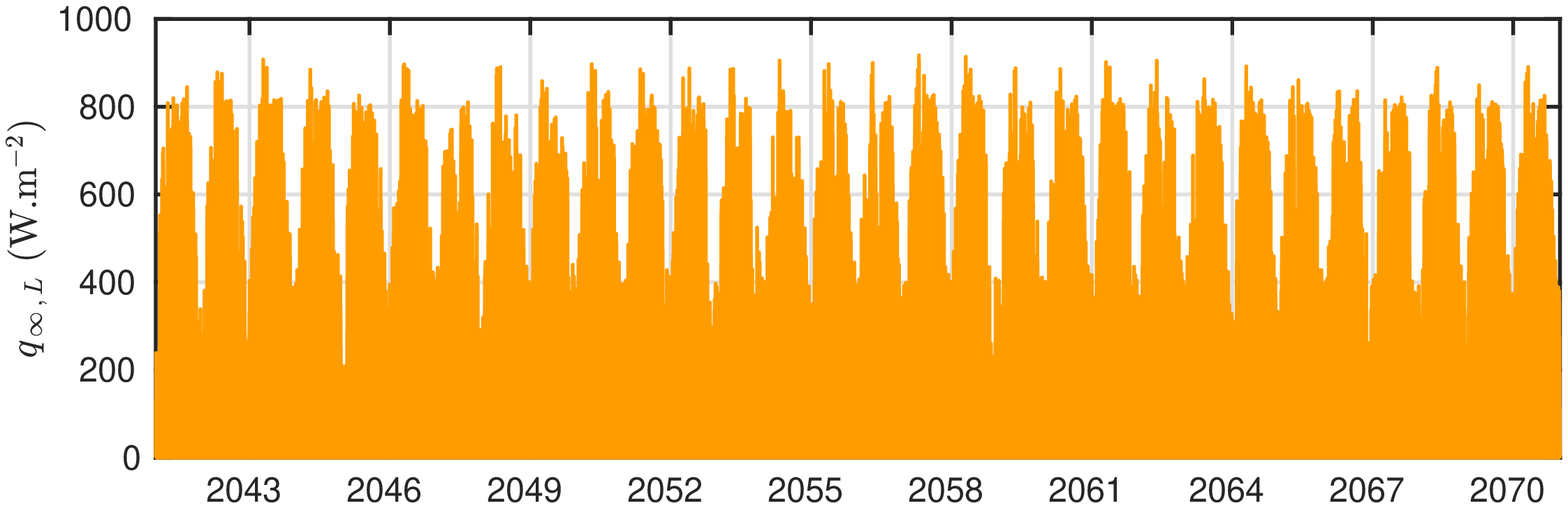}} \\
\subfigure[
\label{fig:real_hL_ft}]{\includegraphics[width=.95\textwidth]{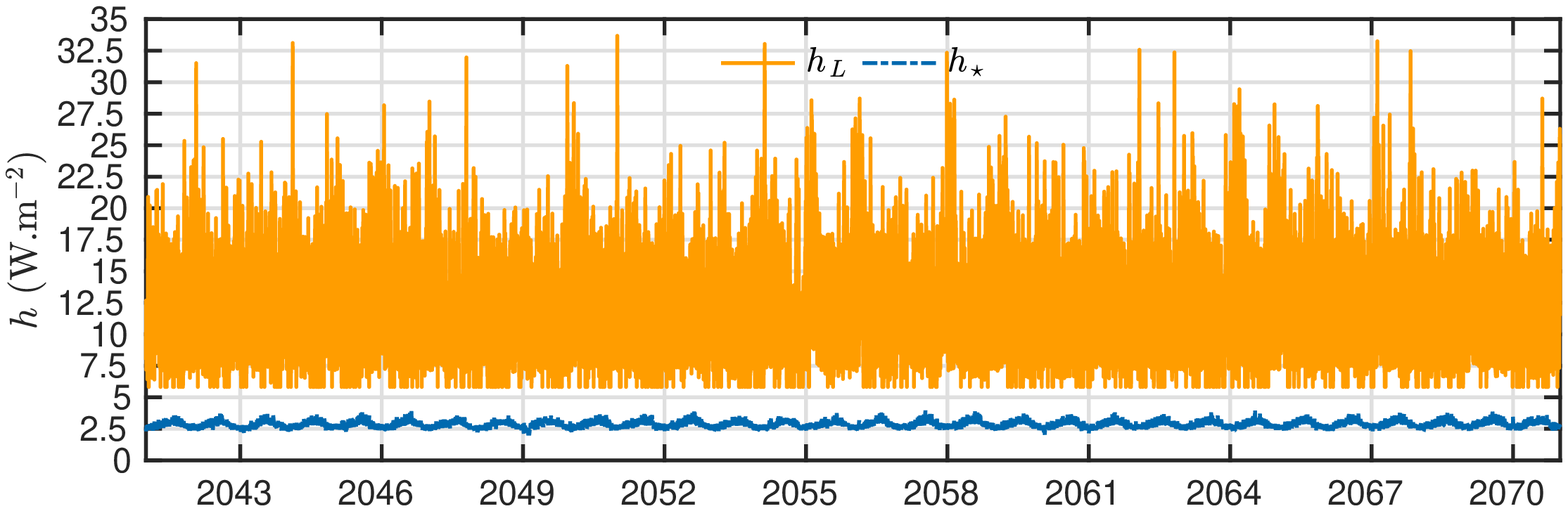}}
\caption{Time evolution of the outside and inside temperature \emph{(a)}, the short wave radiation heat flux \emph{(b)} and the outside surface transfer coefficient (for $\epsilon \egal 1$) \emph{(c)}.}
\end{center}
\end{figure}

\begin{figure}[h!]
\begin{center}
\subfigure[
\label{fig:real_pdf_Tinf}]{\includegraphics[width=.45\textwidth]{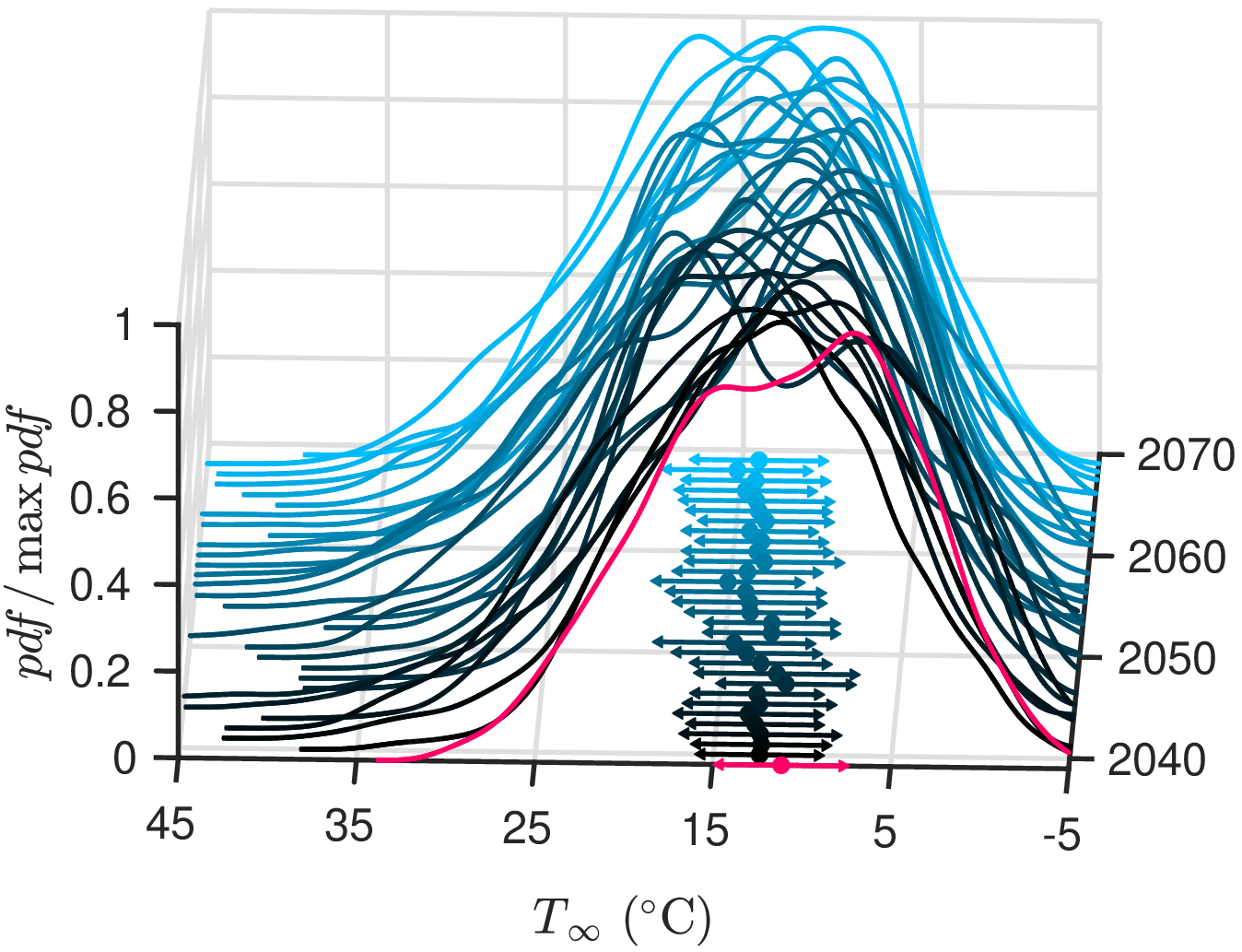}} \hspace{0.2cm}
\subfigure[
\label{fig:real_pdf_Qinf}]{\includegraphics[width=.45\textwidth]{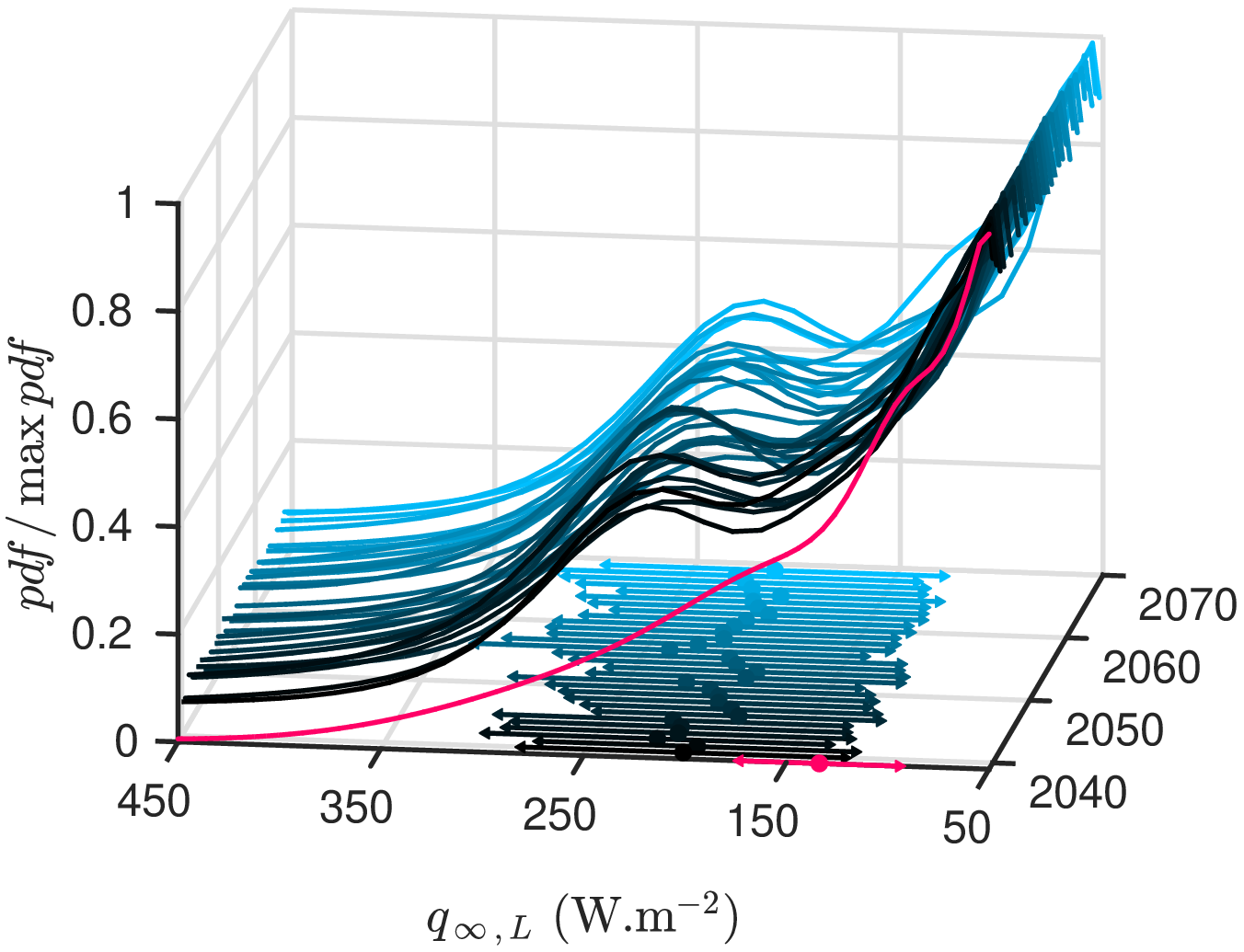}} 
\caption{Scaled probability density function of the outside \emph{(a)} and the short wave radiation heat flux over $50 \ \mathsf{W\,.\,m^{\,-2}}$\emph{(b)} according to the different years. The mean and standard deviation is presented on the $(\,x\,,\,y\,)$ plan for each Figure. The red line corresponds to the standard value.}
\end{center}
\end{figure}

Concerning the outside boundary condition, the climate of Paris is considered with its evolution for the next $30 \ \mathsf{years}$ according to climate change, starting from 2041. The weather file generation methodology is presented in \cite{machard_methodology_2020}. The outside temperature is presented in Figure~\ref{fig:real_Tinf_ft}. The incident vertical short wave radiation is computed using the global horizontal, direct normal and diffuse horizontal radiation data combined with the method described in \cite{evseev_assessment_2009} and the \textsc{Hay} model \cite{hay_study_1979} with a West wall orientation. The results are shown in Figure~\ref{fig:real_qinf_ft} with the time variation of the total short wave radiation flux. For the long wave radiation heat flux, the sky temperature is defined according to the ISO 13790 for temperate areas \cite{evangelisti_sky_2019}. For the convective surface transfer coefficient, the following coefficients are used: $h_{\,L\,,\,0} \egal 5.82 \ \mathsf{W\,.\,m^{\,-2}\,.\,K^{\,-1}}\,$, $h_{\,L\,,\,1} \egal 1.39 \ \mathsf{W\,.\,m^{\,-2}\,.\,K^{\,-1}}$ and $v_{\,0} \egal 1 \ \mathsf{m\,.\,s^{\,-1}}\,$. It arises from \cite{mcadams_heat_1942} for a wall height of $2.5 \ \mathsf{m}\,$. The time variation of the  radiative equivalent and the air velocity dependent surface transfer coefficients are given in Figure~\ref{fig:real_hL_ft}. It can be observed that the convective surface transfer coefficient has higher magnitude than the radiative one. The former varies between $7.5$ and $20 \ \mathsf{W\,.\,m^{\,-2}\,.\,K^{\,-2}}$ while the latter remains stable around $2.5 \ \mathsf{W\,.\,m^{\,-2}\,.\,K^{\,-1}}\,$. The influence of climate change can be noticed in Figures~\ref{fig:real_pdf_Tinf} and \ref{fig:real_pdf_Qinf}. The mean temperature increases slowly over the years. Similarly, occurrences of high radiation heat flux becomes higher over the years. 

Last, regarding the inside boundary conditions, the heat transfer coefficient is set to $h_{\,R} \egal 10 \ \mathsf{W\,.\,m^{\,-2}\,.\,K^{\,-1}}\,$. The inside ambient temperature is defined according to the outside temperature following the ASHRAE Standard 90.1-2013:
\begin{align*}
T_{\,\infty\,,\,R} \egal 
\begin{cases}
\ T_{\,\set}^{\,\min} \ \mathsf{\degC}  \,, 
&  T_{\,\infty\,,\,L} \, \leqslant \, T_{\,\infty\,,\,L}^{\,\min} \,, \\
\ T_{\,\set}^{\,\min} \plus \frac{T_{\,\infty\,,\,L} \moins  T_{\,\infty\,,\,L}^{\,\min} }
{ T_{\,\infty\,,\,L}^{\,\max} \moins  T_{\,\infty\,,\,L}^{\,\min}} 
\ \Bigl(\, T_{\,\set}^{\,\max} \moins T_{\,\set}^{\,\min} \,\Bigr) \,, 
& T_{\,\infty\,,\,L}^{\,\min} \, \leqslant \, T_{\,\infty\,,\,L} \, \leqslant \, T_{\,\infty\,,\,L}^{\,\max}  \,, \\
\ T_{\,\set}^{\,\max} \ \mathsf{\degC} \,, 
& T_{\,\infty\,,\,L} \, \geqslant \, T_{\,\infty\,,\,L}^{\,\max} \,,
\end{cases}
\end{align*}
with $T_{\,\infty\,,\,L}^{\,\min} \egal 17  \ \mathsf{\degC}\,$, $T_{\,\infty\,,\,L}^{\,\max} \egal 27  \ \mathsf{\degC} \,$, $T_{\,\set}^{\,\max} \egal 24 \ \mathsf{\degC}$ and $T_{\,\set}^{\,\min} \egal 19 \ \mathsf{\degC}\,$. It enables smooth transition between winter and summer temperature set points of heating and cooling systems. Figure~\ref{fig:real_Tinf_ft} gives the inside temperature variations.

\begin{table}
\centering
\caption{Values of the parameters considered for the parametric computations. }
\label{tab:real_p_space}
\setlength{\extrarowheight}{.5em}
\begin{tabular}[l]{@{} c cccc}
\hline
\hline
& $k_{\,2} \ \unit{W\,.\,m^{\,-1}\,.\,K^{\,-1}}$
& $c_{\,2} \ \unit{MJ\,.\,m^{\,-3}\,.\,K^{\,-1}}$
& $d_{\,2} \ \unit{m}$
& $\epsilon\ \unit{-}$ \\
Lower bound
& $0.04$
& $0.21$
& $0.05$
& $0.60$ \\
Upper bound
& $0.10$
& $1.02$
& $0.30$
& $0.95$ \\
Best configuration
& $0.04$
& $0.94$
& $0.29$
& $0.63$ \\
Worst configuration
& $0.08$
& $0.96$
& $0.05$
& $0.92$ \\
\hline
\hline
\end{tabular}
\end{table}

\subsection{Results}

\begin{figure}[h!]
\begin{center}
\subfigure[
\label{fig:real_Pspace1}]{\includegraphics[width=.45\textwidth]{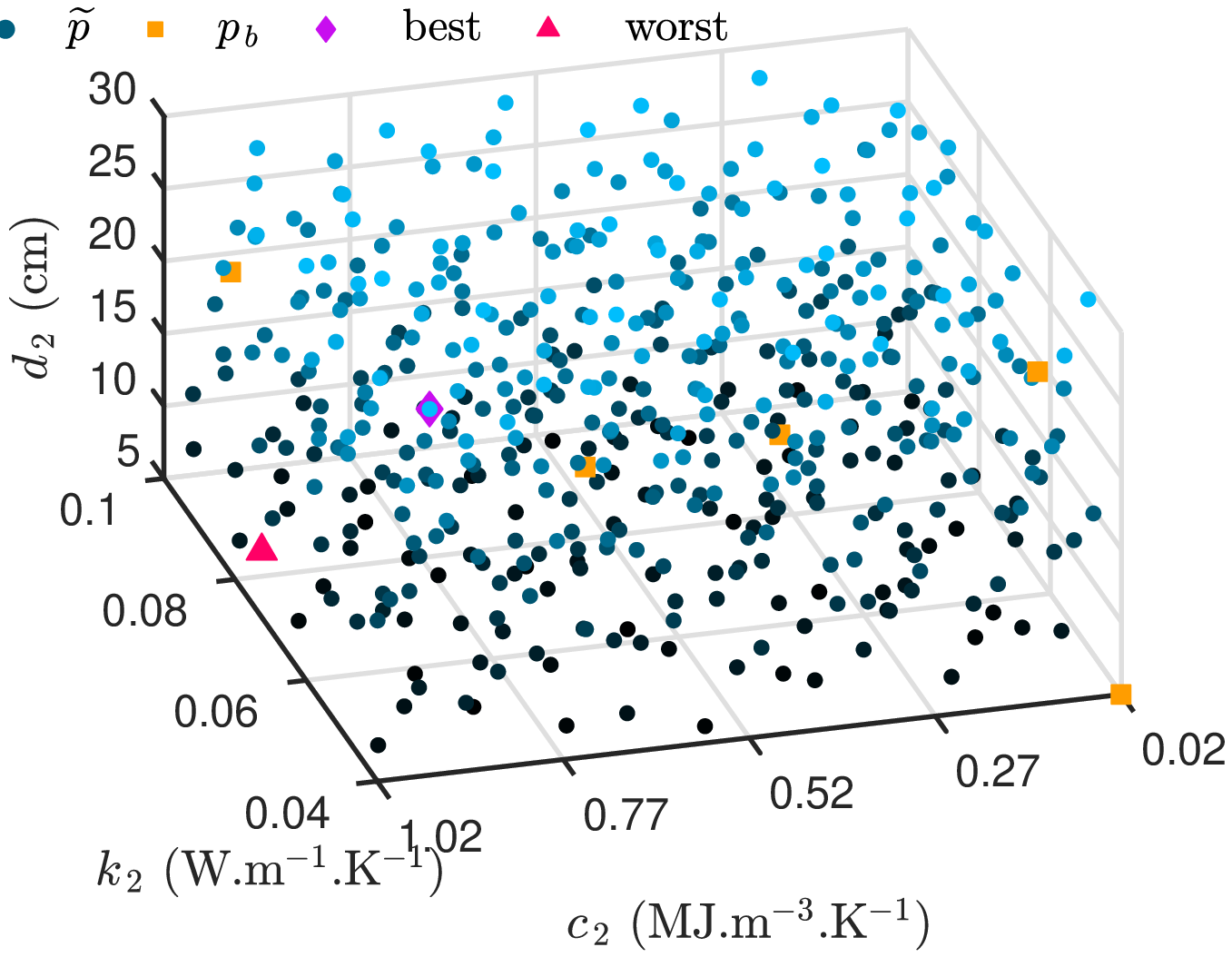}}  \hspace{0.2cm}
\subfigure[
\label{fig:real_Pspace2}]{\includegraphics[width=.45\textwidth]{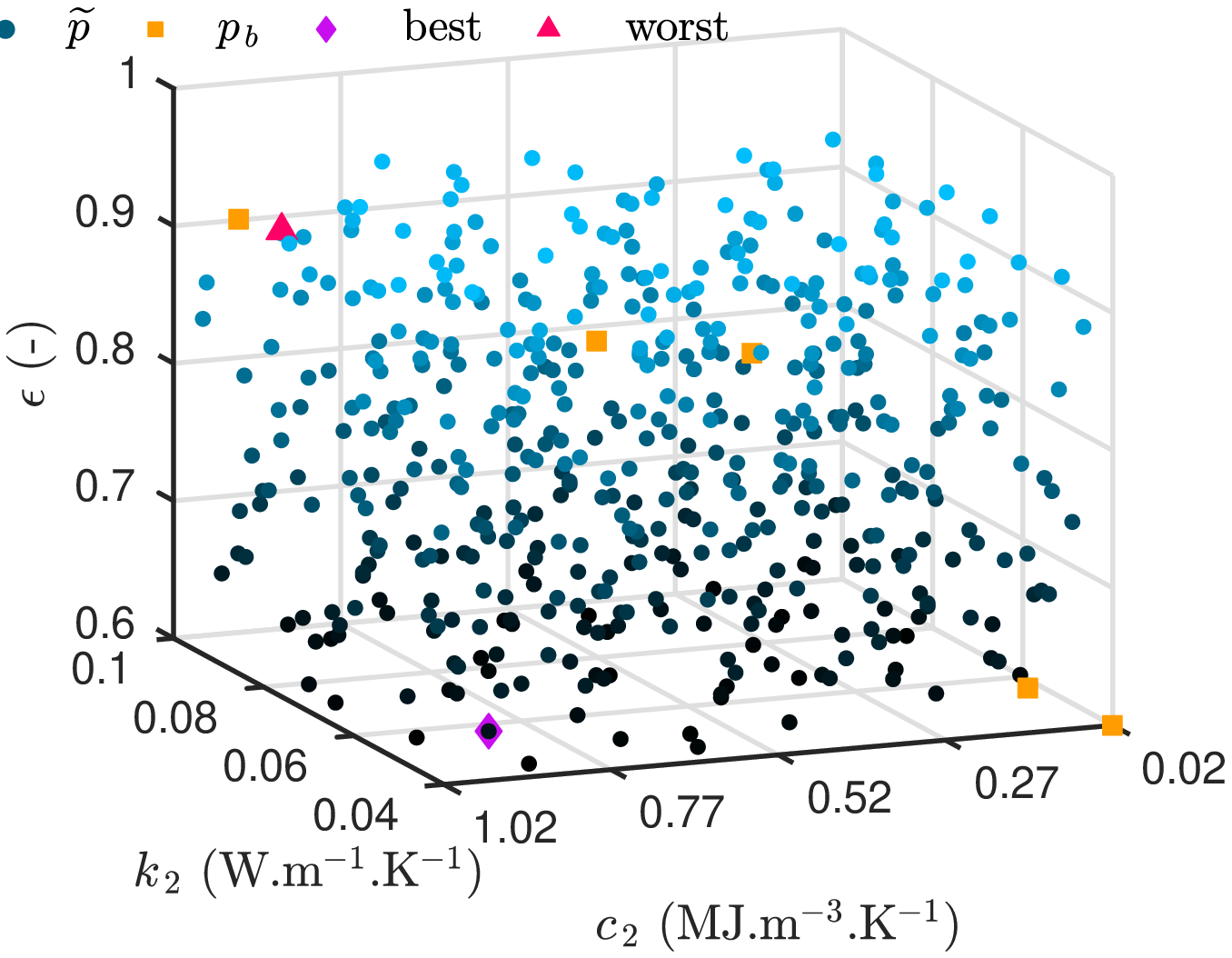}}
\caption{Samples in the parameter space $\Omega_{\,p} \egal \Omega_{\,k_{\,2}} \, \bigcup \, \Omega_{\,c_{\,2}} \, \bigcup \, \Omega_{\,d_{\,2}} \, \bigcup \, \Omega_{\,\epsilon} \,$.}
\label{fig:real_Pspace}
\end{center}
\end{figure}

The computations are done using the \revision{reduced-order} model PODx. For each layer, the mode number is set to $N \egal 10\,$. The tolerances solver are $\mathrm{Tol} \egal 10^{\,-4}\,$. A total of $N_{\,b} \egal 5$ basis are generated using the COM model and projected on a time grid with $\Delta \tau \egal 1\,$. The parameters used for the basis generation are illustrated in Figure~\ref{fig:real_Pspace}. Note that those parameters have been selected randomly in the parameter space. Then, computations are carried for $N_{\,p} \egal 500$ randomly generated parameters $\widetilde{p}$ in the space using \texttt{Halton} sequence generation. The parameter space is presented in Figure~\ref{fig:real_Pspace}. For each parameter $\widetilde{p}$ the reduced basis is interpolated based on the \textsc{Grassmann} manifold and the multiquadric RBF (for the interpolation in the tangent space). In terms of computational cost, results are presented in Table~\ref{tab:cpu_time}. The complete model requires $1475 \ \mathsf{s}$ for computing solution for one parameter for the $30$ years. For the reduced model, the off-line phase includes the preliminary simulations of the COM to generate the $N_{\,b}$ basis ($t_{\,\cpu} \egal N_{\,b} \cdot 1475 \ \mathsf{s}$) and the the computations of the RBF weights ($t_{\,\cpu} \egal 30 \ \mathsf{s}$). Thus, the computational cost of this phase is driven by the COM one. The on-line phase integrates the basis interpolation for the new parameter and the \revision{reduced-order} model computation. For this, the computational time scales with $1.5 \ \mathsf{s}$ for one parameter. The computational cut of the reduced model is about $0.1 \ \%$ of the complete model. Given these values, the computational cost of the reduced model can be given by the following equation for this case study:
\begin{align*}
t_{\,\cpu}^{\,ROM} \egal  t_{\,\cpu}^{\,LOM} \cdot \bigl(\, N_{\,b} \plus 10^{\,-3} \cdot N_{\,p} \,\bigr) \,.
\end{align*}
So, the computational efficiency of the model is proven from the moment one needs to perform parametric computations for more than $N_{\,b}$ parameters.

\begin{figure}[h!]
\begin{center}
\subfigure[
\label{fig:real_Wf_dp}]{\includegraphics[width=.45\textwidth]{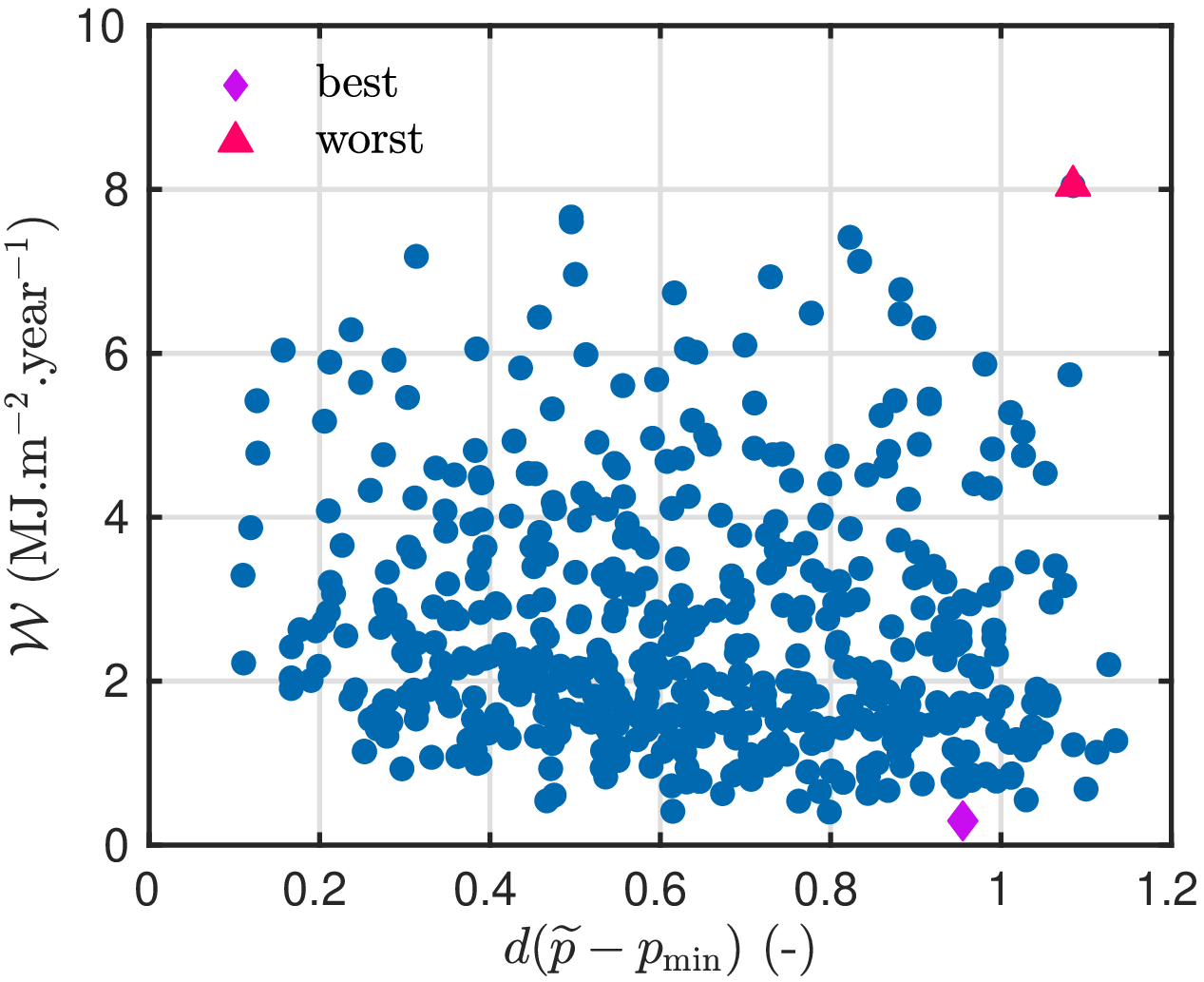}}  \hspace{0.2cm}
\subfigure[
\label{fig:real_pdf_w}]{\includegraphics[width=.45\textwidth]{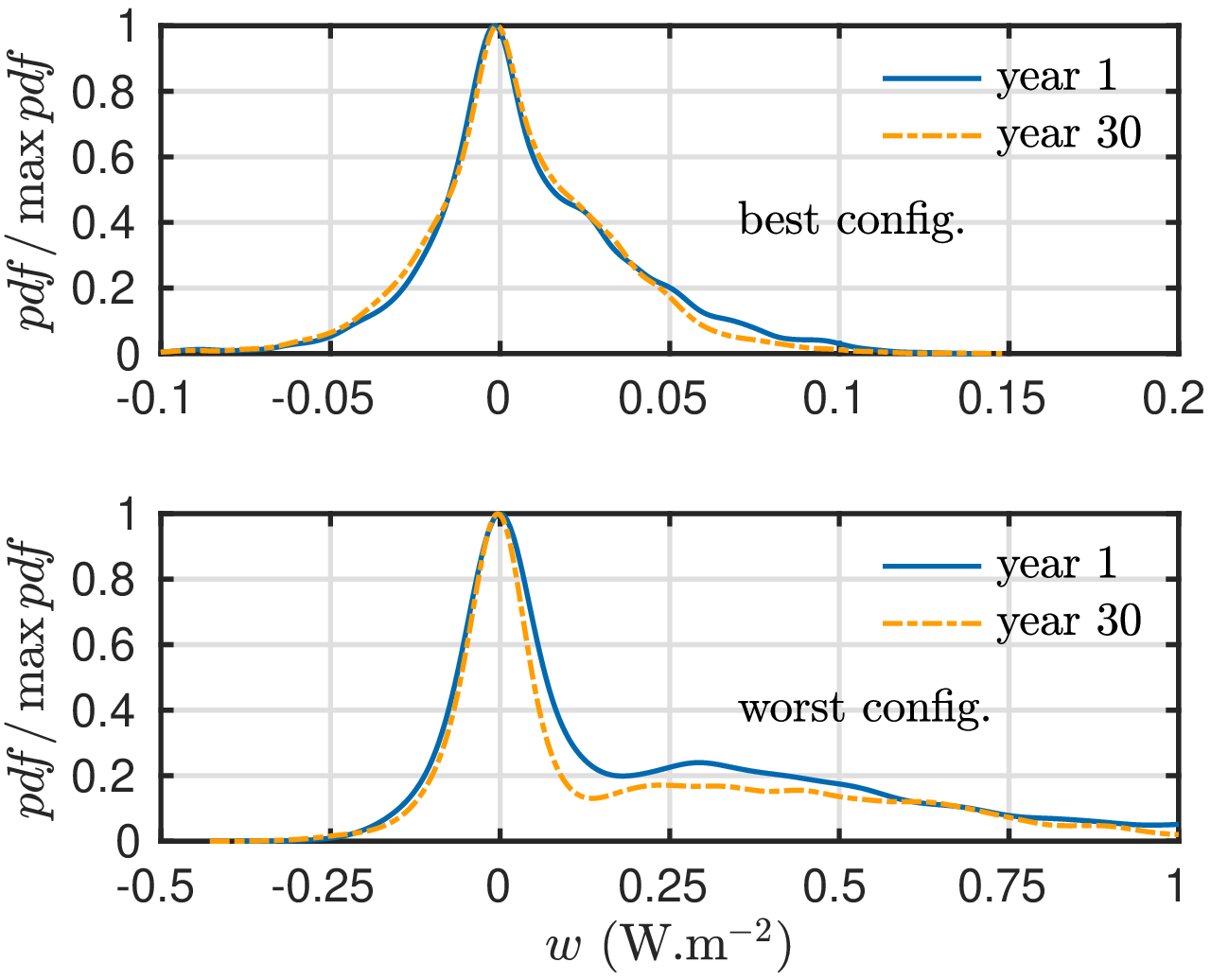}}
\caption{Evolution of the consumed work according to the distance with $p_{\,\min}$ \emph{(a)}. Scaled probability density function of the work rate for the best and worst configurations \emph{(b)}.}
\end{center}
\end{figure}

Figure~\ref{fig:real_Wf_dp} illustrates the variation of the wall consumed work according to the distance of the parameters $p_{\,\min} \egal \bigl(\,k_{\,2} \egal 0.04\,,\, c_{\,2} \egal 0.02 \,,\, d_{\,2} \egal 0.05\,,\, \epsilon \egal 0.6 \,\bigr)\,$. The consumed work varies between $0$ and $12 \ \mathsf{MJ\,.\,m^{\,-2}\,.\,year^{\,-1}}\,$. Since the work is positive, it indicates that for all solutions, in average the wall operates more as a heat pomp or air conditioner.

The best solution is the one with the minimum consumed work. The probability density functions of the work rate for the worst and best configurations are presented in Figure~\ref{fig:real_pdf_w}. As expected, the work rate variation range is lower for the best configuration. For the works configuration, the work range has more occurrences of positive value, indicating a wall working as heat pump or air conditioner. There is no significant variations of the work rate over the years. For the worst configuration, the occurrences in the positive range increases a bit. The parameters related to those two solutions are given in Table~\ref{tab:real_p_space}. As expected, the best configuration is for a low thermal conductivity. For this configuration, the size of the insulation layer is high ($29 \ \mathsf{cm}$) and the emissivity of the outside coating low ($0.63$). The worst configuration corresponds to an insulation layer with small thickness and high thermal conductivity. 

\begin{figure}[h!]
\begin{center}
\subfigure[
\label{fig:real_jbest_ft}]{\includegraphics[width=.95\textwidth]{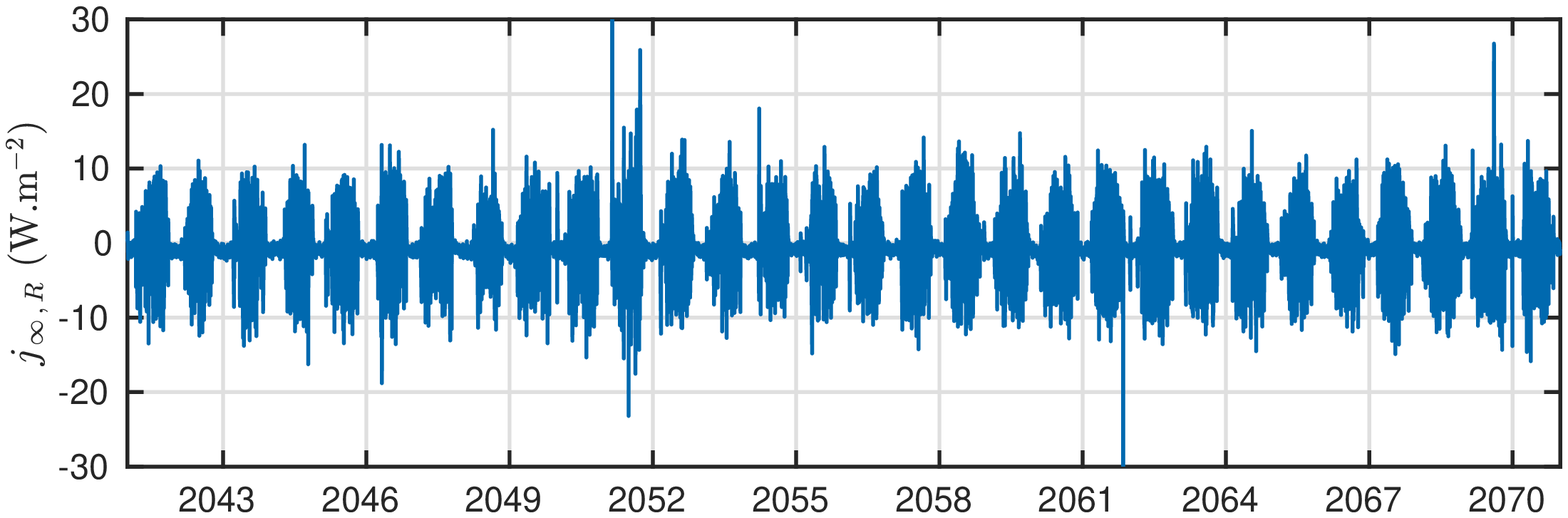}} \\
\subfigure[
\label{fig:real_wbest_ft}]{\includegraphics[width=.95\textwidth]{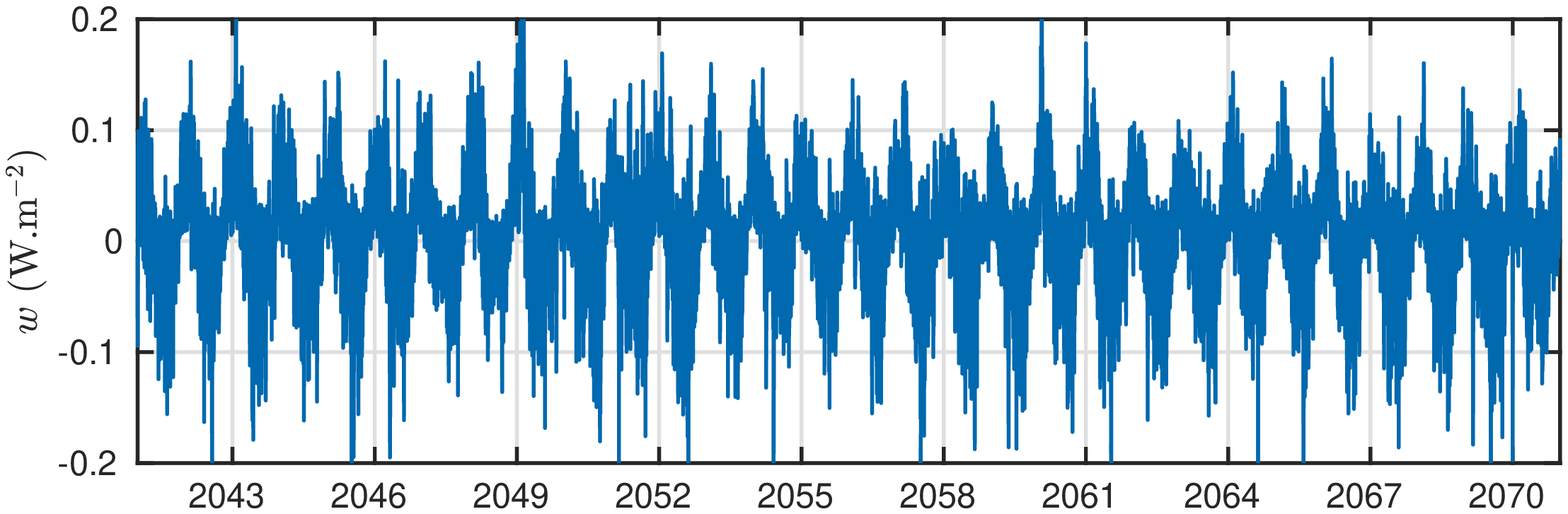}}
\caption{Time evolution of the inside heat flux \emph{(a)} and the work rate  \emph{(b)} for the best configuration.}
\label{fig:real_comp_solution_best}
\end{center}
\end{figure}

\begin{figure}[h!]
\begin{center}
\subfigure[
\label{fig:real_jwrst_ft}]{\includegraphics[width=.95\textwidth]{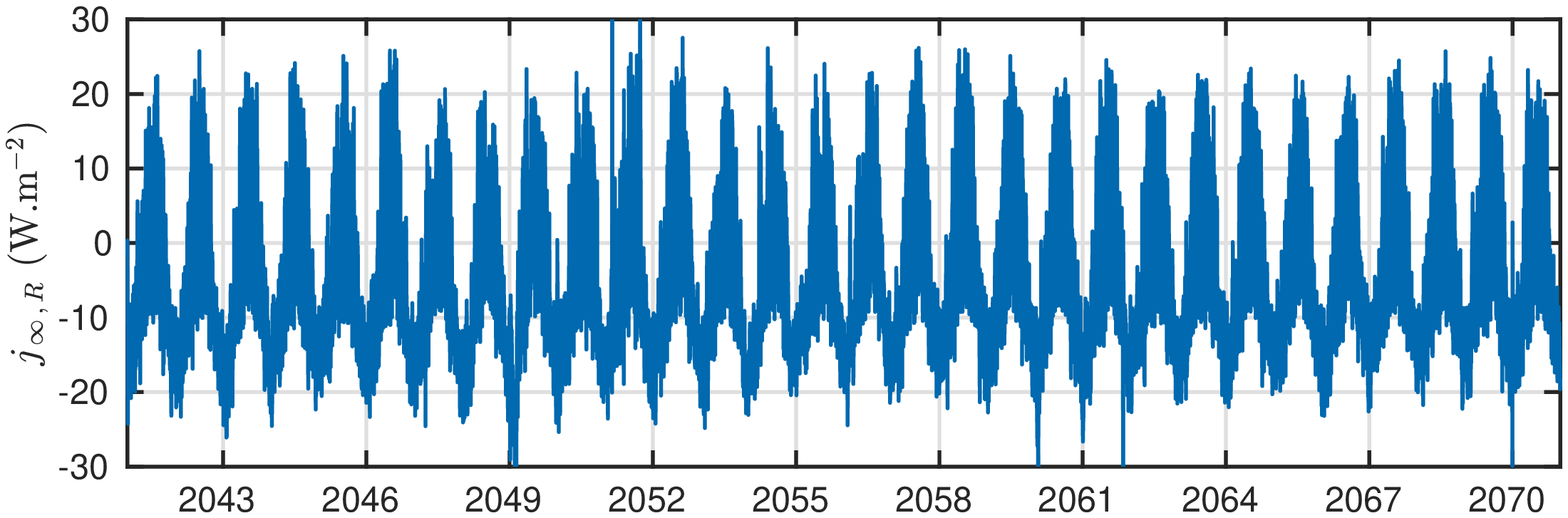}} \\
\subfigure[
\label{fig:real_wwrst_ft}]{\includegraphics[width=.95\textwidth]{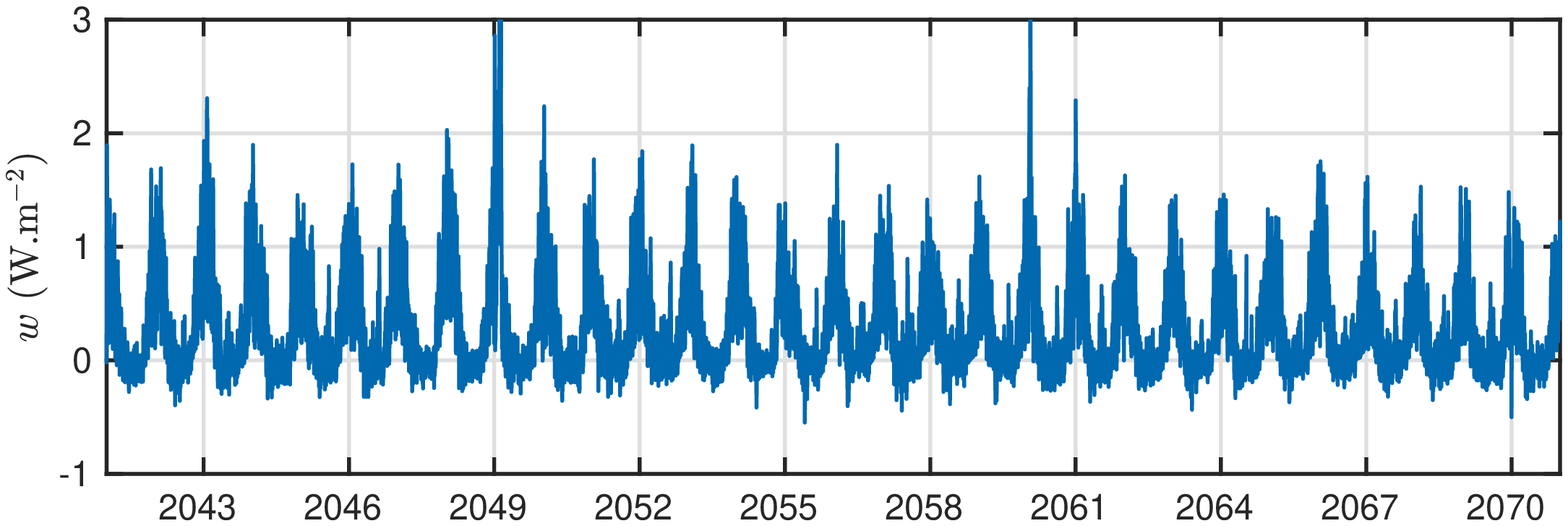}}
\caption{Time evolution of the inside heat flux \emph{(a)} and the work rate  \emph{(b)} for the worst configuration.}
\label{fig:real_comp_solution_wrst}
\end{center}
\end{figure}

A comparison of the solutions for the best and worst configuration is carried in Figures~ \ref{fig:real_comp_solution_best} and \ref{fig:real_comp_solution_wrst}. First, the inside heat flux is much smaller for the best solution than for the worst one. It varies around $-10$  and $10 \ \mathsf{W\,.\,m^{\,-2}}$ for the latter and $-20$  and $20 \ \mathsf{W\,.\,m^{\,-2}}$ for the former. As a consequence, the work rate is higher and more oscillating for the worst configuration. One one hand, in summer condition, the heat flux is high and so the work rate is positive scaling with $2  \ \mathsf{W\,.\,m^{\,-2}}$ for the worst configuration. As a consequence, the wall works as an heat pomp and heat the inside ambient zone. For the best configuration, the heat flux occurs to be negative so the work rate is negative. The wall operates as a heat pomp, cooling \emph{naturally} the inside ambient air zone. On the other hand, the heat flux is negative in winter conditions. The work rate is still positive for the worst configuration so the wall, considered as an air conditioner, is cooling the inside ambient zone.

\begin{table}
\centering
\caption{Computational time of the complete and \revision{reduced-order} models.}
\label{tab:cpu_time}
\setlength{\extrarowheight}{.5em}
\begin{tabular}[l]{@{} c ccc}
\hline
\hline
\multirow{2}{*}{\textit{Model}} 
& \multirow{2}{*}{\textit{Complete Model}}
& \multicolumn{2}{c}{\textit{\revision{Reduced-order} Model}} \\
 & & off-line & on-line  \\
$t_{\,\cpu} \ \unit{s}$
& $1475 \cdot N_{\,p}$ 
& $1475 \cdot N_{\,b} \plus 30$  
& $1.5 \cdot N_{\,p}$\\
\hline 
\hline
\end{tabular}
\end{table}

\newpage 

\newpage

\subsection{Further remarks}

The proposed model enables performing accurate predictions with a reduced computational cost for simulation over $30$ years. Climate change model data are linked to three types of uncertainties: socio-economic scenario uncertainty, model uncertainty and climate natural variability \cite{Hawkins_2009}. In this paper, the latter uncertainty is addressed, while it is often neglected in research papers \cite{Giorgi_2019}. By the middle of this century, the two main climate change uncertainties are \revision{model} uncertainty and \revision{climate} natural variability uncertainty. In the context of this paper, our interest is not to investigate different signals of climate change projections (which would be linked to climate model uncertainty). It is rather to appreciate the variability of the wall response to the climate natural variability, to capture the different weather \revision{variables'} mean change but also their standard deviation change, incorporating the most extreme situations. This is why a multi-year long-term $30$ years period is used, which is also recommended by climatologists for climate change studies \cite{Giorgi_2019}.

\begin{figure}[h!]
\begin{center}
\subfigure[
\label{fig:j1yr_summer_ft}]{\includegraphics[width=.45\textwidth]{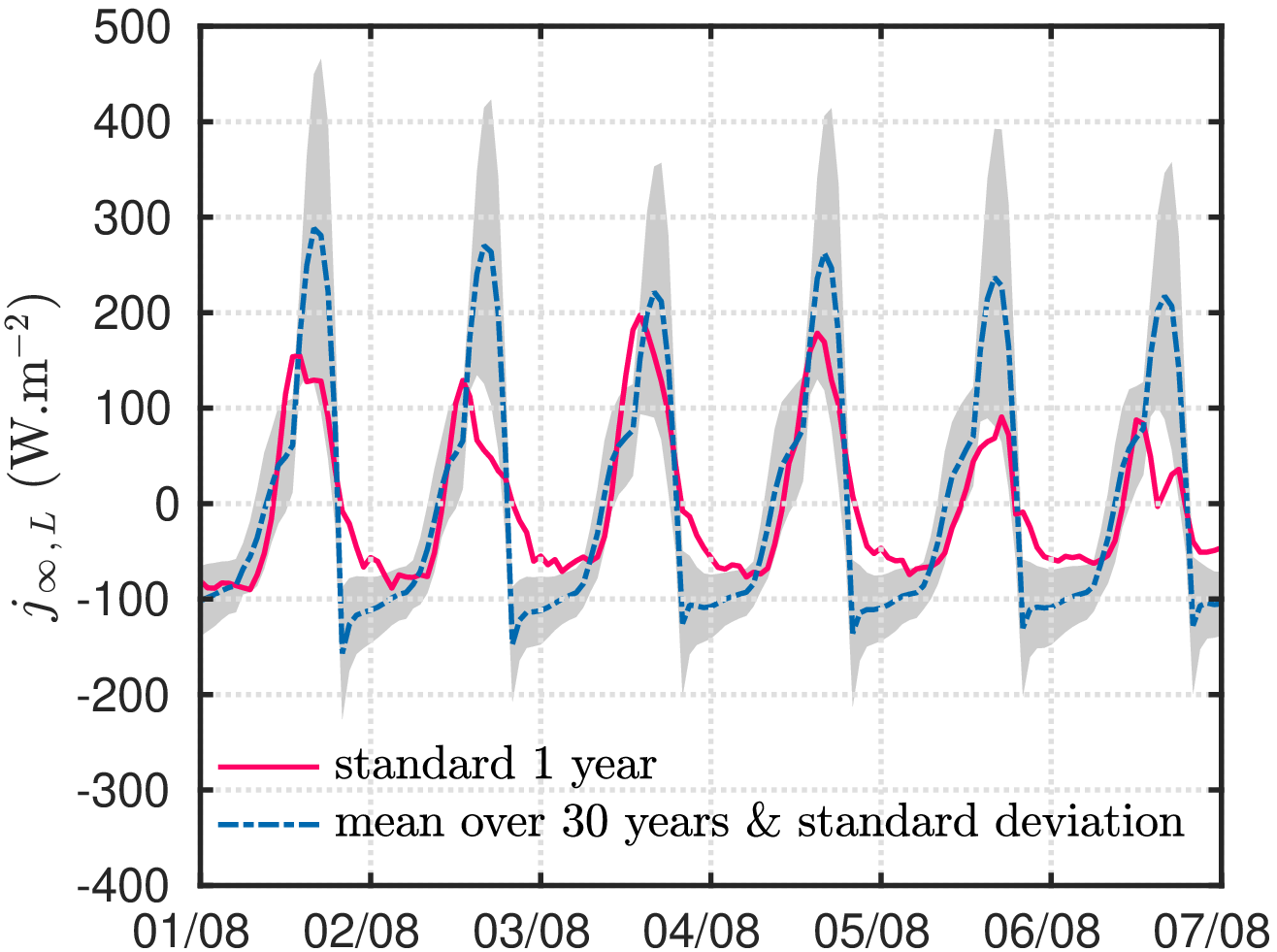}} \hspace{0.2cm}
\subfigure[
\label{fig:Wf_dp_comp}]{\includegraphics[width=.42\textwidth]{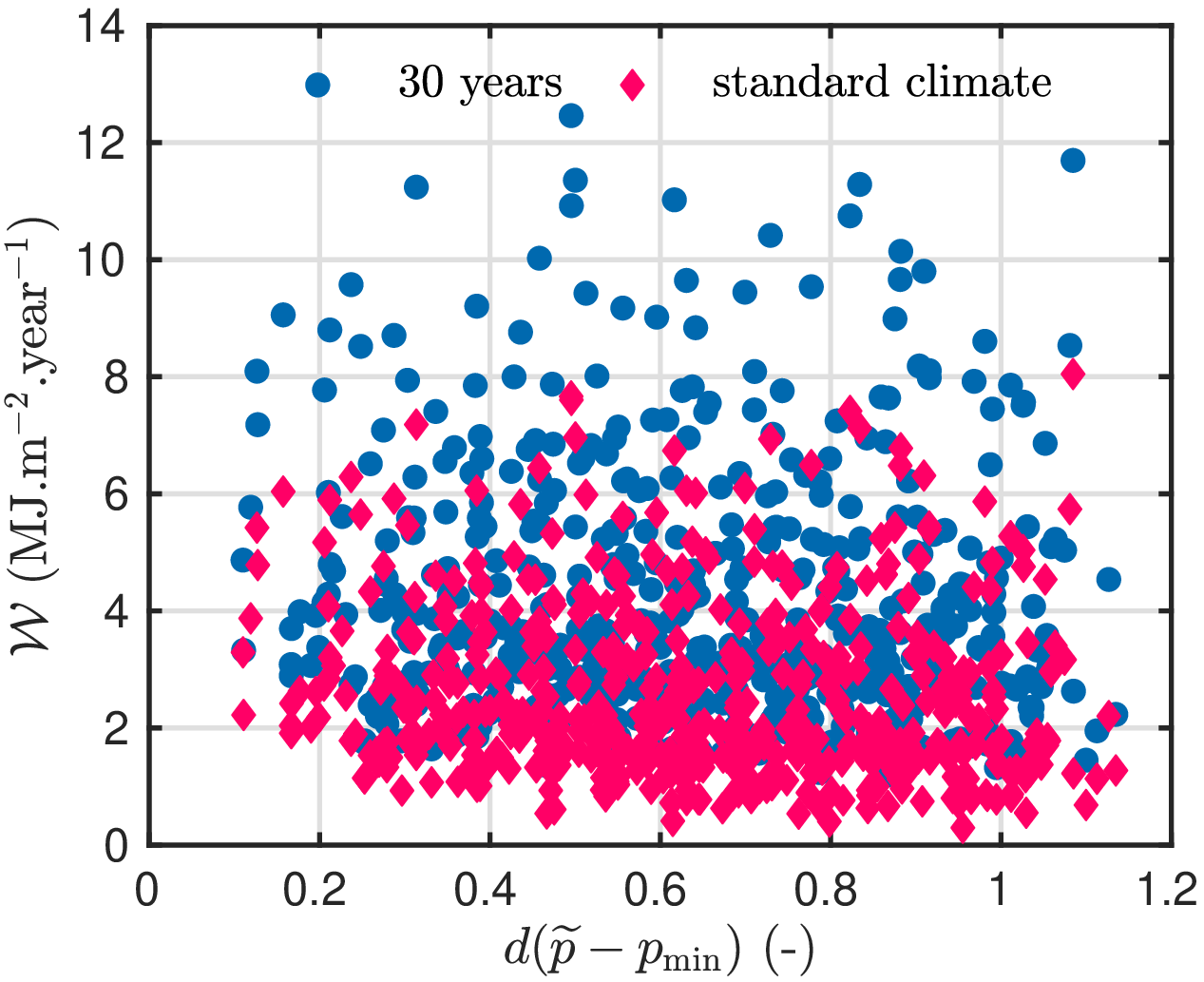}} 
\caption{Time evolution of the inside heat flux for summer period according to the computation carried with $1$ year of standard climate or $30$ years considering climate changes \emph{(a)}. Variation of the consumed work according to the distance with $p_{\,\min}$ \emph{(b)}.}
\end{center}
\end{figure}

With climate change, such simulations are necessary to design resilient buildings, capable of adapting to future extreme heat. Figures~\ref{fig:real_pdf_Tinf} and \ref{fig:real_pdf_Qinf} give a comparison of the probability density function for the $30$ years considering climate change with the standard values. It highlights that standard climate \revision{has} lower mean temperature and radiation flux. The differences are significant for incident radiation flux. The boundary \revision{conditions} are different when considering the climate from standards or \revision{the} model predicting changes due to global warming. Thus, model predictions are varying depending on both approaches. Figure~\ref{fig:j1yr_summer_ft} compares the heat flux predicted with $1$ year of standard climate and the average value computed with $30$ years considering climate changes. The differences are greater at the extremes (minimum or maximum) of the flux. Different predictions may lead to the wrong design of the building wall to face climate change. Figure~\ref{fig:Wf_dp_comp} confronts the consumed work computed with both approaches. The discrepancies are significant and the choice of the design is affected, particularly for walls with low inertia and small insulation length (designs in the top area of the figure). From this further analysis, one may conclude that this level of accuracy in modeling is required to design efficient walls \revision{for} the buildings that will face climate change.

\revision{Intending to elaborate reduced-order} model to cut the computational cost while keeping a satisfying accuracy of the predicted phenomena, a usual approach consists in searching for \revision{decomposed solutions}. Several works in the literature proposed to build previously a spatial basis \cite{berger_evaluating_2019,berger_intelligent_2018}. This approach is certainly interesting for cutting the computational cost of \revision{the} model considering two- or three-dimensional transfer.  However, when considering \revision{long-term} simulation with \revision{a} one-dimensional diffusion process, it is better to preliminary construct a time basis to approximate the solution as illustrated in Figure~\ref{fig:Rom_secret_sauce}. Results in this article highlight that the computational ratio is lower. To build the ROM, integration over the time interval \revision{is} required to compute the coefficients $\beta\,$, $\delta$ and $\gamma$ from Eq.~\eqref{eq:coeff_beta_delta_gamma}. Such integration \revision{is} carried out numerically so it is important to pay attention to the time grid used to generate the time basis $\psi\,$. With \revision{a} coarse time grid, the ROM may lack \revision{accuracy}.

Works in the literature point out the lack of accuracy of the POD \revision{reduced-order} model to perform parametric solutions  \cite{berger_evaluating_2019,Hou_2020}. Here, this problem is handled by proposing an interpolation of \revision{reduced basis}. First, $N_{\,b}$ \revision{bases} are built with preliminary solutions computed for different parameters values. Then, for new parameter values, the basis is computed by using the interpolation method on the tangent space of the \textsc{Grassmann} manifolds. The number of basis preliminary built may be important for the accuracy of the method. In our numerical experiments, $5$ \revision{bases} proved to be sufficient. However, it is suggested to verify the accuracy of the parametric solution with \revision{the} reference solution.

\begin{figure}[h!]
\begin{center}
\includegraphics[width=.9\textwidth]{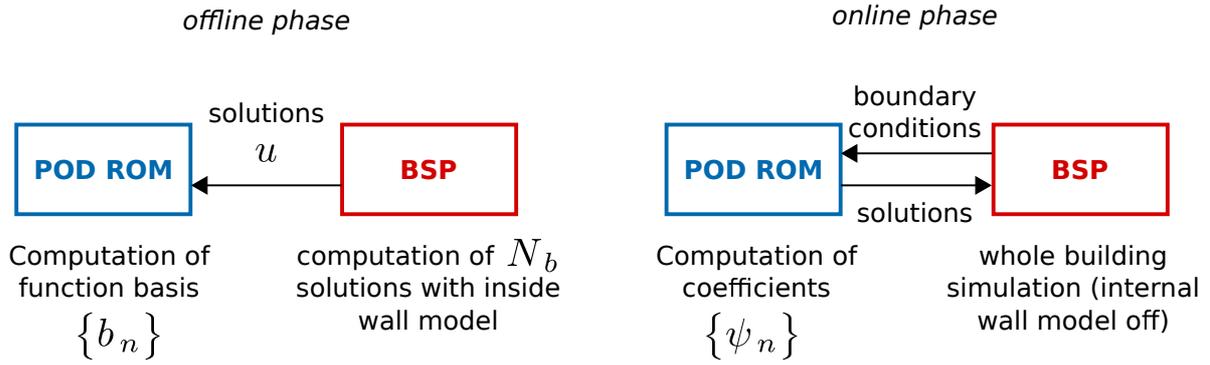}
\caption{Illustration of a integration of ROM with a building simulation program through a co-simulation environment.}
\label{fig:cosimulation}
\end{center}
\end{figure}

An interesting outlook is to integrate such advanced ROM in building simulation programs (BSP). For this, a promising approach is to use the standard Functional Mock-up Interface (FMI) \cite{fmiStandard} to perform the information exchanges between the dynamic BSP and the \revision{reduced-order} model. An illustration of the operation is shown in Figure~\ref{fig:cosimulation}. This approach requires two phases as described in Section~\ref{sec:numerical_model}. The \revision{offline} one intends to compute the POD basis for a small set $N_{\,b}$ of parameters. For this, the solution can be computed by the BSP alone with its existing model of wall heat transfer. Then, during the \revision{online} phase, the information exchanged between two models occurs at each time step. The BSP provides the climatic data (temperatures, radiation flux and wind velocity) and inside ambient conditions at \revision{the} current time instant. Then, the ROM computes the interpolated basis and the solution. It returns to the BSP the temperature profiles and/or the inward heat flux. Note that for this phase, the internal BSP wall model should be turned off. \revision{An example} of such coupling approach using ROM can be \revision{found} in \cite{berger_intelligent_2018}.

\newpage 

\section{Conclusion}

The objective of this article is to investigate model reduction approaches to cut the computational cost of building simulation programs. The method proposed relies on two important points. First, it considers a \revision{reduced basis} for time coordinate built using \revision{the} preliminary computed solution. The spatial coefficients are retrieved by solving a boundary value problem. This approach enables to decrease the solution degree of freedom by focusing on \revision{time-domain} scales. Most of the works in the literature regarding POD \revision{use} a space \revision{reduced basis} and \revision{solve} an initial value problem. The second point concerns the elaboration of a parametric solution. \revision{A new} \revision{reduced basis} is obtained by using the interpolation method on the tangent space on \textsc{Grassmann} manifolds.

The \revision{reduced-order} model efficiency is verified considering reference solutions for three case studies. The first one compares the model to a classical approach and \revision{highlights} on the computational efficiency of the proposed approach. Namely, the model is five times faster than traditional ones. For the second case, an analytical solution is built to compute the temperature field in a two-layer wall. The comparison with the \revision{reduced-order} model shows a very satisfying accuracy. The error scales with $\mathcal{O}(\,10^{\,-3}\,)\,$ for both the solution and its partial derivative This feature is very interesting for the design of the wall energy efficiency since it requires \revision{assessing} the heat flux. The last case focuses on the interpolation of bases. Results highlight \revision{very good} reliability of the parametric \revision{reduced-order} model. It is possible to compute a parametric solution with an error of $\mathcal{O}(\,10^{\,-3}\,)$ at a computational cost scaling with $1.5\%$ of the complete model.

Last, the model is used to perform predictions and design the wall energy efficiency for a case considering climate change over $30$ years. It aims at finding \revision{a} solution that minimizes the work rate of the wall in the parameter space involving the load-bearing material emissivity, the heat capacity, the thermal conductivity and the thickness of the insulation layer. Outside boundary conditions take into account convective heat transfer, short and long wave radiation exchanges. Simulations are carried out for $30$ years with a computational ratio of $0.1\%$ compared to standard approaches.

Future works should focus on the extension of the model reduction for higher dimension heat transfer as well as problems considering coupled heat and mass transfer. It requires \revision{considering} the high \revision{non-linearity} of the material properties. 

\section*{Nomenclature and symbols}

\begin{tabular}{|cll|}
\hline
\multicolumn{3}{|c|}{\emph{Physical parameters}} \\ \hline
\multicolumn{3}{|c|}{Latin letters} \\ 
$c$ & volumetric heat capacity & $\unit{J\,.\,m^{\,-3}\,.\,K^{\,-1}}$ \\
$h_{\,L}\,,\, h_{\,L\,,\,0} \,,\, h_{\,L\,,\,1}\,,\, h_{\,\sky}\,,\,h_{\,R}$ & surface heat transfer coefficient & $\unit{W\,.\,m^{\,-2}\,.\,K^{\,-1}}$ \\
$j\,,\,j_{\,\infty\,,\,L}$ & heat flux & $\unit{W\,.\,m^{\,-2}}$ \\
$k$ & thermal conductivity & $\unit{W\,.\,m^{\,-1} \,.\, K^{\,-1}}$ \\
$\ell$ & wall length & $\unit{m}$ \\
$q_{\,\infty\,,\,L}\,,\,q_{\,\infty\,,\,R}$ & radiation flux & $\unit{W\,.\,m^{\,-2}}$ \\
$x$ & space coordinate & $\unit{m}$ \\
$t\,,\, \tf$ & time & $\unit{s}$ \\
$t_{\,\mathrm{cpu}}$ & CPU time & $\unit{s}$ \\
$T\,,\,T_{\,0}$ & wall temperature & $\unit{K}$ \\
$T_{\,\infty\,,\,L}\,,\,T_{\,\infty\,,\,R}$ & ambient air temperature & $\unit{K}$ \\
$T_{\,\sky}$ & sky temperature & $\unit{K}$ \\
$v_{\,\infty}$ & air velocity & $\unit{m\,.\,s^{\,-1}}$ \\
$T^{\,\max}\,,\,T^{\,\min}$ & reference temperature for dimensionless formulation & $\unit{K}$ \\
$w$ & work rate & $\unit{W\,.\,m^{\,-2}}$ \\
$\mathcal{W}$ & consumed work & $\unit{J\,.\,m^{\,-2}}$ \\
\hline
\multicolumn{3}{|c|}{Greek letters} \\
$\Omega_{\,t}$ & time domain & $\unit{s}$ \\
$\Omega_{\,x}\,,\,\Omega_{\,x}^{\,i}$ & space domain & $\unit{m}$ \\
$\Gamma_{\,L}\,,\,\Gamma_{\,R}\,,\,\Gamma_{\,i}$ & space domain boundary & $\unit{m}$ \\
$\epsilon$ & emissivity & $\unit{-}$ \\
$\sigma$ & \textsc{Stefran}--\textsc{Boltzmann} constant & $\unit{W\,.\,m^{\,-2}\,.\,K^{\,-4}}$ \\
\hline
\end{tabular}

\hspace{2cm}

\begin{tabular}{|cll|}
\hline
\multicolumn{3}{|c|}{\emph{Mathematical notations}} \\
\hline
\multicolumn{3}{|c|}{Latin letters} \\
$a$ & temporal coefficient of ROM & \\
$b$ & spatial coefficient of ROM & \\
$C$ & correlation matrix & \\
$\mathfrak{D}$ & degree of freedom of a solution & \\
$\Bi_{\,L}\,,\,\Bi_{\,R}\,,\,\Bi_{\,\sky}$ & \textsc{Biot} number & \\
$\mathcal{G}$ & basis interpolation application & \\
$\Fo$ & \textsc{Fourier} number & \\
$N$ & order of the reduced model & \\
$N_{\,\ell}$ & number of layers& \\
$N_{\,\chi}\,,\,N_{\,\tau}$ & number of elements for space and time grids & \\
$p$ & parameter in the parameter space & \\
$u\,,\,\tilde{u}\,,\,u_{\,\infty\,,\,L}\,,\,u_{\,\infty\,,\,R}\,,\,u_{\,\sky}$ & dimensionless temperature & \\
$y$ & field of interest & \\
\hline
\multicolumn{3}{|c|}{Greek letters} \\
$\alpha$ & model reduction numerical coefficients &  \\
$\beta^{\,L}\,,\,\beta^{\,R}$ & model reduction numerical coefficients &  \\
$\gamma^{\,L}\,,\,\gamma^{\,R}$ & model reduction numerical coefficients &  \\
$\delta^{\,L}\,,\,\delta^{\,R}$ & model reduction numerical coefficients &  \\
$\chi$ & dimensionless space &  \\
$\tau \,,\, \tau_{\,\mathrm{f}}$ & dimensionless time &  \\
$\varrho_{\,\infty\,,\,L}\,,\,\varrho_{\,\infty\,,\,R}$ & dimensionless radiation flux & \\
$\kappa$ & interface ratio & \\
$\Delta \chi$ & space mesh & \\
$\phi$ & space basis function &  \\
$\psi$ & time basis function &  \\
$\boldsymbol{\Psi}$ & Matrix composed of vector basis &  \\
$\boldsymbol{\Gamma}$ & tangent space matrix &  \\
$\Omega_{\,p}$ & parameter space &  \\
$\varepsilon_{\,2}\,,\,\varepsilon_{\,\infty}\,,\,\varepsilon_{\,\bowtie}$ & error & \\
\hline
\end{tabular}

~\\
\newpage 
\appendix

\section{Analytical solution for heat transfer in a two-layers slab}
\label{sec:analytical_solution}

The details of the analytical solution construction are presented. 

\subsection{Mathematical problem}

The problem is recalled according to the description of the physical model given in Section~\ref{sec:mathematical_model}. The dimensionless temperature is solution of the following governing equation: 
\begin{align*}
\pd{u_{\,i}}{\tau} \moins 
\Fo_{\,i} 
\, \pd{^{\,2} u_{\,i}}{\chi_{\,i}^{\,2}} \egal 0 \,,
\qquad \bigl(\,\chi_{\,i} \,,\,\tau \,\bigr) \, \in \, \bigl[\,0 \,,\, 1 \,\bigr] \, \times \, \bigl[\,0 \,,\, \tau_{\,\mathrm{f}} \,\bigr]\,, \qquad \forall i \, \in \, \bigl\{\, 1 \,,\, 2 \,\bigr\} \,.
\end{align*}
with the initial condition:
\begin{align}
\label{eq:initial_condition}
u_{\,1}(\,\chi_{\,1}  \,,\,\tau \egal 0\,) \egal \ f_{\,1}(\,\chi_{\,1}\,) \,, \qquad
u_{\,2}(\,\chi_{\,2}  \,,\,\tau \egal 0\,) \egal \ f_{\,2}(\,\chi_{\,2}\,) \,. 
\end{align}
\textsc{Dirichlet} boundary conditions are considered at the interfaces $\Gamma_{\,L}$ and $\Gamma_{\,R}$. We have: 
\begin{subequations}
\begin{align}
u_{\,1}(\,\chi_{\,1} \egal 0 \,,\,\tau\,) & \egal 0 \,, \label{eq:sa_BC_L} \\[4pt]
u_{\,1}(\,\chi_{\,1} \egal 1  \,,\,\tau\,) & \egal u_{\,2}(\,\chi_{\,2} \egal 0 \,,\,\tau\,) \,.
\label{eq:sa_IC_R}
\end{align}
\end{subequations}
for $\Omega_{\,x}^{\,1}$ and:
\begin{subequations}
\begin{align}
\kappa_{\,2} \, \pd{u_{\,2}}{\chi_{\,2}}\,\biggr|_{\,\chi_{\,2} \egal 0} & \egal \pd{u_{\,1}}{\chi_{\,1}}\,\biggr|_{\,\chi_{\,1} \egal 1} \,, \label{eq:sa_IC_L} \\[4pt]
u(\,\chi_{\,2} \egal 1  \,,\,\tau\,) & \egal 0 \,,  \label{eq:sa_BC_R}
\end{align}
\end{subequations}
for $\Omega_{\,x}^{\,2}\,$. 

\noindent \textbf{Remark:} The initial condition~\eqref{eq:initial_condition} must verify the interface conditions. More precisely, functions $f_{\,1}$ and $f_{\,2}$ verify:
\begin{align*}
f_{\,1}(\,\chi_{\,1} \egal 1 \,) & \egal f_{\,2}(\,\chi_{\,2} \egal 0 \,) \,, \qquad 
\pd{f_{\,1}}{\chi_{\,1}}\,\biggl|_{\,\chi_{\,1} \egal 1}  \egal \kappa_{\,2} \, \pd{f_{\,2}}{\chi_{\,2}}\,\biggl|_{\,\chi_{\,2} \egal 0} \,.
\end{align*}
\begin{flushright}
$\square$
\end{flushright}

\subsection{Decomposed solution}

We search the solutions according to the following form: 
\begin{align}
\label{eq:decomposed_solution}
u_{\,i}(\,\chi_{\,i}\,,\,\tau\,) \egal \sum_{\,n\egal 1}^{\,\infty} c_{\,n} \, \phi_{\,n}^{\,i}(\,\chi_{\,i}\,) \, \exp \bigl(\, - \Fo_{\,i} \, \beta_{\,i\,n}^{\,2} \, \tau \,\bigr) \,, \qquad i \egal \bigl\{\, 1 \,,\,2 \,\bigr\} \,, 
\end{align}
with
\begin{align*}
\phi_{\,n}^{\,i}(\,\chi_{\,i}\,) \egal A_{\,i\,n} \, \cos (\, \beta_{\,i\,n} \, \chi_{\,i}\,) \plus B_{\,i\,n} \, \sin (\, \beta_{\,i\,n} \, \chi_{\,i}\,) \,.
\end{align*}

\subsubsection{Transcendental equation}

Using this assumed solution, Eq.~\eqref{eq:sa_IC_R} gives: 
\begin{align*}
c_{\,n} \, \phi_{\,n}^{\,1}(\,1\,) \, \exp \bigl(\, - \Fo_{\,1} \, \beta_{\,1\,n}^{\,2} \, \tau \,\bigr) 
\egal
c_{\,n} \, \phi_{\,n}^{\,2}(\,0\,) \, \exp \bigl(\, - \Fo_{\,2} \, \beta_{\,2\,n}^{\,2} \, \tau \,\bigr) \,, 
\end{align*}
which holds $\forall \, \tau \, \geqslant \, 0\,$. Thus, it implies that
\begin{align*}
- \Fo_{\,1} \, \beta_{\,1\,n}^{\,2} \egal - \Fo_{\,2} \, \beta_{\,2\,n}^{\,2} \,, 
\end{align*}
and therefore
\begin{align*}
\beta_{\,2\,n} \egal \overline{\Fo} \, \beta_{\,1\,n} \,, \qquad \overline{\Fo} \egal \sqrt{\frac{\Fo_{\,1}}{\Fo_{\,2}}}\,.
\end{align*}
Then, using Eq.~\eqref{eq:sa_BC_L}, \eqref{eq:sa_BC_R}, \eqref{eq:sa_IC_L} and \eqref{eq:sa_IC_R} in this specific order, we have:
\begin{subequations}
\label{eq:system_Ai_Bi}
\begin{align}
A_{\,1\,n} & \egal 0  \,, \\[4pt]
A_{\,2\,n} \, \cos(\,  \overline{\Fo} \, \beta_{\,1\,n} \,) \plus B_{\,2\,n} \, \sin(\, \overline{\Fo} \, \beta_{\,1\,n} \,) & \egal 0 \,, \label{eq:system_Ai_Bi_2} \\[4pt]
\kappa_{\,2} \, \overline{\Fo} \, \beta_{\,1\,n} \, B_{\,2\,n} \plus 
A_{\,1\,n} \,  \beta_{\,1\,n} \,\sin(\, \beta_{\,1\,n} \,) \moins 
B_{\,1\,n} \,  \beta_{\,1\,n} \,\cos(\, \beta_{\,1\,n} \,) & \egal 0 \,, \\[4pt]
A_{\,1\,n} \, \cos(\, \beta_{\,1\,n} \,) \plus B_{\,1\,n} \, \sin(\, \beta_{\,1\,n} \,) \moins A_{\,2\,n} & \egal 0 \,, 
\end{align}
\end{subequations}
System of Eq.~\eqref{eq:system_Ai_Bi} can be written in the following form:
\begin{align*}
M \cdot 
\begin{bmatrix}
A_{\,1\,n} \\[4pt]
B_{\,1\,n} \\[4pt]
A_{\,2\,n} \\[4pt]
B_{\,2\,n} 
\end{bmatrix} \egal \boldsymbol{0} \,,
\end{align*}
with
\begin{align*}
M \egal
\begin{bmatrix}
1 & 0 & 0 & 0 \\[4pt]
0 & 0 & \cos(\,  \overline{\Fo} \, \beta_{\,1\,n} \,) & \sin(\,  \overline{\Fo} \, \beta_{\,1\,n} \,) \\[4pt]
\beta_{\,1\,n} \,\sin(\, \beta_{\,1\,n} \,)
& - \, \beta_{\,1\,n} \,\cos(\, \beta_{\,1\,n} \,) 
& 0 
& \kappa_{\,2} \, \overline{\Fo} \, \beta_{\,1\,n} \\[4pt]
\cos(\, \beta_{\,1\,n} \,) &
\sin(\, \beta_{\,1\,n} \,) &
-\,1 & 0 
\end{bmatrix} \,.
\end{align*}
To ensure that Eqs.~\eqref{eq:system_Ai_Bi} has a unique solution, the determinant of matrix $M$ needs to be null. It gives the following condition:
\begin{align*}
\beta_{\,1\,n} \, \cos(\, \beta_{\,1\,n} \,) \, \sin(\, \overline{\Fo} \, \beta_{\,1\,n} \,) 
\plus
\kappa_{\,2} \, \overline{\Fo} \, \beta_{\,1\,n} 
\sin(\, \beta_{\,1\,n} \,) \, \cos(\, \overline{\Fo}\,  \beta_{\,1\,n} \,) 
\egal 0 \,,
\end{align*}
which gives the transcendental equation:
\begin{align}
\label{eq:transcendental_eq}
\tan(\, \overline{\Fo} \, \beta_{\,1\,n} \,) 
\plus
\kappa_{\,2} \, \overline{\Fo} \, 
\tan(\, \beta_{\,1\,n} \,) 
\egal 0 \,.
\end{align}
Eq.~\eqref{eq:transcendental_eq} is solved using a numerical approach by searching the roots for each interval of periodicity $\displaystyle  \biggl[\,\frac{-\,\pi}{2} \,,\, \frac{\,\pi}{2}\,\biggr] \plus n\,\pi$ and $\displaystyle \biggl[\,\frac{-\,\pi}{2\,\overline{\Fo}} \,,\, \frac{\,\pi}{2\,\overline{\Fo}}\,\biggr] \plus n\,\pi\,$.

\subsubsection{Space functions coefficients}

From the system of Eqs.~\eqref{eq:system_Ai_Bi}, it can be obtained that: 
\begin{align*}
A_{\,1\,n}  \egal 0 \,, \qquad
B_{\,1\,n}  \egal \frac{A_{\,2\,n}}{\sin(\, \beta_{\,1\,n} \,)} \,,\qquad
B_{\,2\,n}  \egal \frac{A_{\,2\,n}}{\kappa_{\,2} \, \overline{\Fo} } \ \frac{1}{\tan(\, \beta_{\,1\,n} \,)} \,, 
\end{align*}
Substituting the coefficients $B_{\,2\,n}$ into Eq.~\eqref{eq:system_Ai_Bi_2}, it yields to:
\begin{align*}
A_{\,2\,n} \, \cos(\,  \overline{\Fo} \, \beta_{\,1\,n} \,) \plus \frac{A_{\,2\,n}}{\kappa_{\,2} \, \overline{\Fo} } \ \frac{1}{\tan(\, \beta_{\,1\,n} \,)} \, \sin(\, \overline{\Fo} \, \beta_{\,1\,n} \,) & \egal 0 \,,
\end{align*}
which can be reformulated as:
\begin{align*}
A_{\,2\,n} \, \cos(\,  \overline{\Fo} \, \beta_{\,1\,n} \,)  \, 
\Biggl(\,
 1 \plus \frac{\tan(\,\overline{\Fo} \, \beta_{\,1\,n} \,)}
 {\kappa_{\,2} \, \overline{\Fo} \, \tan(\,\beta_{\,1\,n} \,)}
\,\Biggr) & \egal 0 \,,
\end{align*}
where $\displaystyle 1 \plus \frac{\tan(\,\overline{\Fo} \, \beta_{\,1\,n} \,)} {\kappa_{\,2} \, \overline{\Fo} \, \tan(\,\beta_{\,1\,n} \,)} \egal 0$ according to the transcendental equation~\eqref{eq:transcendental_eq}. Thus, the choice of the coefficient $A_{\,2\,n}$ is free. Note that the choice is free since the coefficient $c_{\,n}$ is here to verify the initial condition. 

\subsubsection{Initial condition coefficients}

It remains to determine the coefficients $c_{\,n}$ in Eq.~\eqref{eq:decomposed_solution}, which enables to verify the initial condition. At $\tau \egal 0\,$, we have:
\begin{align}
\label{eq:decomposed_solution_initial_condition}
\sum_{\,i\egal 1}^{\,2} \, \sum_{\,n\egal 1}^{\,\infty} c_{\,n} \, \phi_{\,n}^{\,i}(\,\chi_{\,i}\,) 
\egal
\sum_{\,i\egal 1}^{\,2} \, f_{\,i} (\,\chi_{\,i}\,) \,.
\end{align}
It is important to note that the basis $\bigl\{\,\phi_{\,n}^{\,i}\,\bigr\}$ is not necessarily orthogonal on $\Omega_{\,x}^{\,1} \, \cup \, \Omega_{\,x}^{\,2}$ \cite{ozisik_boundary_1989,tittle_boundary_1965}. Thus, we define a modified basis:
\begin{align*}
\varphi_{\,n}(\,\chi_{\,1}\,,\,\chi_{\,2}\,)  \egal w_{\,1} \, \phi_{\,n}^{\,1}(\,\chi_{\,1}\,)  \plus w_{\,2} \, \phi_{\,n}^{\,2}(\,\chi_{\,2}\,)  \,,
\end{align*}
which is orthogonal if the following condition is verified:
\begin{align*}
w_{\,2} \moins \overline{\Fo}^{\,2} \, \kappa_{\,2} \, w_{\,1} \egal 0 \,.
\end{align*}
Following \textsc{Tittle}'s results, the weight $w_{\,i}$ are chosen as:
\begin{align*}
w_{\,1} \egal 1 \,, \qquad w_{\,2} \egal  \sqrt{\, \frac{c_{\,2} \, (1 \moins \ell_{\,2})}{c_{\,1} \, \ell_{\,2}} \, } \,.
\end{align*}
Then, each function $f_{\,i}$ is decomposed according to:
\begin{align*}
f_{\,i}(\,\chi_{\,i}\,)  \egal \sum_{\,m \egal 1}^{\,\infty} d_{\,i\,m} \, \varphi_{\,n}(\,\chi_{\,1}\,,\,\chi_{\,2}\,) \,,  \qquad i \egal \bigl\{\, 1 \,,\,2 \,\bigr\} \,, 
\end{align*}
which, by construction  of $f_{\,i}\,$, is equivalent to:
\begin{align}
\label{eq:approximation_fi}
f_{\,i}(\,\chi_{\,i}\,)  \egal \sum_{\,m \egal 1}^{\,\infty} d_{\,i\,m} \, w_{\,i} \, \phi_{\,m}^{\,i}(\,\chi_{\,i}\,) \,,  \qquad i \egal \bigl\{\, 1 \,,\,2 \,\bigr\} \,.
\end{align} 
Then, the coefficients $d_{\,i\,m}$ are given by:
\begin{align*}
d_{\,i\,m} & \egal
\int_{\,0}^{\,1} w_{\,i} \, f_{\,i}(\,\chi_{\,i}\,) \, \phi_{\,m}^{\,i}(\,\chi_{\,i}\,) \, \mathrm{d}\chi_{\,i} \, 
\Biggl(\, \sum_{\,i\egal 1}^{\,2}
\int_{\,0}^{\,1} w_{\,i}^{\,2} \, \phi_{\,m}^{\,i}(\,\chi_{\,i}\,)^{\,2} \, \mathrm{d}\chi_{\,i}
\,\Biggr)^{\,-1} \,.
\end{align*}
Using Eqs.~\eqref{eq:decomposed_solution_initial_condition} and \eqref{eq:approximation_fi} we obtain that:
\begin{align*}
\sum_{\,i\egal 1}^{\,2} \, \sum_{\,n\egal 1}^{\,\infty} c_{\,n} \, \phi_{\,n}^{\,i}(\,\chi_{\,i}\,) 
\egal
\sum_{\,i\egal 1}^{\,2} \, \sum_{\,m\egal 1}^{\,\infty} \, d_{\,i\,m} \, w_{\,i} \, \phi_{\,m}^{\,i}(\,\chi_{\,i}\,) \,.
\end{align*}
By identification term by term the coefficients $c_{\,n}$ are:
\begin{align*}
c_{\,n} & = \
\sum_{\,i\egal 1}^{\,2} \int_{\,0}^{\,1} w_{\,i}^{\,2} \, f_{\,i}(\,\chi_{\,i}\,) \, \phi_{\,n}^{\,i}(\,\chi_{\,i}\,) \, \mathrm{d}\chi_{\,i} \, 
\Biggl(\, \sum_{\,i\egal 1}^{\,2}
\int_{\,0}^{\,1} w_{\,i}^{\,2} \, \phi_{\,n}^{\,i}(\,\chi_{\,i}\,)^{\,2} \, \mathrm{d}\chi_{\,i}
\,\Biggr)^{\,-1} \,.
\end{align*}

\bibliographystyle{unsrt}  
\bibliography{bib_exported2}

\begin{thebibliography}{10}

\bibitem{us_energy_information_administration_annual_2021}
U.S. Energy~Information Administration.
\newblock {\em Annual {Energy} {Outlook} 2021}.
\newblock Washington, eia edition, 2021.

\bibitem{mendes_numerical_2019}
N.~Mendes, M.~Chhay, J.~Berger, and D.~Dutykh.
\newblock {\em Numerical {Methods} for {Diffusion} {Phenomena} in {Building}
  {Physics}: {A} {Practical} {Introduction}}.
\newblock Springer International Publishing, 2019.

\bibitem{mirzaei_recent_2015}
P.~A. Mirzaei.
\newblock Recent challenges in modeling of urban heat island.
\newblock {\em Sustainable Cities and Society}, 19:200--206, 2015.

\bibitem{Delgado_2010}
J.M.P.Q. Delgado, Nuno M.~M. Ramos, E.~Barreira, and V.~P. de~Freitas.
\newblock A critical review of hygrothermal models used in porous building
  materials.
\newblock {\em Journal of Porous Media}, 13(3):221--234, 2010.

\bibitem{tariku_transient_2010}
F~Tariku, K.~Kumaran, and P.~Fazio.
\newblock Transient model for coupled heat, air and moisture transfer through
  multilayered porous media.
\newblock {\em International Journal of Heat and Mass Transfer},
  53(15):3035--3044, 2010.

\bibitem{mendes_method_2005}
N.~Mendes and P.~C. Philippi.
\newblock A method for predicting heat and moisture transfer through
  multilayered walls based on temperature and moisture content gradients.
\newblock {\em International Journal of Heat and Mass Transfer}, 48(1):37--51,
  2005.

\bibitem{kalagasidis_international_2007}
A.~Kalagasidis, P.~Weitzmann, T.~R. Nielsen, R.~Peuhkuri, C.E. Hagentoft, and
  C.~Rode.
\newblock The {International} {Building} {Physics} {Toolbox} in {Simulink}.
\newblock {\em Energy and Buildings}, 39(6):665--674, 2007.

\bibitem{MAIA_2021}
J.~Maia, M.~Pedroso, N.M.M. Ramos, P.F. Pereira, I.~Flores-Colen, M.~Glória
  Gomes, and L.~Silva.
\newblock Hygrothermal performance of a new thermal aerogel-based render under
  distinct climatic conditions.
\newblock {\em Energy and Buildings}, 243:111001, 2021.

\bibitem{Guimaraes_2017}
A.~S. Guimarães, I.~M. Ribeiro, and V.~P. de~Freitas.
\newblock A tool to predict water absorption in porous building materials.
\newblock {\em Journal of Porous Media}, 20(2):127--141, 2017.

\bibitem{berger_comparison_2020}
J.~Berger, S.~Gasparin, D.~Dutykh, and N.~Mendes.
\newblock On the comparison of three numerical methods applied to building
  simulation: {Finite}-differences, {RC} circuit approximation and a spectral
  method.
\newblock {\em Building Simulation}, 13(1):1--18, 2020.

\bibitem{berger_evaluation_2019-1}
J.~Berger and D.~Dutykh.
\newblock Evaluation of the reliability of building energy performance models
  for parameter estimation.
\newblock {\em Journal Computational Technologies}, 3(24), 2019.

\bibitem{berger_review_2017}
J.~Berger, N.~Mendes, S.~Guernouti, M.~Woloszyn, and F.~Chinesta.
\newblock Review of {Reduced} {Order} {Models} for {Heat} and {Moisture}
  {Transfer} in {Building} {Physics} with {Emphasis} in {PGD} {Approaches}.
\newblock {\em Archives of Computational Methods in Engineering},
  24(3):655--667, 2017.

\bibitem{berger_estimation_2020}
J.~Berger and B.~Kadoch.
\newblock Estimation of the thermal properties of an historic building wall by
  combining modal identification method and optimal experiment design.
\newblock {\em Building and Environment}, 185:107065, 2020.

\bibitem{Petit_1997}
D.~Petit, R.~Hachette, and D.~Veyret.
\newblock {A Modal Identification Method To Reduce A High-Order Model:
  Application To Heat Conduction Modelling}.
\newblock {\em International Journal of Simulation Modelling}, 17(4):242--250,
  1997.

\bibitem{Girault_2003}
M.~Girault, D.~Petit, and E.~Videcoq.
\newblock {The Use of Model Reduction and Function Decomposition for
  Identifying Boundary Conditions of A Linear Thermal System}.
\newblock {\em Inverse Problems in Engineering}, 11(5):425--455, 2003.

\bibitem{Girault_2004a}
M.~Girault and D.~Petit.
\newblock {Resolution of linear inverse forced convection problems using model
  reduction by the Modal Identification Method: application to turbulent flow
  in parallel-plate duct}.
\newblock {\em International Journal of Heat and Mass Transfer},
  47(17-18):3909--3925, 2004.

\bibitem{Girault2005a}
M.~Girault and D.~Petit.
\newblock {Identification methods in nonlinear heat conduction. Part I: Model
  reduction}.
\newblock {\em International Journal of Heat and Mass Transfer},
  48(1):105--118, 2005.

\bibitem{Girault2005b}
M.~Girault and D.~Petit.
\newblock Identification methods in nonlinear heat conduction. part ii: inverse
  problem using a reduced model.
\newblock {\em International Journal of Heat and Mass Transfer}, 48(1):119 --
  133, 2005.

\bibitem{Gerardin_2017}
J.~G{\'e}rardin, M.H. Aumeunier, M.~Firdaouss, J.L. Gardarein, and F.~Rigollet.
\newblock {R{\'e}duction de mod{\`e}le thermique par M{\'e}thode
  d'Identification Modale (MIM) pour d{\'e}terminer la temp{\'e}rature de
  surface des composants de machine de fusion}.
\newblock In {\em {Congr{\`e}s Fran{\c c}ais de Thermique 2017}}, volume~1 of
  {\em Actes du Congr{\`e}s Fran{\c c}ais de Thermique 2017, ISBN :
  978-2-905267-92-4}, pages 567--574. {IUSTI \& les laboratoires de la
  f{\'e}d{\'e}ration Fabri de Peiresc, Aix-Marseille Universit{\'e}}, 2017.

\bibitem{berger_proper_2015}
J.~Berger, M.~Chhay, S.~Guernouti, and M.~Woloszyn.
\newblock Proper generalized decomposition for solving coupled heat and
  moisture transfer.
\newblock {\em Journal of Building Performance Simulation}, 8(5):295--311,
  2015.

\bibitem{berger_2d_2016}
J.~Berger, W.~Mazuroski, N.~Mendes, S.~Guernouti, and M.~Woloszyn.
\newblock {2D} whole-building hygrothermal simulation analysis based on a {PGD}
  reduced order model.
\newblock {\em Energy and Buildings}, 112:49--61, 2016.

\bibitem{gasparin_spectral_2019}
S.~Gasparin, D.~Dutykh, and Nathan Mendes.
\newblock A spectral method for solving heat and moisture transfer through
  consolidated porous media.
\newblock {\em International Journal for Numerical Methods in Engineering},
  117(11):1143--1170, 2019.

\bibitem{berger_evaluating_2019}
J.~Berger, S.~Guernouti, and M.~Woloszyn.
\newblock Evaluating model reduction methods for heat and mass transfer in
  porous materials: proper orthogonal decomposition and proper generalized
  decomposition.
\newblock {\em Journal of Porous Media}, 22(3), 2019.
\newblock Publisher: Begel House Inc.

\bibitem{Hou_2020}
T.~Hou, S.~Roels, and H.~Janssen.
\newblock Model order reduction for efficient deterministic and probabilistic
  assessment of building envelope thermal performance.
\newblock {\em Energy and Buildings}, 226:110366, 2020.

\bibitem{hou_poddeim_2020}
T.~Hou, K.~Meerbergen, S.~Roels, and H~Janssen.
\newblock {POD}–{DEIM} model order reduction for nonlinear heat and moisture
  transfer in building materials.
\newblock {\em Journal of Building Performance Simulation}, 13(6):645--661,
  2020.

\bibitem{berger_innovative_2017}
J.~Berger and N.~Mendes.
\newblock An innovative method for the design of high energy performance
  building envelopes.
\newblock {\em Applied Energy}, 190:266--277, 2017.

\bibitem{azam_mixed_2018}
M.-H. Azam, S.~Guernouti, M.~Musy, J.~Berger, P.~Poullain, and A.~Rodler.
\newblock A mixed {POD}–{PGD} approach to parametric thermal impervious soil
  modeling: {Application} to canyon streets.
\newblock {\em Sustainable Cities and Society}, 42:444--461, 2018.

\bibitem{berger_intelligent_2018}
J.~Berger, W.~Mazuroski, R.~C.L.F. Oliveira, and N.~Mendes.
\newblock Intelligent co-simulation: neural network vs. proper orthogonal
  decomposition applied to a {2D} diffusive problem.
\newblock {\em Journal of Building Performance Simulation}, 11(5):568--587,
  2018.

\bibitem{mosquera_generalization_2021}
R.~Mosquera, A.~El~Hamidi, A.~Hamdouni, and A.~Falaize.
\newblock Generalization of the {Neville}–{Aitken} interpolation algorithm on
  {Grassmann} manifolds: {Applications} to reduced order model.
\newblock {\em International Journal for Numerical Methods in Fluids},
  93(7):2421--2442, 2021.

\bibitem{mosquera_pod_2019}
R.~Mosquera, A.~Hamdouni, A.~Hamidi, and C.~Allery.
\newblock {POD} basis interpolation via {Inverse} {Distance} {Weighting} on
  {Grassmann} manifolds.
\newblock {\em Discrete \& Continuous Dynamical Systems - S}, 12(6):1743, 2019.

\bibitem{freno_using_2013}
B.~A. Freno, T.~A. Brenner, and P.~G.~A Cizmas.
\newblock Using proper orthogonal decomposition to model off-reference flow
  conditions.
\newblock {\em International Journal of Non-Linear Mechanics}, 54:76--84, 2013.

\bibitem{oulghelou_parametric_2021}
M.~Oulghelou, C.~Allery, and R.~Mosquera.
\newblock Parametric {Reduced} {Order} models based on a {Riemannian}
  {Barycentric} {Interpolation}.
\newblock {\em International Journal for Numerical Methods in Engineering},
  n/a(n/a), 2021.

\bibitem{amsallem_interpolation_2008}
D~Amsallem and C~Farhat.
\newblock Interpolation {Method} for {Adapting} {Reduced}-{Order} {Models} and
  {Application} to {Aeroelasticity}.
\newblock {\em AIAA Journal}, 46(7):1803--1813, 2008.

\bibitem{gasparin_adaptive_2018}
S.~Gasparin, J.~Berger, D.~Dutykh, and N.~Mendes.
\newblock An adaptive simulation of nonlinear heat and moisture transfer as a
  boundary value problem.
\newblock {\em International Journal of Thermal Sciences}, 133:120--139, 2018.

\bibitem{oreskes_verification_1994}
N.~Oreskes, K.~Shrader-Frechette, and K.~Belitz.
\newblock Verification, {Validation}, and {Confirmation} of {Numerical}
  {Models} in the {Earth} {Sciences}.
\newblock {\em Science}, 263(5147):641--646, 1994.

\bibitem{sreedhar_development_2016}
S.~Sreedhar and K.~P. Biligiri.
\newblock Development of pavement temperature predictive models using
  thermophysical properties to assess urban climates in the built environment.
\newblock {\em Sustainable Cities and Society}, 22:78--85, 2016.

\bibitem{mcadams_heat_1942}
{McAdams}.
\newblock {\em Heat transmission / by {William} {H}. {McAdams},...}
\newblock McGraw-Hill, New York London, 2nd edition revised and enlarged
  edition, 1942.

\bibitem{sacadura_transferts_1980}
J.F. Sacadura.
\newblock {\em Transferts thermiques}.
\newblock Techniques et documentation edition, 1980.

\bibitem{strub_second_2005}
F.~Strub, J.~Castaing-Lasvignottes, M.~Strub, M.~Pons, and F.~Monchoux.
\newblock Second law analysis of periodic heat conduction through a wall.
\newblock {\em International Journal of Thermal Sciences}, 44(12):1154--1160,
  2005.

\bibitem{bejan_advanced_2016}
A.~Bejan and A.~Jones.
\newblock {\em Advanced {Engineering} {Thermodynamics}}.
\newblock Wiley Online, fourth edition, 2016.

\bibitem{berger_dimensionless_2021}
J.~Berger, C.~Legros, and M.~Abdykarim.
\newblock Dimensionless formulation and similarity to assess the main phenomena
  of heat and mass transfer in building porous material.
\newblock {\em Journal of Building Engineering}, 35:101849, 2021.

\bibitem{gasparin_innovative_2019}
S.~Gasparin, J.~Berger, D.~Dutykh, and N.~Mendes.
\newblock An innovative method to determine optimum insulation thickness based
  on non-uniform adaptive moving grid.
\newblock {\em Journal of the Brazilian Society of Mechanical Sciences and
  Engineering}, 41(4):173, 2019.

\bibitem{thomas_fully_1980}
H.~R. Thomas, K.~Morgan, and R.~W. Lewis.
\newblock A fully nonlinear analysis of heat and mass transfer problems in
  porous bodies.
\newblock {\em International Journal for Numerical Methods in Engineering},
  15(9):1381--1393, 1980.

\bibitem{berezin_computing_1965}
I.S. Berezin and N.P. Zhidkov.
\newblock {\em Computing {Methods}. {Volume} 2.}
\newblock Pergamon Press, 1965.

\bibitem{shampine_matlab_1997}
L.~Shampine and M.W. Reichelt.
\newblock The {MATLAB} {ODE} {Suite}.
\newblock {\em SIAM Journal on Scientific Computing}, 18(1):1--22, 1997.
\newblock Publisher: Society for Industrial and Applied Mathematics.

\bibitem{tallet_pod_2015}
A.~Tallet, C.~Allery, and F.~Allard.
\newblock {POD} approach to determine in real-time the temperature distribution
  in a cavity.
\newblock {\em Building and Environment}, 93:34--49, 2015.

\bibitem{tallet_fast_2017}
A.~Tallet, E.~Liberge, and C.~Inard.
\newblock Fast {POD} method to evaluate infiltration heat recovery in building
  walls.
\newblock {\em Building Simulation}, 10(1):111--121, 2017.

\bibitem{sempey_fast_2009}
A.~Sempey, C.~Inard, C.~Ghiaus, and C.~Allery.
\newblock Fast simulation of temperature distribution in air conditioned rooms
  by using proper orthogonal decomposition.
\newblock {\em Building and Environment}, 44(2):280--289, 2009.

\bibitem{sirovich_turbulence_1987}
L.~Sirovich.
\newblock Turbulence and the dynamics of coherent structures. {I} - {Coherent}
  structures. {II} - {Symmetries} and transformations. {III} - {Dynamics} and
  scaling.
\newblock {\em Quarterly of Applied Mathematics}, 45:573--582, 1987.

\bibitem{holmes_turbulence_1996}
P.~Holmes, J.~L. Lumley, and G.~Berkooz.
\newblock {\em Turbulence, {Coherent} {Structures}, {Dynamical} {Systems} and
  {Symmetry}}.
\newblock Cambridge {Monographs} on {Mechanics}. Cambridge University Press,
  Cambridge, 1996.

\bibitem{meyer_time-discrete_2015}
Gunter~H Meyer.
\newblock {\em The {Time}-{Discrete} {Method} of {Lines} for {Options} and
  {Bonds}: {A} {PDE} {Approach}}.
\newblock WORLD SCIENTIFIC, 2015.

\bibitem{kierzenka_bvp_2001}
J.~Kierzenka and L.~Shampine.
\newblock A {BVP} solver based on residual control and the {Maltab} {PSE}.
\newblock {\em ACM Transactions on Mathematical Software}, 27(3):299--316,
  2001.

\bibitem{oulghelou_non-intrusive_2020}
M.~Oulghelou and C.~Allery.
\newblock Non-intrusive reduced genetic algorithm for near-real time flow
  optimal control.
\newblock {\em International Journal for Numerical Methods in Fluids},
  92(9):1118--1134, 2020.

\bibitem{oulghelou_non_2021}
M.~Oulghelou and C.~Allery.
\newblock Non intrusive method for parametric model order reduction using a
  bi-calibrated interpolation on the {Grassmann} manifold.
\newblock {\em Journal of Computational Physics}, 426:109924, 2021.

\bibitem{oulghelou_fast_2018}
M.~Oulghelou and C.~Allery.
\newblock A fast and robust sub-optimal control approach using reduced order
  model adaptation techniques.
\newblock {\em Applied Mathematics and Computation}, 333:416--434, 2018.

\bibitem{asme_vv_asme_2009}
ASME V\&V and Committee Members.
\newblock {\em {ASME} {V}\&{V} 20-2009 {Standard} for {Verification} and
  {Validation} in {Computational} {Fluid} {Dynamics} and {Heat} {Transfer}
  ({V}\&{V20} {Committee} {Chair} and principal author)}, volume~20.
\newblock New York, american society of mechanical engineering edition, 2009.

\bibitem{driscoll_chebfun_2014}
T.~A. Driscoll, N.~Hale, and L.~N. Trefethen.
\newblock {\em Chebfun {Guide} » {Chebfun}}.
\newblock Oxford, UK, pafnuty publications edition, 2014.

\bibitem{machard_methodology_2020}
A.~Machard, C.~Inard, J.M. Alessandrini, C.~Pelé, and J.~Ribéron.
\newblock A {Methodology} for {Assembling} {Future} {Weather} {Files}
  {Including} {Heatwaves} for {Building} {Thermal} {Simulations} from the
  {European} {Coordinated} {Regional} {Downscaling} {Experiment}
  ({EURO}-{CORDEX}) {Climate} {Data}.
\newblock {\em Energies}, 13(13):3424, 2020.

\bibitem{evseev_assessment_2009}
E.~G. Evseev and A.~I. Kudish.
\newblock The assessment of different models to predict the global solar
  radiation on a surface tilted to the south.
\newblock {\em Solar Energy}, 83(3):377--388, 2009.

\bibitem{hay_study_1979}
J.~E. Hay.
\newblock {\em Study of shortwave radiation on non-horizontal surfaces.},
  volume~79.
\newblock Atmospheric Environment Service, Downsview, canada atmospheric
  environment service edition, 1979.
\newblock OCLC: 816567244.

\bibitem{evangelisti_sky_2019}
L.~Evangelisti, C.~Guattari, and F.~Asdrubali.
\newblock On the sky temperature models and their influence on buildings energy
  performance: {A} critical review.
\newblock {\em Energy and Buildings}, 183:607--625, 2019.

\bibitem{Hawkins_2009}
E.~Hawkins and R.~Sutton.
\newblock The potential to narrow uncertainty in regional climate predictions.
\newblock {\em Bulletin of the American Meteorological Society},
  90(8):1095--1108, 2009.

\bibitem{Giorgi_2019}
F.~Giorgi.
\newblock Thirty years of regional climate modeling: Where are we and where are
  we going next?
\newblock {\em Journal of Geophysical Research: Atmospheres},
  124(11):5696--5723, 2019.

\bibitem{fmiStandard}
FMI-Standard.
\newblock {FMI-Standard} {Functional Mock-up Interface}.
\newblock \url{fmi-standard.org}, 2017.
\newblock Accessed: 2017-07-03.

\bibitem{ozisik_boundary_1989}
M.N. Ozisik.
\newblock {\em Boundary {Value} {Problems} of {Heat} {Conduction}}.
\newblock Dover Publications, New york, dover publications edition, 1989.
\newblock Google-Books-ID: CSwbAQAAIAAJ.

\bibitem{tittle_boundary_1965}
C.~W. Tittle.
\newblock Boundary {Value} {Problems} in {Composite} {Media}:
  {Quasi}‐{Orthogonal} {Functions}.
\newblock {\em Journal of Applied Physics}, 36(4):1486--1488, 1965.

\end{thebibliography}


\end{document}